\pdfoutput=1 
\documentclass[aps,prl,twocolumn,superscriptaddress,amsfont,graphicx,nofootinbib,preprintnumbers,reprint]{revtex4-2}
\usepackage{color,graphicx,epsfig}
\usepackage{ifpdf}
\usepackage{amsmath}
\usepackage{bm}
\usepackage{color}
\usepackage[english]{babel}
\usepackage{graphicx}
\usepackage{amsfonts}
\usepackage{amssymb}
\usepackage{braket}
\usepackage{hyperref}
\usepackage{xspace}
\usepackage{enumerate}
\usepackage{mathrsfs}
\usepackage{dsfont}
\usepackage{soul}

\definecolor{nicered}{rgb}{0.7,0.1,0.1}
\definecolor{nicegreen}{rgb}{0.1,0.5,0.1}
\hypersetup{colorlinks,citecolor= nicegreen,linkcolor= nicered}

\def\beq{\begin{equation}}
\def\eeq{\end{equation}}
\def\bea{\begin{eqnarray}}
\def\eea{\end{eqnarray}}
\def\bsp#1\esp{\begin{split}#1\end{split}}

\newcommand{\rd}{\mathrm{d}}

\newcommand{\eps}{\epsilon}

\newcommand{\ord}{\mathcal{O}}

\newcommand{\nil}{\mathfrak{v}}

\definecolor{NavyBlue}{cmyk}{0.94,0.54,0,0}

\definecolor{AIMagenta}{rgb}{0.75,0,0.55}

\newcommand{\bI}{{\bf{I}}}
\newcommand{\bA}{{\bf{A}}}
\newcommand{\bJ}{{\bf{J}}}

\newcommand{\bP}{{\bf{P}}}
\newcommand{\bU}{{\bf{U}}}
\newcommand{\bM}{{\bf{M}}}
\newcommand{\bN}{{\bf{N}}}
\newcommand{\bQ}{{\bf{Q}}}
\newcommand{\bZ}{{\bf{Z}}}
\newcommand{\bOmega}{{\bf{\Omega}}}
\newcommand{\bx}{{\bf{x}}}
\newcommand{\af}{\mathsf{a}}
\newcommand{\sv}{\textrm{sv}}
\newcommand{\bxb}{\overline{\bf{x}}}

\newcommand{\bPtilde}{\mathbf{\widetilde{P}}}
\newcommand{\bUcan}{\mathbf{\widetilde{U}}}
\newcommand{\bUeqv}{\mathbf{U}}
\newcommand{\bZcan}{\mathbf{\widetilde{Z}}}
\newcommand{\bZeqv}{\mathbf{Z}}
\newcommand{\bReqv}{\mathbf{R}_{\textrm{eqv}}}
\newcommand{\bRcan}{\mathbf{R}_{\textrm{can}}}

\DeclareMathOperator{\dlog}{dlog}
\newcommand{\mot}{\mathfrak{m}}

\newcommand{\UHP}{\mathbb{H}}

\newcommand{\bCcan}{\mathbf{\widetilde{C}}}

\newcommand{\bPhat}{\mathbf{\widehat{P}}}
\newcommand{\dimeps}{\varepsilon}

\begin{document}

\preprint{BONN-TH-2026-22} 
\preprint{UUITP-28/26}

\title{Feynman integrals, coactions and Lie-algebra representations}

\author{Emiel Claasen}
\affiliation{Max-Planck-Institut f\"ur Gravitationsphysik (Albert-Einstein-Institut), Am M\"uhlenberg 1, 14476 Potsdam, Germany}

\author{Mehregan Doroudiani}
\affiliation{School of Physics \& Astronomy, University of Southampton, University Road, SO17 1BJ, Southampton, U.K.}

\author{Claude Duhr}
\affiliation{Bethe Center for Theoretical Physics and Cluster of Excellence ``Color meets Flavor'', Universit\"at Bonn, 53115 Bonn, Germany}

\author{Axel~Kleinschmidt}
\affiliation{Max-Planck-Institut f\"ur Gravitationsphysik (Albert-Einstein-Institut), Am M\"uhlenberg 1, 14476 Potsdam, Germany}

\author{Franca Lippert}
\affiliation{Bethe Center for Theoretical Physics, Universit\"at Bonn, 53115 Bonn, Germany}

\author{Oliver Schlotterer}
\affiliation{Department of Physics and Astronomy, Uppsala University, Box 516, 75120 Uppsala, Sweden}
\affiliation{Department of Mathematics, Centre for Geometry and Physics, Uppsala University, Box 480, 75106 Uppsala, Sweden}

\author{Yoann Sohnle}
\affiliation{Department of Physics and Astronomy, Uppsala University, Box 516, 75120 Uppsala, Sweden}
\affiliation{SLAC National Accelerator Laboratory, Stanford University, Stanford, CA 94309, USA}

\date{\today}

\begin{abstract}
    We outline a link between canonical differential equations of Feynman integrals and matrix representations of flat connections taking values in certain Lie algebras.  Leveraging insights from string theory and properties of periods associated with
    these Lie algebras, we deduce single-valued solutions to the differential equations and coactions for the integrals. We introduce a novel
    form of the differential equations
    which connects with the representation theory of the Lie algebras and may depart from $\dimeps$-factorisation in a controlled way. 
    We illustrate the formalism through relevant examples.
\end{abstract}

\maketitle

\section{Introduction}

\noindent
Feynman integrals are integrals over the internal momenta in a Feynman diagram, and they are a cornerstone of most precision computations in high-energy and gravitational-wave physics. 
They often diverge, and we work in dimensional regularisation~\cite{tHooft:1972tcz,Cicuta:1972jf,Bollini:1972ui}, heuristically defined by integrating in a non-integer number of spacetime dimensions $D=D_0-2\dimeps$, where $D_0$ is an (even) integer. It was realised very recently that the appropriate mathematical framework to study dimensionally-regulated integrals is twisted de Rham theory~\cite{AomotoKita,Mastrolia:2018uzb}. 
The coefficients of the Laurent expansion around $\dimeps=0$ are (numerical) \emph{periods}~\cite{MR1852188,Bogner:2007mn}, i.e., complex numbers obtained via absolutely convergent integrals of an algebraic function over a domain defined by polynomial inequalities. They are also of interest in modern algebraic geometry and number theory. 

The most prominent example of (families of) periods in both mathematics and physics are multiple zeta values (MZVs) and multiple polylogarithms (MPLs)~\cite{Lappo:1953m, Remiddi:1999ew,Goncharov:1998kja,Goncharov:2001iea,Waldschmidt:2002}, with the latter defined as the iterated integrals 
\beq\label{eq:MPL_def}
G_{a_1,\ldots,a_n}(x) = \int_{0}^{x}\frac{\rd t}{t-a_1}\,G_{a_2,\ldots,a_n}(t) \,,
\eeq
where $a_i, x\in\mathbb{C}$. MZVs correspond to $a_i\in\{0,1\}$ and evaluation at $x=1$. 
We always regularise endpoint divergences via tangential base-points \cite{Deligne1989TheGR, Brown:2014pnb, Panzer:2015ida, Abreu:2022mfk}.
There are efficient tools for the numerical evaluation~\cite{Gehrmann:2001pz,Gehrmann:2001jv,Vollinga:2004sn,Buehler:2011ev,Frellesvig:2016ske,Ablinger:2018sat,Naterop:2019xaf} and algebraic manipulation~\cite{Ablinger:2013cf,Duhr:2014woa,Panzer:2014caa,Duhr:2019tlz,Kardos:2025klp,hyperlogprocedures} of MPLs. 
Their symbol and coaction\footnote{More precisely, their motivic uplifts are equipped with a coaction.} calculus ~\cite{Goncharov:2005sla,Goncharov:2001iea,Goncharov:2010jf,Brown:2011ik,brownmixedZ,Duhr:2011zq, Duhr:2012fh} inform direct-integration~\cite{Brown:2008um,Bogner:2014mha,Anastasiou:2013srw,Ablinger:2014yaa,Panzer:2014caa,Kardos:2025klp} and bootstrap methods (see,~e.g.,~\cite{Caron-Huot:2020bkp} and references therein).

The arguably most powerful (indirect) method to compute Feynman integrals is the differential equations method~\cite{Kotikov:1990kg,Kotikov:1991pm,Kotikov:1991hm,Remiddi:1997ny,Gehrmann:1999as} augmented by the idea of a canonical basis~\cite{Henn:2013pwa}, itself inspired by iterated integrals and properties of MPLs, and recently extended beyond MPLs~\cite{Pogel:2022yat,Pogel:2022ken,Pogel:2022vat,Gorges:2023zgv,Duhr:2025lbz,Maggio:2025jel,e-collaboration:2025frv,Bree:2025tug,e-collaboration:2026aup,Forner:2026vby}. Furthermore, many Feynman integrals~\cite{Usyukina:1992jd,Usyukina:1993ch,Dixon:2012yy,Drummond:2012bg,Chavez:2012kn,Drummond:2013nda,Brown:2015ztw,Schnetz:2016fhy,DelDuca:2016lad,Schnetz:2021ebf,Haug:2025ava} evaluate to single-valued MPLs (SVMPLs), which are combinations of holomorphic and antiholomorphic MPLs such that all branch cuts cancel~\cite{brownSVHPLs,brownSV,Schnetz:2013hqa,Brown:2013gia, DelDuca:2016lad,Broedel:2016kls}. 

It has long been known, however, that not all Feynman integrals can be expressed in terms of MPLs (see, e.g.,~ \cite{Bourjaily:2022bwx}). The first example is the two-loop sunrise integral with three massive propagators, whose maximal cut (obtained by putting all propagators on shell) computes the periods of a family of elliptic curves, cf.,~e.g.,~\cite{Sabry,Bauberger:1994by,Laporta:2004rb}. Beyond the maximal cut, the sunrise integral can be expressed in terms of generalisations of polylogarithms to elliptic curves~\cite{Bloch:2013tra,Brown:2011wfj,Broedel:2014vla,Broedel:2017kkb,Broedel:2017siw,Enriquez2023EllipticH}, as well as (in the equal-mass case) iterated integrals of modular forms~\cite{ManinModular,Brown:2014pnb, Broedel:2015hia, Ablinger:2017bjx,Adams:2017ejb,Duhr:2019rrs, Matthes2022,Broedel:2021zij}. 
Now\-adays we know many examples of Feynman integrals that evaluate to functions defined on more complicated geometries, including higher-genus and del Pezzo surfaces, as well as Calabi--Yau and Fano varieties~\cite{Brown:2010bw,Huang:2013kh,Bloch:2014qca,Hauenstein:2014mda,Bloch:2016izu,Primo:2017ipr,Bourjaily:2018ycu,Bourjaily:2018yfy,Bourjaily:2019hmc,Klemm:2019dbm,Bonisch:2020qmm,Bonisch:2021yfw,Pogel:2022yat,Duhr:2022pch,Pogel:2022ken,Pogel:2022vat,Marzucca:2023gto,Duhr:2023eld,Frellesvig:2023bbf,Klemm:2024wtd,Duhr:2024hjf,Jockers:2024uan,Frellesvig:2024zph,Driesse:2024feo,Duhr:2024uid,Frellesvig:2024rea,Schimmrigk:2024xid,Duhr:2025tdf,Duhr:2025ppd,Brammer:2025rqo,Duhr:2025kkq,Pogel:2025bca,Duhr:2025ouy,delaCruz:2025szs,Bargiela:2025vwl,Yang:2026rgb}. 
While 
in principle a general theory for (motivic uplifts of) these periods exists (cf.,~e.g.,~\cite{Brown:2015fyf, Brown:2015ylf}), 
their understanding has not yet reached the same level as for MPLs,
which is a main bottleneck for amplitudes beyond MPLs.

To make progress, it is important to look for other areas where periods arise. The low-energy or $\alpha'$-expansion of string amplitudes cleanly isolates Riemann surfaces of a given genus and offers a particularly convenient laboratory for studying the associated periods~\cite{Berkovits:2022ivl}. At tree level, the $\alpha'$-expansions of open- and closed-string amplitudes solely involve MZVs and their single-valued counterparts, respectively, and showcase the link between single-valued integration and double copy 
\cite{Schlotterer:2012ny, Stieberger:2013wea, Stieberger:2014hba, Schlotterer:2018zce, Brown:2018omk, Vanhove:2018elu, Brown:2019wna, Mafra:2022wml, Baune:2024uwj}.
At loop level, open-string amplitudes suggest organising principles for MPLs on higher-genus surfaces \cite{Broedel:2014vla, Broedel:2017jdo, DHoker:2023khh, DHoker:2025jgb}, whereas closed-string amplitudes introduce non-holomorphic modular forms \cite{DHoker:2015gmr,DHoker:2015wxz, DHoker:2016mwo, DHoker:2017pvk, DHoker:2020uid}, which are expected to be obtained via a single-valued/double-copy construction~\cite{Zerbini:2015rss, DHoker:2015wxz, Broedel:2018izr, Panzertalk, Gerken:2020xfv}.
At one-loop one encounters \emph{equivariant} analogues of iterated Eisenstein integrals \cite{Gerken:2020yii, Dorigoni:2022npe, Dorigoni:2024oft}, which are combinations of holomorphic and anti-holomorphic functions closely related to single-valued periods \cite{Brown:2014pnb, Brown:2017qwo, Brown:2020mhp}.

The main goal of this paper is to initiate a program that builds on synergies between scattering amplitudes in string theory and QFT and that aims at leveraging the advances from one field and to apply it to the other, in particular in the context of iterated integrals and periods.

\section{Canonical differential equations}

\noindent
We consider families of scalar Feynman integrals\footnote{Using tensor reduction we can express Feynman integrals with non-trivial tensor numerators in terms of scalars.} that share the same propagators.  Using integration-by-parts (IBP) relations~\cite{Tkachov:1981wb,Chetyrkin:1981qh} or intersection numbers~\cite{Mastrolia:2018uzb,Frellesvig:2019kgj}, one can find a finite number~\cite{Smirnov:2010hn,Bitoun:2017nre} of basis integrals, typically referred to as \emph{master integrals} (MIs), which we collect in the vector $\bI = (I_1,\ldots, I_N)^T$. The MIs are functions of the \emph{external scales} $\bx=(x_1,\ldots,x_s)$ (i.e., scalar products between external momenta and propagator masses), and they satisfy a first-order linear system of differential equations~\cite{Kotikov:1990kg,Kotikov:1991pm,Kotikov:1991hm,Remiddi:1997ny,Gehrmann:1999as},
\beq\label{eq:DEQ_general}
\rd \bI(\bx,\dimeps) = \bOmega(\bx,\dimeps)\,\bI(\bx,\dimeps)\,,
\eeq
where $\rd = \partial+\overline{\partial}= \sum_{k}(\rd x_k\,\partial_{x_k}+\rd\bar{x}_k\,\partial_{\bar{x}_k})$ is the total differential with respect to the external scales (and their complex conjugates),\footnote{We interpret the external scales $\bf{x}$ as holomorphic variables, and we 
have $\overline{\partial}\bI(\bx,\dimeps)=0=\overline{\partial}\bJ(\bx,\dimeps)$ for the MIs and their gauge transforms in (\ref{eq:canon_DEQ}), but not necessarily for their counterparts (\ref{toeqvg}) in equivariant gauge below.} and $\bOmega(\bx,\dimeps)$ is a matrix of rational functions in $\dimeps$ and rational one-forms in $\bx$ subject to the flatness condition $\rd \bOmega = \bOmega\wedge\bOmega$. 

In order to solve~\eqref{eq:DEQ_general}, it is useful to change to a so-called \emph{canonical basis} $\bJ(\bx,\dimeps) = \bRcan(\bx,\dimeps)\bI(\bx,\dimeps)$. 
In spite of the rapidly growing intuition and arsenal of algorithms to determine suitable rotation matrices $\bRcan(\bx,\dimeps)$ in the case of MPLs~\cite{Henn:2014qga,Lee:2014ioa,Meyer:2016slj,Prausa:2017ltv,Gituliar:2017vzm,Meyer:2017joq,Dlapa:2020cwj,Lee:2020zfb,Henn:2020lye,Chen:2020uyk}, there is no general consensus for
what a canonical basis is beyond MPLs, except that one typically requires the differential equations to be $\dimeps$-factorised,
\beq\label{eq:canon_DEQ}
\rd\bJ(\bx,\dimeps) = \dimeps\,\bA(\bx)\,\bJ(\bx,\dimeps)\,,\quad \bA(\bx) = \sum_i\bA_i \,\omega_i(\bx)\,.
\eeq
We use the operational definition of~\cite{Gorges:2023zgv,Duhr:2025lbz,Forner:2026vby}, in which the one-forms $\omega_i$ have at most logarithmic singularities (locally) and define independent cohomology classes. This is closely related to~\cite{Pogel:2022yat,Pogel:2022ken,Pogel:2022vat,Maggio:2025jel,e-collaboration:2025frv,Bree:2025tug,e-collaboration:2026aup} (see also~\cite{Dlapa:2022wdu,Yang:2025ofz,Chaubey:2025adn,Chaubey:2026egr}).

As an illustrative example, for massless four-point functions (at least up to three loops) the matrix $\bA(\bx)$ takes the form~\cite{Henn:2013pwa,Henn:2016jdu,Henn:2020lye}
\beq\label{eq:box_deq}
\bA(x) = \bA_0\,\dlog x +  \bA_1\,\dlog (1-x)\,,
\eeq
where $\bA_0$ and $\bA_1$ are constant matrices and $x=-s/t$, where $s$ and $t$ are the usual Mandelstam variables.

We consider a fundamental solution matrix $\bPtilde(\bx,\dimeps)$ to~\eqref{eq:canon_DEQ}, which may be written as a path-ordered exponential whose $\dimeps$-expansion involves iterated integrals~\cite{chen1977iterated}, 
\beq\label{eq:Pexp}
\bPtilde(\bx,\dimeps) = \bUcan(\bx,\dimeps)\bN(\bx_0,\dimeps)\,,\quad \bUcan(\bx,\dimeps)=\operatorname{Pexp}\Big[\dimeps\int_{\gamma}\bA\Big]\,,
\eeq
where $\gamma$ is a path from an initial point $\bx_0$ to a generic point $\bx$ in the space of kinematics.
$\bN(\bx_0,\dimeps)$ is the value at the point $\bx_0$,
and encodes a choice of basis of integration cycles (e.g., a basis of cut contours). In the following it will be important that all cycles are genuine geometric twisted cycles \cite{Aomoto:1994}. We assume that $\bN(\bx_0,\dimeps)$ admits an expansion in $\dimeps$ in terms of MPLs evaluated at special values, like (cyclotomic) MZVs.

\section{The main correspondence}

\noindent
In string theory and mathematics, one often studies integrable connections $\nabla$ that take values in some Lie algebra $\nil$. The most prominent example is the universal Knizhnik–Zamolodchikov (KZ) equation~\cite{Knizhnik:1984nr,Etingof:1998}, which in the case of a single variable may be cast in the~form
\beq\label{eq:KZ}
\nabla \mathbb{G}(x) = 0\,,\quad \nabla = \rd - \af_0\,\dlog x - \af_1\,\dlog(1-x)\,.
\eeq
Here $\af_0$ and $\af_1$ are formal non-commutative variables which generate a free Lie algebra. The solution to~\eqref{eq:KZ} is an infinite series of words in the letters $\af_i$ whose coefficients are MPLs~(\ref{eq:MPL_def}),
\beq
\mathbb{G} = 1+\af_0\,G_0 + \af_1\,G_1 + \af_0^2 G_{0,0}+ \af_0\af_1 G_{0,1}+\ldots\,,
\label{MPLser}
\eeq
and we dropped the arguments of the MPLs for readability. 
For generalisations of the KZ connection (\ref{eq:KZ}) to an arbitrary number of points on the sphere \cite{Knizhnik:1984nr, Drinfeld:1991}, to elliptic curves \cite{Levrac, Calaque:2009, EnriquezEllAss, Brown:2011wfj} and to higher-genus curves \cite{Bernard:1988, Enriquez:2011, DHoker:2023vax, Baune:2024biq, DHoker:2026lgg, DHoker:2026mxk}, the Lie algebra 
is no longer free. The flatness of $\nabla$ then relies on non-trivial commutation relations of the braid algebra at genus $g=0$ \cite{jimbo1989introduction, kassel2012quantum} and the Drinfeld--Kohno algebra $\mathfrak{t}_{g,n}$ for $n$-punctured genus $g\geq 1$ surfaces \cite{Bezru:1994, Calaque:2009, Enriquez:2011}. 

By comparing~\eqref{eq:KZ} to~\eqref{eq:canon_DEQ} and~\eqref{eq:box_deq}, we may interpret the canonical differential equation as a finite-dimensional representation $\rho$ of the abstract Lie algebra $\nil$ generated by the $\af_i$, i.e., $\rho(\af_i)=\dimeps\bA_i$. If we extend the representation to the universal enveloping algebra, we
identify $\rho(\mathbb{G}) = \bUcan$ for the series (\ref{MPLser}) in MPLs as a simple example of~$\bUcan$ in (\ref{eq:Pexp}).
This idea is not entirely new, and has already appeared in different forms in the physics literature, cf., e.g.,~\cite{Henn:2013tua}. Here, however, we will go several steps further.

First, the KZ-type connections studied in mathematics and string theory are typically based on \emph{pro-nilpotent} Lie algebras, defined as inverse limits of nilpotent Lie algebras. A Lie algebra $\mathfrak{g}$ is nilpotent if its lower central series, defined inductively by $\mathfrak{g}_{n} = [\mathfrak{g},\mathfrak{g}_{n-1}] $ and $\mathfrak{g}_1=\mathfrak{g}$, terminates. The pro-nilpotent nature of the Lie algebra $\nil$ generated by $\rho(\af_i)=\dimeps\bA_i$ can be seen from the fact that $[\rho(\af_i),\rho(\af_j)]=0\!\!\mod \dimeps^2$. We stress that the $\bA_i$ themselves are not required to be nilpotent, but it is the grading induced by $\dimeps$ that makes this Lie algebra \emph{pro}-nilpotent! We have analysed various canonical differential equations for Feynman integrals associated to modular and/or elliptic curves, and we found that in all cases the connections of the canonical differential equations are genuine representations of the Calaque--Enriquez--Etingof (CEE) connection \cite{Calaque:2009} (or its ellipsi\-tomic generalisation \cite{Calaque:2019ell, Calaque:2020ell, Hopper:2021ell, Hopper:2022ell} in case of non-trivial level structure), and in particular the matrices $\dimeps\bA_i$ satisfy precisely the expected commutation relations.

Second, while the connection matrix $\bOmega$ is always rational in $\bx$, and thus single-valued, the matrix $\bA$ is usually multi-valued. This is due to $\bRcan$ involving the periods of an underlying geometric object. These come with their own monodromy group $S$ that is typically reductive. The matrix $\bA(\bx)$ changes by a gauge transformation if we move $\bx$ along a closed curve $\Gamma$. This is also true in the case of MPLs which involve square root arguments; cf., e.g., $\dlog\frac{1-\sqrt{x}}{1+\sqrt{x}}$ if
$\Gamma$ encircles the origin. The solution matrix changes under analytic continuation as
\beq\label{eq:anal_cont}
\mathcal{M}_\Gamma\big[\bUcan(\bx,\dimeps)\big] = \bZcan_{\Gamma}(\bx,\dimeps)\bUcan(\bx,\dimeps)
\bZcan_{\Gamma}(\bx_0,\dimeps)^{-1}
\bUcan_\Gamma(\dimeps)\,.
\eeq
Here $\mathcal{M}_\Gamma$ is the operator that encodes analytic continuation along $\Gamma$, the matrix $\bZcan_\Gamma$ represents the gauge transformation, and the matrix $\bUcan_\Gamma(\dimeps)$ is defined as in (\ref{eq:Pexp}) with $\Gamma$, anchored at $\bx_0$, in the place of $\gamma$. 
This can be interpreted as follows: 
The monodromy group $G$ of the connection has a pro-unipotent subgroup $U\subset G$ 
associated with the pro-nilpotent Lie algebra $\nil$.
The in general $\bx$-dependent $\bZcan_\Gamma(\bx,\dimeps)$ satisfies the cocycle property for the group $S$. The transformation law of $\bUcan_\Gamma(\dimeps)$ in turn includes transport with the gauge transformations~$\bZcan_\Gamma$ at the endpoints $\bx$ and~$\bx_0$. They can be combined into a representation 
$\bCcan_\Gamma(\bx_0,\dimeps)=\bZcan_\Gamma(\bx_0,\dimeps)^{-1} \bUcan_\Gamma(\dimeps)$
of~$U$. Two examples of these structures are provided in the supplemental material. 
This setup is very reminiscent of the \emph{relative completion} from~\cite{Hain:1998}, which is the mathematical framework for a treatment of iterated integrals of Eisenstein series and/or elliptic MPLs as periods~\cite{Hain:2014, Brown:2014pnb,Brown:2015ylf}. 
These two observations combined allow us to propose that:
\vskip 2mm
\noindent\fbox{%
    \parbox{\dimexpr\linewidth-3\fboxsep-3\fboxrule}{%
Canonical differential equations are finite-dimensional representations of flat connections taking values in pro-nilpotent Lie algebras.  }%
}
\vskip 2mm

This proposal is our main result. 
It establishes a correspondence between canonical differential equations and modern treatments of iterated integrals in mathematics and string theory. 
We stress that we expect this correspondence to hold for all cases where a canonical basis can be identified, and not just in cases related to genus-0 or genus-1 geometries. In particular, in line with the recent proposal in~\cite{Forner:2026vby}, we expect that the functions that solve canonical differential equations extend the definitions of \emph{pure functions} from MPLs~\cite{Arkani-Hamed:2010pyv} and genus-1 functions~\cite{Broedel:2018qkq} to arbitrary geometries.

\section{Consequences of the correspondence}

\noindent
Our correspondence opens the possibility for many new results and applications.

\paragraph{\bf{Equivariant~gauges.}}~The matrix $\bZcan_\Gamma(\bx,\dimeps)$ in the monodromy (\ref{eq:anal_cont}) may depend on both $\bx$ and $\dimeps$. We propose the existence of a distinguished,  \textit{equivariant gauge} defined by 
\beq
\bUeqv(\bx,\dimeps) = \bReqv(\bx,\dimeps)\bUcan(\bx,\dimeps)\bReqv(\bx_0,\dimeps)^{-1}\,,
\label{toeqvg}
\eeq
and characterised by the property that the cocycle 
\beq
\bZeqv_{\Gamma}(\dimeps) = \bReqv(\Gamma\cdot\bx,\dimeps)\bZcan_{\Gamma}(\bx,\dimeps)\bReqv( \bx,\dimeps)^{-1} \, ,
\label{zequiv}
\eeq  
which encodes the monodromy action of the reductive group $S$, is $\bx$-independent and  defines a genuine representation of~$S$ ($\Gamma\cdot\bx$ refers to the image of $\bx$ under transport along~$\Gamma$). 
The price to pay is that the connection in such a gauge may be neither
$\dimeps$-factorised nor meromorphic, but still such that the connection only has logarithmic singularities and its differential forms define independent cohomology classes.
We also require that
the connection still takes values in a pro-nilpotent Lie algebra, preserving that only a finite number of terms contribute to a given order in $\dimeps$ of the MIs.
We have constructed  equivariant gauges for a large number of examples, including integrals that evaluate to MPLs with square-root arguments (where the non-trivial entries of $\bZeqv_{\Gamma}$ are all $\pm1$), integrals attached to families of elliptic curves, and also the three-loop equal-mass banana integral, associated with a K3 geometry. We are not aware of any counterexample, but do not know of a general proof of the existence of an equivariant gauge.

By comparing to the existing literature on MPLs and iterated Eisenstein integrals~\cite{Brown:2014pnb, Brown:2020mhp, Dorigoni:2024oft}, we see that the existence of an equivariant gauge has an intriguing consequence for the initial condition $\bN(\bx_0,\dimeps)$ of (\ref{eq:Pexp}), namely it 
must take the factorised form 
\beq\label{eq:N_fac}
\bN(\bx_0,\dimeps) = \bReqv(\bx_0,\dimeps)^{-1}\bM(\bx_0,\dimeps)\bQ(\bx_0,\dimeps)\,.
\eeq
In the remainder of this work, we shall no longer display the dependence on $\bx_0$ for ease of notation.
The first factor captures the rotation of the initial condition to the equivariant gauge, cf.\ (\ref{toeqvg}). The last factor $\bQ(\dimeps)$ is a Laurent series in $\dimeps$ and satisfies $\overline{\bQ(-\dimeps)}=\bQ(\dimeps)$, where $\bar\dimeps=\dimeps$. Its Laurent coefficients  only involve algebraic numbers and powers of $2\pi i$ (subject to a trivial coaction~\cite{brownmixedZ}). The middle factor $\bM(\dimeps)=\mathds{1}+\mathcal{O}(\dimeps)$ of (\ref{eq:N_fac}) only involves MPLs of algebraic numbers that admit a non-trivial coaction, like $\zeta_{2k+1}$. 
The form of $\bM(\dimeps)$ is highly constrained, and provides matrix representations $\rho(\sigma)$ of the \emph{zeta generators} $\sigma$~\cite{Drinfeld:1991, Furusho2000TheMZ, DG:2005, Brown:depth3, hainmatsumoto2020, Furusho:2020, Dorigoni:2024iyt} including their cyclotomic incarnations \cite{Gonch:cyclo, Glanois:2016, Hopper:2022ell, Hirose:2023}, which are derivations acting on the pro-nilpotent Lie algebra $\nil$ and inherit the representation~$\rho$ of $\nil$.
We note that this implies a \emph{coaction principle} \cite{Brown:2015fyf, Panzer:2016snt}, because the way higher-depth MPLs arise in $\bM(\dimeps)$ is fixed by the primitive elements in the Hopf algebra of (de Rham) MPLs, which conjecturally have depth 1, cf.~e.g.,~\cite{Dupont:2024} (see (\ref{knownco}) below). A detailed derivation of~\eqref{eq:N_fac} will be presented in~\cite{mat:progress}, and we have checked that~\eqref{eq:N_fac} holds for various hypergeometric functions and Feynman integrals for which the full period matrix, which includes also cuts, can be constructed, including various one- and two-loop integrals from~\cite{Abreu:2017mtm,Abreu:2021vhb}. Examples with genus-0 and genus-1 geometries can be found in the supplemental material.

\paragraph{\bf{Single-valued analogues.}} 

In~\cite{Frost:2023stm,Frost:2025lre} an alternative way to define SVMPLs based on the properties of the KZ connection and zeta generators was presented. This approach was also applied at genus one to derive single-valued and equivariant iterated Eisenstein integrals for $\operatorname{SL}_2(\mathbb{Z})$~\cite{Brown:2020mhp,Dorigoni:2024oft} and single-valued elliptic MPLs in one variable \cite{Schlotterer:2025qjv}. Inspired by this, we may write down a single-valued analogue of $\bUeqv$ in equivariant gauge (\ref{toeqvg}): 
\beq\label{eq:Usv}
\bUeqv^{\textrm{sv}}(\bx,\dimeps) = \bUeqv(\bx,\dimeps)\bM^{\sv}(\dimeps)
 \overline{\bUeqv(\bx,-\dimeps)^{-1}} 
\bM^{\sv}(\dimeps)^{-1}\, ,
\eeq
where $\bM^{\sv}(\dimeps)=\bM(\dimeps)\overline{\bM(-\dimeps)^{-1}}$ is obtained from the factor $\bM(\dimeps)$ in (\ref{eq:N_fac}) by replacing all MPLs and MZVs by their single-valued versions, cf.~\cite{Schnetz:2013hqa, Brown:2013gia,brownSVHPLs,brownSV, DelDuca:2016lad,Broedel:2016kls}. Note that $\bQ(\dimeps)$ has canceled in~\eqref{eq:Usv}, i.e., $\bUeqv^{\textrm{sv}}(\bx,\dimeps) = \mathds{1}+\mathcal{O}(\dimeps)$,
and $\bUeqv^{\textrm{sv}}(\bx,\dimeps)$ cannot be holomorphic, due to the explicit appearance of $\bxb$.
Instead, $\bUeqv^{\textrm{sv}} $ is single-valued in the following sense: From~\eqref{eq:anal_cont}, it is easy to check that
\begin{equation}\begin{split}\label{eq:Usv_monodromy}
\mathcal{M}_\Gamma&\big[\bUeqv^{\textrm{sv}}(\bx,\dimeps)\big]=\\
&=\bZeqv_\Gamma(\dimeps)\bUeqv^{\textrm{sv}}(\bx,\dimeps)\bM^{\sv}(\dimeps)
\overline{\bZeqv_\Gamma(-\dimeps)^{-1}}
\bM^{\sv}(\dimeps)^{-1}\,.
\end{split}
\end{equation}
We see that $\bUeqv^{\textrm{sv}}$ still is multi-valued, but $\bUcan_\Gamma$ in (\ref{eq:anal_cont}) cancels. 
This is actually completely analoguous to the situation encountered for MPLs with square root arguments: The single-valued analogue of $\log\frac{1-\sqrt{x}}{1+\sqrt{x}}$ is $\log\big|\frac{1-\sqrt{x}}{1+\sqrt{x}}\big|^2$. This
is single-valued with respect to the (unipotent) monodromy of the logarithm, but still changes sign if we change the branch of the square root.
In this sense, the entries of $\bU^{\sv}$ generalise SVMPLs -- just like the multivalued solutions of general canonical differential equations can be seen as pure functions that generalise MPLs.

We have applied~\eqref{eq:Usv} to construct single-valued solution matrices for the Gauss
hypergeometric function associated to the Legendre family of elliptic curves~\cite{Broedel:2018rwm}, and also for the equal-mass two-loop kite family~\cite{Adams:2018yfj}.\footnote{Two examples of Gauss hypergeometric functions are discussed in the supplemental material.} They can be expanded in terms of iterated Eisenstein integrals of levels 2 and 6, respectively. We have checked that the $\dimeps$-expansions of the corresponding single-valued solutions only involve equivariant iterated Eisenstein integrals, which are examples of non-holomorphic modular forms~\cite{Brown:2017qwo}.
These are the first instances of such equivariant integrals beyond the case of length one \cite{Drewitt:2023con, Duhr:2025lvr} for the congruence subgroups $\Gamma(2)$ and $\Gamma_1(6)$.
A companion paper \cite{lv2:progress} provides a representation-agnostic construction of equivariant iterated Eisenstein integrals for level~2, extending the level-1 construction of~\cite{Brown:2020mhp,Dorigoni:2024oft}.

 \paragraph{\bf{Coactions.}} 
Inspired by the formalism of~\cite{Frost:2023stm,Frost:2025lre,Kleinschmidt:2025dtk} for coactions for motivic MPLs and iterated Eisenstein integrals for the full modular group,
 we propose the coaction on the (motivic version of) $\bUeqv$ to be
 \beq\label{eq:coaction}
 \Delta\big( \bUeqv^{\mot}(\bx,\dimeps)\big) = \bUeqv^{\nil}(\bx,\dimeps)\bM^{\nil}(\dimeps) \bUeqv^{\mot}(\bx,\dimeps)\bM^{\nil}(\dimeps)^{-1}\,.
 \eeq
 Here, the superscript `$\mot$' indicates that the iterated integrals in $\bUeqv^{\mot}$ should be understood as the \emph{motivic uplifts} of the periods rather than numbers, assuming that these uplifts exist. Similarly, the superscript `$\nil$' refers to de Rham periods for the pro-unipotent group $U$, and we use the  shorthand for the tensor product 
\beq\label{eq:XuYm}
X^{\mot}Y^{\nil} = Y^{\nil}X^{\mot} \equiv X^{\mot}\otimes Y^{\nil}\,.
\eeq
Motivic periods can be represented by integrals using the period map, while for de Rham periods there is in general no such map. 
Instead, de Rham periods can be represented by single-valued integrals,
\beq
\label{utosv}
\bUeqv^{\nil} \to \bUeqv^{\sv}\,,\quad \bM^{\nil} \to \bM^{\sv}\,,
\eeq
see~\cite{Brown:2015ylf, Brown:2018omk} for
a detailed account. 
We expect the coaction (\ref{eq:coaction}) to commute with expanding in $\dimeps$, which yields coactions (for the pro-unipotent group $U$) on motivic periods order by order in~$\dimeps$.
This matches the behaviour observed for the coaction  in~\cite{Schlotterer:2012ny, Drummond:2013vz, Abreu:2017enx,Abreu:2017mtm,Abreu:2019wzk,Britto:2021prf,Abreu:2021vhb, Kamlesh:2024cvo}, and we expect this to find some explanation in the context of the theory of Mellin motives currently being developed~\cite{Mellin:progress}, similar to the case of hypergeometric functions proven in~\cite{Brown:2019jng,Kamlesh:2024cvo}. 
At each order in $\dimeps$, it matches the known coaction formula for MPLs
\cite{Goncharov:2001iea, Goncharov:2005sla, Brown:2011ik, brownmixedZ,Frost:2023stm,Frost:2025lre} and otherwise entails proposals for coactions of iterated integrals beyond genus zero (as in \cite{Kleinschmidt:2025dtk} for Eisenstein integrals of ${\rm SL}_2(\mathbb Z)$).\footnote{Also see \cite{Broedel:2018iwv, Wilhelm:2022wow, Forum:2022lpz} for earlier coaction proposals beyond MPLs.} Following our earlier comments, the factor of $\bM(\dimeps)$ obeys 
\beq
\Delta \bM^{\mot}(\dimeps) = \bM^{\nil}(\dimeps) \bM^{\mot}(\dimeps)\,,
\label{knownco}
\eeq
which explains why the entire information about the expansion of $\bM^{\mot}(\dimeps)$ is encoded in the primitive elements.

One can check that~\eqref{eq:coaction} has all the properties expected from the coaction. 
Since $\bUeqv^{\mot}(\bx,\dimeps)$ and $\bUeqv^{\nil}(\bx,\dimeps)$ are taken to obey the same differential equations (see also
sec.\ 7.4 of \cite{Brown:2015ylf}), total derivatives ${\rm d}$ only act in the de Rham~slot, 
\beq
\Delta \, {\rm d} = 
(\mathds{1}\otimes {\rm d})\Delta\,, 
\eeq
just like in the case of MPLs~\cite{Duhr:2012fh,Brown:2015ylf}. Instead, analytic continuation acts in both tensor slots,
\beq\label{eq:DeltaMC}
\Delta\mathcal{M}_{\Gamma} = (\mathcal{M}_\Gamma\otimes \mathcal{M}_\Gamma)\Delta\,,
\eeq
where $\mathcal{M}_\Gamma$ acts in the de Rham slot 
as in \eqref{eq:Usv_monodromy} with $^{\rm sv}$ replaced by $^{\nil}$ on both sides. 
At first sight, this seems to be at odds with the known behaviour of the coaction on MPLs, where monodromy only acts in the motivic (first) slot~\cite{Duhr:2012fh,Brown:2015ylf}. There is no contradiction:
for rational arguments, the second entry is fully single-valued,
but in general it depends on the branch of square-root arguments.
This also resolves the tension between modular transformations and the coaction on iterated Eisenstein integrals proposed in~\cite{Kleinschmidt:2025dtk}, once the de Rham versions of iterated Eisenstein integrals are taken to mirror the ${\rm SL}_2(\mathbb Z)$ transformation of their single-valued counterparts up to $\bM^{\nil} \leftrightarrow \bM^{\sv}$.

\paragraph{\bf{Towards a diagrammatic coaction for Feynman integrals.}} The motivic Galois group is expected to act on Feynman integrals~\cite{Brown:2015fyf,Mellin:progress} (see also~\cite{Schnetz:2013hqa,Panzer:2016snt,Schnetz:2017bko}). In~\cite{Abreu:2014cla,Abreu:2015zaa,Abreu:2017enx,Abreu:2017mtm,Abreu:2021vhb} it was conjectured that it is possible to write the coaction on certain dimensionally-regulated Feynman integrals that evaluate to MPLs 
such that the motivic entries are the MIs and the de Rham entries are the cuts of the integral (see also~\cite{Tapuskovic:2019cpr,Mounoud:2026kgu}, and extensions to cosmological correlators are discussed in \cite{McLeod:2026jpz,McLeod:2026kpo}). 

Suppose that we work in an equivariant gauge and construct a fundamental solution matrix $\bP(\bx,\dimeps)$ whose entries are cut integrals~\cite{Abreu:2021vhb,Bonisch:2021yfw,Kreimer:2021jni}. The vector of master integrals in this gauge is $\bJ_{\textrm{eqv}}(\bx,\dimeps)=\bP(\bx,\dimeps) \bQ_F(\dimeps)$, where the Laurent coefficients of the vector $\bQ_F(\dimeps)$ only involve $2\pi i$ and algebraic numbers.  
We can construct the single-valued fundamental solution matrix 
\beq\label{eq:Psv}
\bP^{\sv}(\bx,\dimeps)=\bP(\bx,\dimeps)\overline{\bP(\bx,-\dimeps)^{-1}} = \bUeqv^{\sv}(\bx,\dimeps)\bM^{\sv}(\dimeps)\,,
\eeq
where  $\bReqv(\bx,\dimeps)^{-1}\bP^{\rm sv}(\bx,\dimeps) \overline{ \bReqv(\bx,\dimeps) }$ satisfies the same holomorphic differential equation as $\widetilde{\bP}(\bx,\dimeps)$ in (\ref{eq:Pexp}). 
Via reverse-unitarity~\cite{Anastasiou:2002yz,Anastasiou:2003yy}, we can interpret the entries of $\bP^{\sv}$, or rather its uplift $\bP^{\nil}$ as in~\eqref{utosv}, as \emph{de Rham cuts}. 
From~\eqref{eq:coaction} and~\eqref{knownco},~we see that the coaction of (the motivic uplift) $\bJ_{\textrm{eqv}}^{\mathfrak{m}}$ is given~by
\beq\label{eq:Feynman_coaction}
\Delta\big(\bJ_{\textrm{eqv}}^{\mathfrak{m}}(\bx,\dimeps)\big) = \bP^{\nil}(\bx,\dimeps)\bJ_{\textrm{eqv}}^{\mathfrak{m}}(\bx,\dimeps)\,.
\eeq
The coaction involves the master integrals in the motivic slot, and de Rham-cuts in the de Rham slot. We have checked in a variety of one- and two-loop examples from~\cite{Abreu:2017mtm,Abreu:2021vhb} that~\eqref{eq:Feynman_coaction} reproduces the results in terms of MPLs if the de Rham-cuts are interpreted as their images under de Rham projection, which in particular sends the motivic version of $2\pi i$ to zero~\cite{Brown:2015ylf,Brown:2018omk}. We illustrate this in detail on the example of the one-mass bubble integral in the supplemental material. 

The de Rham projection, however, is only available for Feynman integrals that evaluate to MPLs. Our~\eqref{eq:Feynman_coaction} is more general, and applies also beyond MPLs, where the de Rham-cuts $\bP^{\nil}$ may be represented by the entries of $\bP^{\sv}$ according to (\ref{utosv}). In particular, we can apply~\eqref{eq:Feynman_coaction} to write down a coaction for the equal-mass sunrise integral, which is related to an elliptic curve. It would be interesting to compare this result with \cite{Hidding:2021xot}  and to understand the relationship with the motivic coaction on the sunrise integral in strictly two dimensions from~\cite{Tapuskovic:2023xiu}.
Details and additional examples beyond MPLs will be presented in dedicated publications.

\section{Conclusions}

\noindent
Our understanding of dimensionally-regulated Feynman integrals has tremendously advanced over the last decade.
While a lot of progress was made towards generalising canonical differential equations beyond MPLs, the understanding of the relevant special functions, in particular their coaction, is not yet fully developed. 

The program outlined in this letter takes a major step forwards. Inspired by recent developments in string theory, we argued that canonical differential equations are finite-dimensional matrix representations of connections taking values in pro-nilpotent Lie algebras. 
Constructions of single-valued iterated integrals and coactions beyond genus 0 can then be directly
applied to Feynman integrals. This paves the way for extending results from MPLs to more general classes of functions. 

If the hidden geometry of a Feynman integral is a single Riemann surface, then the literature on the associated flat connections offers a growing body of information on the pro-nilpotent Lie algebra. For Feynman integrals that mix different geometries, the approach of this letter still gives access to finite-dimensional representations, equivariant gauges and coactions, even if the abstract connections may be hard to pinpoint.

We also expect that our approach will inspire novel results of purely mathematical interest. For example, inspired by the canonical differential equation for hypergeometric functions associated to elliptic curves, we could propose a general version of equivariant iterated Eisenstein integrals for the subgroup $\Gamma(2)$~\cite{lv2:progress} and their coaction \cite{mat:progress}. As another application, we foresee that one may be able to construct a symbol map and an $f$-alphabet beyond MPLs, which would allow one to study the coaction properties of certain QFT quantities, like the anomalous magnetic moment of the muon~\cite{Schnetz:2017bko}, and possibly to uncover yet unknown properties of some of the most fundamental quantities in QFT.

\begin{acknowledgments}
\emph{Acknowledgments:}
We are grateful to Samuel Abreu, Ruth Britto, Einan Gardi, Christoph Nega and Lorenzo Tancredi for valuable discussions. We would like to thank the Erwin Schr\"odinger International Institute for Mathematics and Physics (ESI), University of Vienna (Austria), for the opportunity to participate in the Thematic Programme ``Amplitudes and Algebraic Geometry'' in 2026 where a significant part of this work has been accomplished and for the support given. This research was supported by the European Research Council (ERC) under the European Union’s research and innovation programme grant agreements 101043686 (ERC Consolidator Grant LoCoMotive) and 101167287 (ERC Synergy Grant MaScAmp). Views and opinions expressed are however those of the author(s) only and do not necessarily reflect those of the European Union or the European Research Council. Neither the European Union nor the granting authority can be held responsible for them. MD was supported by the ERE grant RF\textbackslash ERE\textbackslash 221103 associated with the Royal Society University Research Fellowship Grant URF\textbackslash R1\textbackslash 221236 and in part by a grant of access to OpenAI models through the ChatGPT for Academic Researchers program.
The research of YS is supported by the Knut and Alice Wallenberg Foundation, Grant KAW 2025.0414. The research of CD was supported in part by the Deutsche Forschungsgemeinschaft (DFG, German Research
Foundation) under Germany’s Excellence Strategy – EXC 3107 – Project-ID 53376636.

\emph{Tool and computational resource disclosure:}
Suggestions by AI tools were used in the research process. However, all the statements of this work, as well as their proofs, are due to the authors. The authors take full responsibility for the content of the publication.

\end{acknowledgments}


\providecommand{\href}[2]{#2}\begingroup\raggedright\endgroup

\clearpage

\onecolumngrid
\appendix*
\allowdisplaybreaks
\renewcommand{\thesubsection}{\Alph{subsection}}
\renewcommand{\theequation}{\Alph{subsection}.\arabic{equation}}
\setcounter{secnumdepth}{2}

\section*{Supplemental material}

\setcounter{equation}{0}

We illustrate the central ideas of this letter on two examples, both of which use the Gauss hypergeometric function~${}_2F_1$, for which we use the following variant
\begin{equation}
\label{eq:2F1def}
\tag{$\star$}
\mathcal{F}(a,b,c; x) = \frac{\Gamma(b)\Gamma(c{-}b)}{\Gamma(c)}
 {}_2F_1(a,b;c; x) = \int_0^1 z^b (1-z)^{c-b} (1- x z)^{-a} \frac{\rd z}{z(1-z)}  \,.
\end{equation}
Here $a$, $b$ and $c$ are in general complex parameters, chosen such that the integral converges. Integrals of this type show up in QFT, cf.,~e.g.,~\cite{Moch:2001zr,Mimachi:2002gi,Huber:2007dx,Kalmykov:2008ofy,Derkachov:2018rot,Abreu:2019wzk,Duhr:2019rrs,Derkachev:2020xib,Blumlein:2021hbq,Duhr:2023bku,Porkert:2025gjd,Sohnle:2026ivr} and string theory~\cite{Virasoro:1969me,Britto:2021prf,Mafra:2022wml}. 
Here, the single variable $x$ corresponds to the vector of external scales $\bx$ in the body of the letter.

We are interested in cases related to dimensionally-regulated Feynman integrals. Hence, we focus on parameters that take the form
\begin{equation*}
a = \bar{a} + \tilde{a}\dimeps\,,\qquad b = \bar{b} + \tilde{b}\dimeps\,,\qquad
c = \bar{c} + \tilde{c}\dimeps\,,
\end{equation*}
where $\bar{a},\bar{b},\bar{c}$ and $\tilde{a},\tilde{b},\tilde{c}$ are rational constants, and we are interested in the Laurent expansion around $\dimeps=0$. Then for $\dimeps=0$ we obtain a genuine period integral, and different choices of the parameters $\bar{a}$, $\bar{b}$ and $\bar{c}$ correspond to different types of geometries. We discuss two cases: one associated with the Riemann sphere and MPLs where $\bar{a},\bar{b},\bar{c}$ are integers (appendix \ref{appA}), and one associated with elliptic curves and iterated integrals of Eisenstein series of the congruence subgroup $\Gamma(2)$ of ${\rm SL}_2(\mathbb Z)$ where $\bar{a},\bar{b},\bar{c}$ include half-integers (appendix \ref{appB}).

\subsection{Example 1: The one-loop bubble integrals with one massive propagator}
\label{appA}
\setcounter{equation}{0}

As a first example we discuss the master integrals for the family of one-loop bubble integrals with one massive propagator in $D=2-2\dimeps$, 
\beq
I_{\nu_1,\nu_2}(p^2,m^2;\dimeps) = \frac{e^{\gamma_E\dimeps}}{{i\pi^{D/2}}}\int\frac{\rd^D k}{(k^2-m^2)^{\nu_1}[(k-p)^2]^{\nu_2}}\,,
\eeq
where $\gamma_E=-\Gamma'(1)$ is the Euler--Mascheroni constant. The only non-trivial functional dependence of these integrals is through the dimensionless ratio $x = \frac{m^2}{-p^2}$, so without loss of generality we put $p^2=-1$ in the following. Note that the integrals are real in the Euclidean region $p^2<0$ and $m^2>0$.

The family has 2 master integrals, which we choose as $\bI(x,\dimeps) = \big(I_{1,0}(-1,x;\dimeps),I_{1,1}(-1,x;\dimeps)\big)^T $. We can rotate to a canonical basis via
$\bJ(x,\dimeps) = \big(J_1(x,\dimeps),J_2(x,\dimeps)\big)^T=\bRcan(x,\dimeps)\bI(x,\dimeps)$, with $\bRcan(x,\dimeps) = \operatorname{diag}\big(1,\tfrac{1}{2}(1+x)\big)$. We obtain the canonical differential equation,
\beq\label{eq:bubble_DEQ}
\rd\bJ(x,\dimeps) = \dimeps\,\bA(x)\,\bJ(x,\dimeps)\,,\qquad \bA(x) = \bA_0\,\dlog x+\bA_1\,\dlog(1+x) = \left(\begin{smallmatrix}
    -1 & 0 \\ 1/2&0\end{smallmatrix}\right) \,\dlog x + \left(\begin{smallmatrix}
    0 & \phantom{-}0 \\ 0&-2\end{smallmatrix}\right)\,\dlog(1+x)\,,
\eeq
Comparing~\eqref{eq:bubble_DEQ} to~\eqref{eq:KZ}, we recognise an instance of the KZ connection (after replacing $x$ by $-x)$. Since the
KZ connection in one variable takes values in the (completed) free Lie algebra $\operatorname{Lie}[\af_0,\af_1]$ 
in two generators, we obtain a representation~$\rho$ of $\operatorname{Lie}[\af_0,\af_1]$ by $2\times2$ matrices, $\rho(\af_i) = \dimeps\,\bA_i$, $i=0,1$, and the solution $\bJ(x,\dimeps)$ lives in a representation of the (completed) Lie group. Since this 
is an (unfaithful) representation of a free Lie algebra, there are no special commutation relations between the matrices to check. We recall that the connection in~\eqref{eq:KZ} is flat, irrespective of what the generators $\af_i$ are.

Note that, as $\bRcan(x,\dimeps)$ is a rational function of $x$, it is single-valued, and thus the cocycle of (\ref{eq:anal_cont}) equals $\widetilde{\bZ}_\Gamma(x,\dimeps)=\mathds{1}$, and the reductive group $S$ is trivial. In particular, the cocycle is constant, so that the canonical basis defines at the same time an equivariant gauge. The element $\Gamma$ is a closed path in $X=\mathbb{P}^1\setminus\{0,-1,\infty\}$, and the monodromy group $G$ is a representation of the pro-unipotent completion of the fundamental group of $X$, which is the free group on two generators (the small loops around $0$ and $-1$). We note that the analytic solution to~\eqref{eq:bubble_DEQ} is well known (see e.g.,~\cite{tHooft:1978jhc,Abreu:2017mtm}), and it can be expressed in terms of Gauss' hypergeometric function in (\ref{eq:2F1def}),
\beq
J_1(x,\dimeps) =-\frac{e^{\gamma_E\dimeps}\Gamma(1+\dimeps)}{\dimeps}\,x^{-\dimeps}\,,\qquad
J_2(x,\dimeps) = -\frac{e^{\gamma_E\dimeps}\Gamma(1+\dimeps)}{2\dimeps}\,(1+x)^{-\dimeps}\,{}_2F_1\left(-\dimeps,1+\dimeps;1-\dimeps;\frac{1}{1+x}\right)\,.
\label{canj1j2}
\eeq
In the following we use these analytic expressions to illustrate the consequences of our main correspondence on a simple example of a concrete Feynman integral

\paragraph{\bf The fundamental solution matrix of cuts.}
We may pick a fundamental solution matrix whose entries consist of cuts of the canonical master integrals, 
\beq\label{eq:bubble_P}
\bP(x,\dimeps) = \begin{pmatrix}     \mathcal{C}_{e_1}J_1(x,\dimeps)&0\\
     \mathcal{C}_{e_1}J_2(x,\dimeps)
    &\mathcal{C}_{e_1,e_2}J_2(x,\dimeps) 
    \end{pmatrix}\,,
    \eeq
where $\mathcal{C}_{e_1}J_k$ and $\mathcal{C}_{e_1,e_2}J_k$ denote the single- and double cuts of $J_k$. Explicit analytic results for the cuts can be found in appendix B of~\cite{Abreu:2017mtm}, and we recall them here for convenience:
\beq\bsp
\mathcal{C}_{e_1}J_1(x,\dimeps) &\,= \frac{e^{\gamma_E\dimeps-i\pi \dimeps}}{\Gamma(1-\dimeps)}\,x^{-\dimeps}\,,\\
\mathcal{C}_{e_1}J_2(x,\dimeps) &\,= -\frac{e^{\gamma_E\dimeps-i\pi\dimeps}\,x^{-\dimeps}}{2\Gamma(1-\dimeps)}\,(1+x)\,{}_2F_1(1,1+\dimeps;1-\dimeps;-x)\,,\\
\mathcal{C}_{e_1,e_2}J_2(x,\dimeps) &\,= \frac{e^{\gamma_E\dimeps+i\pi\dimeps}\,\Gamma(1-\dimeps)}{\Gamma(1-2\dimeps)}\,(1+x)^{-2\dimeps}\,.
\esp
\label{ordcuts}
\eeq
We stress that the results from~\cite{Abreu:2017mtm} are valid only modulo $i\pi$. However, as we will soon see, this is sufficient for all practical purposes.

\paragraph{\bf Initial conditions and zeta generators.}
We find that in terms of the fundamental solution matrix in~\eqref{eq:bubble_P}, the vector of canonical master integrals can be cast into the form
\beq
\bJ(x,\dimeps) = \bP(x,\dimeps)\bQ_F(\dimeps)\,,\textrm{~~~~with~~~}\bQ_F(\dimeps) 
 = \begin{pmatrix}
1\\1\end{pmatrix}
\frac{2 \pi i }{e^{-2 i \pi \dimeps} - 1} 
= -\frac{1}{\dimeps}\begin{pmatrix}
1\\1\end{pmatrix}\,
\left[1+i \pi  \dimeps  -\frac{\pi ^2 }{3}\,\dimeps ^2-\frac{\pi ^4}{45}\, \dimeps ^4 +\ord(\dimeps^6)\right]  
     \,.
\eeq
We see that the solution takes the expected factorised form, where the vector of initial conditions only involves powers of $i\pi$. The dependence on all transcendental constants other than $i\pi$ is encoded in the fundamental solution matrix of cuts, which itself admits a factorised representation (cf.~\eqref{eq:N_fac}),
\beq
\label{eq:initMQ}
\bP(x,\dimeps) = \bU(x,\dimeps)\,\bM(\dimeps)\,\bQ(\dimeps)\,,
\eeq
where $\bU(x,\dimeps)$ is the usual path-ordered exponential of $\dimeps\,\bA(x)=\bA_{\mathrm{eqv}}(x,\dimeps)$ in \eqref{eq:bubble_DEQ}
and the $\dimeps$-expansion of $\bU(x,\dimeps)$ involves harmonic polylogarithms (HPLs)~\cite{Remiddi:1999ew} (cf.~\eqref{eq:Pexp}). The $\dimeps$-expansion of $\bQ(\dimeps) = \sum_{k=0}^\infty \bQ_k\,(i\pi\dimeps)^k$ in turn only involves powers of $i \pi$, with $\bQ_k$ constant matrices with rational entries, e.g.,
\beq
\bQ_0 = \begin{pmatrix}
 1 & 0 \\[1mm]
 -\frac{1}{2} & 1 
\end{pmatrix} = -\bQ_1\,,\qquad \bQ_2 = \begin{pmatrix}
 \frac{7}{12}&0 \\[1mm]
-\frac{7}{24}& \frac{3}{4}  
\end{pmatrix}\,,\quad \bQ_3 = \begin{pmatrix}
  -\frac{1}{4}&0 \\[1mm]
 \frac{1}{8} &-\frac{5}{12} 
\end{pmatrix}\,,\quad \bQ_4 = \begin{pmatrix}
  \frac{121}{1440}&0 \\[1mm]
-\frac{121}{2880}& \frac{5}{32}  
\end{pmatrix}\,.
\label{defqs}
\eeq
Finally, the matrix $\bM(\dimeps)$ in (\ref{eq:initMQ}) encodes all remaining transcendental numbers, and its form it tightly constraint. For example, we have for the first few orders,
\beq
\bM(\dimeps) = \mathds{1} + \dimeps^3\,\zeta_3\,\bM_3 + \dimeps^5\,\zeta_5\,\bM_5 +  \dimeps^6\,\zeta_3^2\,\frac{1}{2}\bM_3^2+\ord(\dimeps^7)\,,
\eeq
with
\beq
\bM_3 = \begin{pmatrix}
 -\frac{1}{3} & 0 \\
 -1 & -\frac{7}{3} 
\end{pmatrix} \textrm{~~~and~~~} \bM_5 = \begin{pmatrix}
 -\frac{1}{5} & 0 \\
 -3 & -\frac{31}{5}
 \end{pmatrix}
 \,.
\label{mmatrx}
\eeq
The coefficient of $\zeta_3^2$ is determined by the coefficient $\bM_3$ of the single power of $\zeta_3$, consistently with the coaction formula (\ref{knownco}) for $\bM(\dimeps)$. More generally, only the coefficients $\bM_{2k+1}$ of the single zetas $\zeta_{2k+1}$ are free, and all other coefficients are fixed in terms of them by (\ref{knownco}).\footnote{The property $[\bM_3 ,\bM_5 ]= 0$ of our particular matrix representations (\ref{mmatrx}) is then consistent with the absence of irreducible MZVs at weight 8, which in fact extends to arbitrary weight by general properties of the $\Gamma$ and $_2F_1$ functions in (\ref{canj1j2}).} This is not a coincidence, but the matrices $\bM_{2k+1}$ provide a representation $\rho$ 
of the Lie algebra of (genus-zero) zeta generators $\sigma_{2k+1}$ attached to the (abstract) KZ connection~\eqref{eq:KZ}, $\bM_{2k+1}\,\dimeps^{2k+1} = \rho(\sigma_{2k+1})$~\cite{Drinfeld:1991, Furusho2000TheMZ, DG:2005}. While  
the Lie algebra of zeta generators $\sigma_{2k+1}$ is free \cite{Levine} (before passing to matrix representations $\rho$) they have fixed commutation relations with the formal variables $\af_i$ appearing in the KZ connection  
\beq
\label{genalg}
[\sigma_{2k+1},\af_0] = 0 \textrm{~~~and~~~} [\sigma_{2k+1},\af_1] =
[\af_1,
W_{2k+1}(\af_0,\af_1)]\,,
\eeq
where $W_{2k+1}(\af_0,\af_1)$ are Lie polynomials of degree $2k+1$ defined through the Drinfeld associator \cite{Frost:2023stm, Dorigoni:2024iyt}. Explicit  expressions of $g_{2k+1}(\af_0,\af_1)=W_{2k+1}(\af_0,-\af_1)$ up to $k=5$ can be found in the ancillary file of~\cite{Dorigoni:2024iyt}, e.g., 
\beq\bsp
W_3(\af_0,\af_1)&\, =  - [ \af_0 + \af_1 , [ \af_0, \af_1 ] ] \,,\\
W_5(\af_0,\af_1) &\,=   - [ \af_0,[\af_0,[\af_0,[\af_0,\af_1 ]] ]]  - \frac{ 3}{2} [ \af_1,[\af_0,[\af_0,[\af_0,\af_1]] ]]
-  \frac{1}{2} [ \af_0,[\af_1,[\af_0,[\af_0,\af_1]] ]]  \\
&\quad \,   - \frac{1}{2} [ \af_1,[\af_0,[\af_1,[\af_0,\af_1]] ]] 
-  \frac{3}{2} [ \af_0,[\af_1,[\af_1,[\af_0,\af_1]] ]] - [\af_1,[\af_1,[\af_1,[\af_0,\af_1]] ]]  \,.
\esp\eeq
We stress that the commutation relations are a property of the abstract KZ connection, and they are not specific to the example of the bubble integral discussed here. They therefore put strong constraints on the matrices $\bM_{2k+1}$. Indeed, passing to our concrete two-dimensional representations of (\ref{eq:bubble_DEQ}) and (\ref{mmatrx}), we see that the following commutation relations hold,
\beq
\label{repalg}
\bsp
[\bM_{2k+1},\bA_0] = 0\,,&\qquad [\bM_{2k+1},\bA_1] = 
[\bA_1,
W_{2k+1}(\bA_0,\bA_1)] 
\esp\eeq
In fact, in our example these commutations relations are strong enough to fix the $\bM_{2k+1}$ completely, up to a multiple of the identity matrix. This remaining degree of freedom can easily be fixed from the analytic result for the tadpole integral, which is easy to obtain. In other words, the part of the initial condition not involving $i\pi$ is entirely fixed from the zeta generators and the normalisation of the tadpole integral. We observe this to be a rather generic feature for Feynman integrals, also for more complicated examples.

Let us conclude this discussion with a comment. We already mentioned that the results for the cuts from~\cite{Abreu:2017mtm} are only defined modulo $i\pi$. This does not affect our conclusions. Indeed, changing the results by terms involving $i\pi$ corresponds to different choice for the matrix $\bQ(\dimeps)$ and the vector $\bQ_F(\dimeps)$, and it does not affect the commutation relations coming from the zeta generators. 

\paragraph{\bf Single-valued analogues of cuts.}
We can construct the matrix $\bP^{\sv}$ of single-valued cuts according to~\eqref{eq:Psv}. We~find 
\beq\label{eq:bubble_Psv}
\bP^{\sv}(x,\dimeps) = \begin{pmatrix}
     \operatorname{Cut}^{\sv}_{e_1}J_1(x,\dimeps)&0\\
\operatorname{Cut}^{\sv}_{e_1}J_2(x,\dimeps)&\operatorname{Cut}^{\sv}_{e_1,e_2}J_2(x,\dimeps) 
    \end{pmatrix}\,,
    \eeq
    where we defined 
\beq\bsp
\operatorname{Cut}^{\sv}_{e_1}J_1(x,\dimeps)&\,= e^{2 \gamma_E  \dimeps } \,\frac{\Gamma (1+\dimeps ) }{\Gamma (1-\dimeps )}\,| x| ^{-2 \dimeps }\,,\\
\operatorname{Cut}^{\sv}_{e_1}J_2(x,\dimeps)&\,= e^{2 \gamma_E  \dimeps }\,\frac{  \Gamma (1-\dimeps )  }{2  \Gamma (1+\dimeps )}\,\frac{   \Gamma (1+2 \dimeps ) }{\Gamma (1-2 \dimeps ) }\,|1+ x| ^{-4 \dimeps } \,(1+\bar{x})\,{}_2F_1\left(1,1-\dimeps ;1+\dimeps ;-\bar{x}\right)\\
&\quad \,-e^{2 \gamma_E  \dimeps }\,\frac{  \Gamma (1+\dimeps ) }{2 \Gamma (1-\dimeps )}\,| x| ^{-2 \dimeps } \,(1+x)\, {}_2F_1(1,\dimeps +1;1-\dimeps ;-x)\,,\\
\operatorname{Cut}^{\sv}_{e_1,e_2}J_2(x,\dimeps)&\,= e^{2 \gamma_E  \dimeps } \,\frac{\Gamma (1-\dimeps )  }{ \Gamma (1+\dimeps )}\,\frac{ \Gamma (1+2 \dimeps ) }{\Gamma (1-2 \dimeps ) }\,| 1+x| ^{-4 \dimeps }\,.
\esp
\label{svcuteq}
\eeq
These functions can be interpreted as \emph{single-valued analogues of cuts} in the following sense.
First, $\bP^{\sv}$ satisfies the same holomorphic differential equation as $\bP$, namely $\partial \bP^{\sv}(x,\dimeps) = \dimeps\,\bA(x)\,\bP^{\sv}(x,\dimeps)$, but its entries are non-holomorphic, single-valued functions of $x$.\footnote{Even though this is not manifest in (\ref{svcuteq}), the right-hand side of the second and third line is expressible in terms of the single-valued versions of $_2F_1$ discussed for instance in
~\cite{Brown:2019jng, Britto:2021prf, Duhr:2023bku,Mimachi:2002gi} } Second, since the Laurent coefficients of $\bP(x,\dimeps)$ can be expressed in terms of HPLs, the $\dimeps$-expansion of $\bP^{\sv}(x,\dimeps)$ contains single-valued HPLs~\cite{brownSVHPLs}. More specifically, we have
\beq\label{eq:svP-Psv}
\sv\big(\bP(x,\dimeps)\big) = \bP^{\sv}(x,\dimeps)\bQ_0\,,
\eeq
where $\sv$ is the map that replaces HPLs and MZVs by their single-valued versions  \cite{Schnetz:2013hqa, Brown:2013gia}, and $\bQ_0$ is defined by (\ref{defqs}). 
Explicitly, we can write the single-valued cuts as linear combinations of images under $\sv$ of ordinary cuts in (\ref{ordcuts}),
\beq\bsp\label{eq:sv_to_svCut}
\sv\big(\mathcal{C}_{e_1}J_1(x,\dimeps)\big)&\,=\operatorname{Cut}^{\sv}_{e_1}J_1(x,\dimeps)\,,\\
\sv\big(\mathcal{C}_{e_1}J_2(x,\dimeps)\big)&\,=\operatorname{Cut}^{\sv}_{e_1}J_2(x,\dimeps)-\frac{1}{2}\operatorname{Cut}^{\sv}_{e_1,e_2}J_2(x,\dimeps) \,,\\
\sv\big(\mathcal{C}_{e_1,e_2}J_2(x,\dimeps)\big)&\,=\operatorname{Cut}^{\sv}_{e_1,e_2}J_2(x,\dimeps) \,.
\esp\eeq
Note that, since $\sv(i\pi) =0$, this construction is again independent of the fact that the cuts from~\cite{Abreu:2017mtm} had been defined modulo $i\pi$. In particular, $\bP^{\sv}$ is well defined, without any ambiguities.

\paragraph{\bf Coaction of the bubble and tadpole integrals.} Finally, let us discuss the coaction $\Delta$. Using~\eqref{eq:Feynman_coaction}, we find
\beq\bsp\label{eq:bubble_coaction}
\Delta\big(J_1^\mathfrak{m}(x,\dimeps)\big) &\,= J_1^\mathfrak{m}(x,\dimeps)\otimes \operatorname{Cut}^{\nil}_{e_1}J_1(x,\dimeps) \,,\\
\Delta\big(J_2^\mathfrak{m}(x,\dimeps)\big) &\,= J_1^\mathfrak{m}(x,\dimeps)\otimes \operatorname{Cut}^{\nil}_{e_1}J_2(x,\dimeps) + J_2^\mathfrak{m}(x,\dimeps)\otimes \operatorname{Cut}^{\nil}_{e_1,e_2}J_2(x,\dimeps) \,.
\esp\eeq
The quantities entering the coaction can be interpreted in the following way. $J_k^\mathfrak{m}(x,\dimeps)$ is the motivic uplift of the Feynman integral $J_k(x,\dimeps)$, and we may represent $J_k^\mathfrak{m}(x,\dimeps)$ by this Feynman integral via the period map 
\beq
J_1^\mathfrak{m}(x,\dimeps) \longrightarrow J_1(x,\dimeps)\,,\qquad J_2^\mathfrak{m}(x,\dimeps) \longrightarrow
J_2(x,\dimeps)\,.
\eeq
The quantities $\operatorname{Cut}^{\nil}_{e_1}J_k(x,\dimeps)$ and $\operatorname{Cut}^{\nil}_{e_1,e_2}J_2(x,\dimeps)$ are de Rham cuts. In general, they cannot be represented by ordinary cut integrals. Instead, we may represent them by the corresponding single-valued integrals from~\eqref{svcuteq},
\beq
\operatorname{Cut}^{\nil}_{e_1}J_k(x,\dimeps) \longrightarrow \operatorname{Cut}^{\sv}_{e_1}J_k(x,\dimeps)\,,\quad
\operatorname{Cut}^{\nil}_{e_1,e_2}J_2(x,\dimeps) \longrightarrow \operatorname{Cut}^{\sv}_{e_1,e_2}J_2(x,\dimeps)\,.
\eeq
The coaction of one-loop integrals in dimensional regularisation was studied in detail in~\cite{Abreu:2017enx,Abreu:2017mtm}. At first glance, our~\eqref{eq:bubble_coaction} looks very different from the coaction in that reference. We now show that~\eqref{eq:bubble_coaction} reproduces the results of~\cite{Abreu:2017enx,Abreu:2017mtm} for the tadpole and bubble integrals. As a starting point, we note that for Feynman integrals whose Laurent expansion only involves MPLs, there is an alternative way to represent de Rham cuts, namely as images of ordinary cuts under the \emph{de Rham projection} $\Pi_{\mathfrak{dr}}$~\cite{Brown:2015ylf,Brown:2018omk}, which acts order by order in $\dimeps$ and can be understood as replacing MPLs by analogues of MPLs modulo their branch cuts. In particular, $\Pi_{\mathfrak{dr}}$ annihilates (the motivic uplift of) $i\pi$, so that the preimages are only defined up to $i\pi$, in agreement with the results of~\cite{Abreu:2017mtm}. In our case, this can be expressed as (cf.~\eqref{eq:svP-Psv}):
\beq
\Pi_{\mathfrak{dr}}\big(\bP^{\mathfrak{m}}(x,\dimeps)\big) = \bP^{\nil}(x,\dimeps)\bQ_0\,.
\eeq
Explicitly, we have (cf.~\eqref{eq:sv_to_svCut})
\beq\bsp\label{eq:bubble_Pidr}
\operatorname{Cut}^{\nil}_{e_1}J_1(x,\dimeps)&\, = \Pi_{\mathfrak{dr}}\big(\mathcal{C}_{e_1}J^{\mathfrak{m}}_1(x,\dimeps)\big)\,,\\
\operatorname{Cut}^{\nil}_{e_1}J_2(x,\dimeps)&\, = \Pi_{\mathfrak{dr}}\!\left(\mathcal{C}_{e_1}J^{\mathfrak{m}}_2(x,\dimeps)+\frac{1}{2}\mathcal{C}_{e_1,e_2}J^{\mathfrak{m}}_2(x,\dimeps)\right)\,,\\
\operatorname{Cut}^{\nil}_{e_1,e_2}J_2(x,\dimeps)&\, = \Pi_{\mathfrak{dr}}\big(\mathcal{C}_{e_1,e_2}J^{\mathfrak{m}}_2(x,\dimeps)\big)\,.
\esp\eeq
Upon inserting~\eqref{eq:bubble_Pidr} into~\eqref{eq:bubble_coaction}, we recover the coaction of the tadpole and bubbles integrals from~\cite{Abreu:2017enx,Abreu:2017mtm}.

\subsection{Example 2: Hypergeometric integrals and elliptic curves}
\label{appB}
\setcounter{equation}{0}

As a second example, we consider the hypergeometric integral~\eqref{eq:2F1def} with the parameter choice \begin{equation}
\label{eq:param2F1}
a= \frac12+\alpha \dimeps\,,\quad 
b=\frac12-\beta \dimeps \,,\quad
c=1-(\beta+ \gamma)\dimeps\,,
\end{equation}
where $\alpha,\beta,\gamma$ are (non-zero) rational constants, and the half-integer shifts lead to an elliptic nature of the problem, as can be seen from the fact that for $\dimeps=0$ the hypergeometric integral in~\eqref{eq:2F1def} reduces to (twice) the complete elliptic integral of the first kind, 
\beq
\operatorname{K}(x)
=\int_0^1 \frac{\rd t}{\sqrt{(1-t^2)(1-x t^2)}}\,.
\eeq
Thus, the geometry associated with the above integral and period matrix is the Legendre family of elliptic curves, 
\begin{equation}
y^2= z (z-1) (z- x^{-1})\,.
\end{equation}
The branch points are at $0,1, x^{-1},\infty$, and $x$ parametrises this family. It is associated with the modular curve $X=\Gamma(2)\backslash \UHP$, where $\UHP=\{ \tau\in \mathbb{C} \, :\, \mathrm{Im}( \tau)>0\}$ is the upper half-plane that is acted upon by the principal congruence subgroup $\Gamma(2)=\{\gamma\in \mathrm{SL}_2(\mathbb{Z}):\gamma=\mathds{1}\!\!\mod 2\}\subset \mathrm{SL}_2(\mathbb{Z})$.
The Laurent expansion in $\dimeps$ of this integral was first considered in~\cite{Broedel:2017kkb}.

\paragraph{\bf Period matrix as fundamental solution matrix.} 
To describe the period matrix for this example, it is useful to take a perspective on hypergeometric functions via twisted de Rham theory~\cite{AomotoKita}.
We use the twist function $\Phi(z)=z^b (1-z)^{c-b} (1-x  z)^{-a}$ and make the following choice of bases of the twisted homology ($\gamma_j$) and twisted cohomology ($\omega_i$) groups~\cite{intersection, AomotoKita, Mizera:2019gea, Brown:2019jng},
\begin{equation}
\gamma_1: 0 \longrightarrow 1\,,\quad 
\gamma_2: 0 \longrightarrow x^{-1}\,, \quad 
\omega_1 = \frac{\rd z}{z(1-z)} \,,\quad
\omega_2 = \frac{\rd z}{1-z}\,.
\end{equation}
The entries of the (twisted) period matrix are $\int_{\gamma_j} \Phi \omega_i$ with $1\leq i,j \leq 2$, and  with respect to our choice of basis the period matrix becomes (after appropriate analytic continuation): 
\begin{equation}
\bPhat( x,\dimeps) =
\begin{pmatrix}
\mathcal{F}(a,b,c; x) &  x^{-b}\,\mathcal{F}\big(1+b-c,b,1-a+b; x^{-1}\big)\\[2mm]
\mathcal{F}(a,b+1,c+1; x) &  x^{-b-1}\,\mathcal{F}\big(1+b-c,b+1,2-a+b; x^{-1}\big)
\end{pmatrix}\,.
\label{defphat}
\end{equation}
The vector of master integrals $\bI(\bx,\dimeps)$ in (\ref{eq:DEQ_general}) is obtained from $\bPhat(x,\dimeps)$ by choosing a specific integration contour $\gamma$, i.e.,~right-multiplying $\bPhat(x,\dimeps)$ by a vector describing $\gamma$ in the basis $\{\gamma_j\}$. We shall consider the full matrix $\bPhat(x,\dimeps)$ here instead of singling out a specific integration contour.

\paragraph{\bf Canonical differential equation.}
Differentiating (\ref{defphat}) with respect to $x$, we find the differential equation~\eqref{eq:DEQ_general} satisfied by $\bPhat( x,\dimeps)$.
The rotation of the differential equation to canonical form~\eqref{eq:canon_DEQ} was first discussed in~\cite{Broedel:2018rwm}, and is achieved by left-multiplying  $\bPhat(x,\dimeps)$ by \begin{equation}
\label{eq:Rcan2F1}
    \bRcan(x,\dimeps) = \begin{pmatrix}
        \frac{1}{2 \operatorname{K}(x)} & 0 \\ \frac{1}{\pi^2 \dimeps} \Big( {-}2 \operatorname{E}(x) +2 \big(1-(\alpha+\gamma) \dimeps -x(\alpha +\beta) \dimeps \big) \operatorname{K}(x)\Big) & \frac{1}{\pi^2 \dimeps} 4 x \operatorname{K}(x) \big({-}\frac{1}{2} + (2\alpha  + \gamma) \dimeps \big)
        \end{pmatrix}
\,,
\end{equation}
where $\operatorname{E}(x)=\int_0^1 \sqrt{\frac{1-x t^2}{1-t^2}}\rd t$ the complete elliptic integral of the second kind. 
The resulting $\bPtilde(x,\dimeps) = \bRcan(x,\dimeps) \bPhat(x,\dimeps)$ satisfies the canonical differential equation~\eqref{eq:canon_DEQ} with the connection
\begin{equation}
\dimeps\, \bA(x)  = \frac{\dimeps \, \rd x}{x(1-x)} \begin{pmatrix}
        \frac{1}{2} \big((x+1)\alpha +\gamma -x\beta\big) & \frac{\pi^2}{8\operatorname{K}(x)^2} \\
        \frac{2}{\pi^2} \Big(\big(\alpha+\gamma+x(\alpha+\beta)\big)^2-4x \beta (2\alpha+\gamma)
        \Big)\operatorname{K}(x)^2 & \frac{1}{2} \big((x+1)\alpha +\gamma -x\beta\big)
\end{pmatrix}\,.
\end{equation}

The modular curve $X$ can be mapped to its covering space, the full upper half-plane $\UHP$, using the Hauptmodul of~$X$~\cite{Zagier:123} and we denote the resulting variable by $\tau= \frac{i \operatorname{K}(1-x)}{\operatorname{K}(x)}\in\UHP$.
In this new variable, the square of the complete elliptic integral $\operatorname{K}(x)^2$ can be expressed as a holomorphic Eisenstein series of weight 2 for the congruence subgroup $\Gamma(2)$ and the connection takes the form 
\begin{equation}
\label{eq:Ahk}
\dimeps\, \bA(\tau)  = -\frac{\pi \dimeps \,\rd\tau}{2 i} \begin{pmatrix}
        h_2(\tau) & 1 \\ h_4(\tau) & h_2(\tau)
        \end{pmatrix}\,,
\end{equation}
where $h_2(\tau)$ and $h_4(\tau)$ are modular forms for $\Gamma(2)$ of weights $2$ and $4$, respectively, that can be determined explicitly in terms of Jacobi theta functions 
\beq\bsp
\label{hfcts}
    h_2(\tau)&=(\alpha-\beta)\,\vartheta_2^4(\tau)+(\alpha+\gamma)\,\vartheta_3^4(\tau)\,,\\
h_4(\tau)&=\big((\alpha-\beta)\,\vartheta_2^4(\tau)+(\alpha+\gamma)\,\vartheta_3^4(\tau)\big)^2-4\alpha\beta\,\vartheta_2^4(\tau)\vartheta_4^4(\tau)\,.
\esp \eeq
Both modular forms have constant terms of transcendentality zero. 

The transformed period matrix as a solution to the canonical differential equation is then
\begin{equation}
\label{eq:pertau}
\bPtilde(\tau,\dimeps) = \bUcan(\tau,\dimeps) \bN(\tau_0,\dimeps)\,, 
\quad \text{with}\quad 
\bUcan(\tau,\dimeps) =  \operatorname{Pexp}\Big[\dimeps\int_{\tau_0}^\tau \bA(\tau_1)\Big]\,,
\end{equation}
as in~\eqref{eq:Pexp}, where we choose as the path of integration $\gamma:  \tau_0 \to \tau$. In the context of iterated integrals of modular forms~\cite{ManinModular, Broedel:2015hia, Ablinger:2017bjx,Adams:2017ejb,Duhr:2019rrs, Matthes2022,Broedel:2021zij},
it is customary to send $\tau_0\to i \infty$  and then the integrals require (tangential base-point) regularisation~\cite{Brown:2014pnb}. We begin by keeping $\tau_0$ generic here, in order to study the properties of the general solution $\bUcan(\tau,\dimeps)$, and we shall come back to the initial conditions $\bN(\tau_0,\dimeps)$  and the limit $\tau_0\to i \infty$ below.

\paragraph{\bf Monodromy and modularity.}
The reductive group $S$ of our general discussion comes from the monodromy group associated with the gauge transformation $\bRcan(x,\dimeps)$ in~\eqref{eq:Rcan2F1}. This monodromy group is an (index-2) subgroup $\Gamma(2)^+\subset \Gamma(2)$, where $\Gamma(2)$ acts (non-faithully) on the covering space $\UHP$ of $X$ by 
\begin{equation}
\tau \mapsto \Gamma\cdot \tau = \frac{a\tau+b}{c\tau+d}\,,
\quad\text{for}\quad \Gamma=\begin{pmatrix} a& b\\c&d\end{pmatrix}\in \Gamma(2) 
\,.
\end{equation}
The subgroup is $\Gamma(2)^+=\ker(\chi)$, where $\chi : \Gamma(2) \to \mathbb{Z}_2$ is a character defined by $\chi(\Gamma)=(-1)^{(d-1)/2}$, and $\Gamma(2)^+$ in particular does not contain $-\mathds{1}$. The character shows up since the elliptic integrals $\operatorname{K}$ and $\operatorname{E}$ appearing in~\eqref{eq:Rcan2F1} are  modular forms for $\Gamma(2)^+$ of weight $1$ that can also be viewed as modular forms for $\Gamma(2)$ transforming with this character. For this reason we shall use $S=\Gamma(2)$ in the following and keep track of the character. Note that $\Gamma(2)^+$ and $\Gamma(2)$ have the same action on $\UHP$.

The character $\chi$ drops out from the transformation of the connection~\eqref{eq:Ahk}, i.e.,~the even-weight functions $h_k(\tau)$ in~\eqref{hfcts} satisfy $h_k(\Gamma\cdot \tau) = j(\Gamma,\tau)^k h_k(\tau)$ with the $\Gamma(2)$-cocycle $j(\Gamma,\tau) = c\tau+d$. 
The connection $\dimeps\, \bA(\tau)$ from~\eqref{eq:Ahk} then transforms~as 
\begin{equation}
\dimeps\, \bA(\Gamma\cdot \tau)%
= \bZcan_\Gamma(\tau,\dimeps) \dimeps\, \bA(\tau)  \bZcan_\Gamma(\tau,\dimeps)^{-1} + \rd \bZcan_\Gamma(\tau,\dimeps)  \bZcan_\Gamma(\tau,\dimeps)^{-1}\,,
\end{equation}
where
\begin{equation}
\label{eq:LamZ}
 \bZcan_\Gamma(\tau,\dimeps) = \begin{pmatrix}
j(\Gamma,\tau)^{-1} & 0 \\ 
\frac{2i c}{\pi \dimeps} & j(\Gamma,\tau)
 \end{pmatrix}\,.
\end{equation}
The matrix $\bZcan_\Gamma(\tau,\dimeps)$ implements the gauge transformation of the connection $\dimeps\, \bA(\tau)$ under modular transformations. It has a simple pole in $\dimeps$ and agrees for consistency with the transformation of the rotation $\bRcan(\tau,\dimeps)$ in~\eqref{eq:Rcan2F1}  under such transformations~\cite{Broedel:2018rwm}, 
\begin{equation}
\bRcan(\Gamma \cdot \tau,\dimeps)  =  \chi(\Gamma)\,\bZcan_\Gamma(\tau,\dimeps) \bRcan(\tau,\dimeps)\,.\end{equation}

The  statement analogous to~\eqref{eq:anal_cont} is the modular transformation
\begin{equation}
\label{eq:mon2F1}
\bUcan(\Gamma \cdot \tau,\dimeps)  =  \bZcan_\Gamma(\tau,\dimeps)\bUcan(\tau,\dimeps) \bZcan_\Gamma(\tau_0,\dimeps)^{-1}
\bUcan_\Gamma (\dimeps) \,,
\end{equation}
where the factor $\bUcan_\Gamma(\dimeps)$  is given by
\begin{equation}
\bUcan_\Gamma(\dimeps) 
=\bUcan (\Gamma\cdot \tau_0,\dimeps)
=\operatorname{Pexp}\Big[\dimeps\int_{\tau_0}^{\Gamma\cdot\tau_0} \bA(\tau_1)\Big]\,.
\label{tilduga}
\end{equation}
This is independent of $\tau$, but depends on $\tau_0$,
and in the limit $\tau_0\to i\infty$ relates to so-called multiple modular values~\cite{Brown:2014pnb}. However, in this limit one has to combine the last two factors in~\eqref{eq:mon2F1} into $\bCcan_\Gamma(\dimeps) = \bZcan_\Gamma(\tau_0,\dimeps)^{-1}\bUcan_\Gamma (\dimeps)$ to obtain a well-defined expression (with a pole in $\dimeps$) since the limit $\tau\to i \infty$ of $\bZcan_\Gamma(\tau,\dimeps)$ is not well-defined on its own. 
More precisely, the matrix $\bZcan_\Gamma(\tau,\dimeps)$ satisfies the cocycle property for the congruence group $\Gamma(2)$, while the matrix $\bUcan_\Gamma (\dimeps)$ transforms as a parallel transport, 
\begin{equation}
\label{eq:ZUtrm2F1}
\bZcan_{\Gamma_1 \Gamma_2} (\tau,\dimeps) = \bZcan_{\Gamma_1}(\Gamma_2 \cdot \tau,\dimeps) \bZcan_{\Gamma_2}(\tau,\dimeps) \,,\quad\quad
\bUcan_{\Gamma_1 \Gamma_2} (\dimeps) =  \bZcan_{\Gamma_1} (\Gamma_2\cdot\tau_0,\dimeps) \bUcan_{\Gamma_2} (\dimeps)  \bZcan_{\Gamma_1} (\tau_0,\dimeps)^{-1} \bUcan_{\Gamma_1} (\dimeps)\,,
\end{equation}
where the second equality follows from concatenation of paths together with the gauge transformation of $\dimeps\,\bA(\tau)$. Note that these two equations imply that the combined $\bCcan_\Gamma(\dimeps)$ discussed above transforms as
\begin{equation}
\bCcan_{\Gamma_1 \Gamma_2} (\dimeps)
= \bZcan_{\Gamma_2}(\tau_0,\dimeps)^{-1}  \bUcan_{\Gamma_2}(\dimeps) \bZcan_{\Gamma_1} (\tau_0,\dimeps)^{-1} \bUcan_{\Gamma_1} (\dimeps)
= \bCcan_{ \Gamma_2} (\dimeps) \bCcan_{\Gamma_1 } (\dimeps)\,.
\end{equation}
Thus, $\bCcan_{\Gamma}(\dimeps)$ is a 
matrix representation (up to swapping the matrix multiplication order),
known as the monodromy representation, that also satisfies a covariant differential equation  $\partial_{\tau_0} \bCcan_\Gamma(\dimeps) = \big[ \dimeps\, \bA(\tau_0),\bCcan_\Gamma(\dimeps)\big]$.

As already mentioned, in the context of our main proposed correspondence, here the reductive group $S$ is the congruence subgroup $\Gamma(2)$. The pro-unipotent group $U$ is given by the group associated with 
the pro-nilpotent Lie algebra $\nil$
associated with $\Gamma(2)$~\cite{Calaque:2019ell, Calaque:2020ell, Hopper:2021ell, Duhr:2025lvr,lv2:progress}, in analogy with the case of the full modular group~\cite{Brown:2014pnb,Brown:2020mhp}. We can view the connection and all objects here as finite-dimensional matrix representations of
the Calaque--Gonzalez connection \cite{Calaque:2019ell, Calaque:2020ell} featuring
 Eisenstein series for $\Gamma(2)$. We next illustrate some of the associated structures and relegate details to~\cite{lv2:progress}.

\paragraph{\bf Equivariant gauge.}
As is evident from~\eqref{eq:LamZ}, for this example the matrix $\bZcan_\Gamma(\tau,\dimeps)$ is $\tau$-dependent and we therefore make a further gauge transformation to arrive at the equivariant gauge~\eqref{toeqvg}. The transformation that achieves this here is 
\begin{equation}
\label{eq:Reqv2F1}
\bReqv(\tau,\dimeps) = \begin{pmatrix}
1 &  \frac{\pi \dimeps \tau}{2 i } \\
0 &1
\end{pmatrix}\,.
\end{equation}
It has the desired property that the quantity (see (\ref{zequiv}))
\begin{equation}
\label{eq:Z2F1}
\bZ_\Gamma(\dimeps) = \bReqv(\Gamma\cdot \tau,\dimeps)\bZcan_\Gamma(\tau,\dimeps) \bReqv( \tau,\dimeps)^{-1} = 
\begin{pmatrix}
a & \frac{b \pi \dimeps}{2 i } \\
\frac{2i c}{\pi  \dimeps} & d
\end{pmatrix}\,,
\end{equation}
becomes $\tau$-independent and turns the cocycle into a representation $\bZ_{\Gamma_1 \Gamma_2}(\dimeps) = \bZ_{\Gamma_1}(\dimeps) \bZ_{\Gamma_2}(\dimeps)$.
We also observe that $\overline{\bZ_\Gamma(-\dimeps)}= \bZ_\Gamma(\dimeps)$.
The resulting connection obtained from (\ref{eq:Ahk})~is 
\begin{equation}
\label{eq:Aeqv2F1}
\bA_{\textrm{eqv}}(\tau,\dimeps)  
= - \frac{\pi \dimeps\,  \rd \tau}{2 i}\begin{pmatrix}
h_2(\tau) +  \frac{\pi \dimeps\,\tau}{2 i}  h_4(\tau) & - \left(\frac{\pi  \dimeps\, \tau}{2 i}\right)^2 h_4(\tau) \\ 
h_4(\tau) & h_2(\tau) - \frac{\pi \dimeps\,\tau}{2 i } h_4(\tau)
\end{pmatrix}\,,
\end{equation}
and transforms $\Gamma(2)$-equivariantly: 
\beq\bA_{\textrm{eqv}}(\Gamma\cdot\tau,\dimeps) =  \bZ_\Gamma(\dimeps) \bA_{\textrm{eqv}}(\tau,\dimeps) \bZ_\Gamma(\dimeps)^{-1}\,,\qquad \Gamma\in\Gamma(2)\,.
\label{eq:A_eqv_trafo}
\eeq
We also see explicitly that the connection is no longer $\dimeps$-factorised, but instead has acquired polynomial corrections in $\dimeps$ (that in this case vanish in the limit $\dimeps\to 0$).
We also have a cocycle $\bU_\Gamma(\dimeps)= \operatorname{Pexp}\big[\int_{\tau_0}^{\Gamma\cdot\tau_0} \bA_{\textrm{eqv}}(\tau_1)\big]$. 

The form~\eqref{eq:Aeqv2F1} is the right one to compare with an abstract connection for iterated Eisenstein integrals for $\Gamma(2)$, generalising the one for $\mathrm{SL}_2(\mathbb{Z})$~\cite{Brown:2014pnb, Brown:2020mhp, Dorigoni:2022npe, Dorigoni:2024oft} and given by 
\begin{equation}
\label{eq:AG2}
\mathbb{A}(\tau) =-\rd \tau \sum_{\mu=3,4} \frac{f^{(2)}(\omega_\mu,\tau)}{2\pi i} \tilde{\eps}_{2|\mu} -  \rd \tau \sum_{k=4}^{\infty} (k{-}1) \sum_{j=0}^{k-2} \frac{(-1)^j}{j!} (2\pi i \tau)^{j} \sum_{\mu = 2}^4 \frac{ f^{(k)}(\omega_\mu,\tau) }{(2\pi i)^{k-1}} \eps_{k | \mu}^{(j)}\,.
\end{equation}
We shall discuss this connection in much more detail in the companion paper~\cite{lv2:progress}. Here it suffices to say that there are derivation-algebra-valued generators $\eps_{k|\mu}$ for every holomorphic Eisenstein series for $\Gamma(2)$. For modular weight $k=2$, there are only two Eisenstein series, and we denote the corresponding derivations by $\tilde\eps_{2|\mu}$ for reasons explained in~\cite{lv2:progress}.
The $\tilde\eps_{2|\mu}$ together with the $\mathfrak{sl}_2$-companions $\eps_{k|\mu}^{(j)}= \mathrm{ad}_{\eps_0}^j \eps_{k|\mu}$ ($j=0,1,\ldots,k-2$) of $\eps_{k|\mu}$ for $k\geq 4$  generate the pro-nilpotent algebra $\mathfrak{u}$, which is a subalgebra of the pro-nilpotent algebra $\nil$ in the body of the paper that in addition contains the (nilpotent) $\mathfrak{sl}_2$-generator $\eps_0$.
The functions $f^{(k)}(z,\tau)$ appearing in (\ref{eq:AG2}) are the expansion coefficients of the doubly-periodic Kronecker--Eisenstein series \cite{Brown:2011wfj, Broedel:2014vla} evaluated at specific $2$-torsion points $z=\omega_\mu \in \{ \frac{1}{2} , \, \frac{1{+}\tau}{2}, \, \frac{\tau}{2}\}$ and they furnish a basis of the space of the Eisenstein series. We can rewrite the modular forms $h_2(\tau)$ and $h_4(\tau)$ defined in (\ref{hfcts}) and appearing in~\eqref{eq:Aeqv2F1} in this basis according to
\begin{align}
 h_2(\tau)&=\frac{2}{\pi^2}\Big[(\beta+\gamma)\,f^{(2)}(\omega_3,\tau)+(3\alpha-\beta+2\gamma)\,f^{(2)}(\omega_4,\tau)\Big]\,,\nonumber
\\[1mm]
h_4(\tau) &= \frac{8}{\pi^4} \Big[ d_2  f^{(4)}(\omega_2,\tau) + d_3  f^{(4)}(\omega_3,\tau)+ d_4  f^{(4)}(\omega_4,\tau)  \Big] \, ,
\end{align}
where 
\beq
\bsp
d_2 &=18\alpha^2-7\alpha\beta+7\beta^2+15\alpha\gamma-7\beta\gamma+4\gamma^2\,, \\
d_3 &=18\alpha^2-13\alpha\beta+4\beta^2+21\alpha\gamma-7\beta\gamma+7\gamma^2\,, \\
d_4 &=9\alpha^2+5\alpha\beta+4\beta^2+9\alpha\gamma-\beta\gamma+4\gamma^2\,.
\esp
\label{defds}
\eeq

By identifying \eqref{eq:Aeqv2F1} as the matrix representation $\bA_{\textrm{eqv}}(\tau,\dimeps) = \rho(\mathbb{A}(\tau))$ of (\ref{eq:AG2}),
we can then read off the following non-vanishing matrix representatives for our example
\begin{align}
\label{eq:epsrep2F1}
\rho(\eps_0)  &= \frac{\dimeps}4\begin{pmatrix} 0 &1\\ 0 & 0 \end{pmatrix}\,,& 
\rho(\tilde\eps_{2|3} )&=2\dimeps(\beta+\gamma)\,\mathds{1}\,,&
\rho(\tilde\eps_{2|4}) &=2\dimeps(3\alpha-\beta+2\gamma)\,\mathds{1}\,, \notag
\\
\rho(\eps^{(0)}_{4|\mu})&=-\frac{32\dimeps}{3}\,d_\mu\begin{pmatrix}0&0\\1&0\end{pmatrix} \,,&
\rho(\eps^{(1)}_{4|\mu})&=-\frac{8\dimeps^2}{3}\,d_\mu\begin{pmatrix}1&0\\0&-1\end{pmatrix} \,,&
\rho(\eps^{(2)}_{4|\mu}) &=\frac{4\dimeps^3}{3}\,d_\mu\begin{pmatrix}0&1\\0&0\end{pmatrix}\,.
\end{align}
The last three equations are consistent with the adjoint of action of $\eps_0$, i.e.,
\beq
\big[\rho(\eps_0),\rho(\tilde\eps_{2|\mu})\big]=0\,,\quad \rho(\eps_{4|\mu}^{(1)}) = \big[\rho(\eps_0), \rho(\eps_{4|\mu}^{(0)})\big]\,,\quad  \rho(\eps_{4|\mu}^{(2)}) = \Big[\rho(\eps_0), \big[\rho(\eps_0), \rho(\eps_{4|\mu}^{(0)})\big]\Big]\,,
\eeq
as well as nilpotency, \beq
\big[\rho(\eps_0),\rho(\eps_{4|\mu}^{(2)})\big]=0\,.
\eeq
We also note that in this representation the change to equivariant gauge is achieved by $\bReqv(\tau,\dimeps)= q^{-\rho(\eps_0)}$ in (\ref{eq:Reqv2F1}) for $q=e^{2\pi i \tau}$.
The representation is highly unfaithful and infinitely many generators $\eps_{k|\mu}^{(j)}$ are represented trivially. 

The abstract algebra $\mathfrak{u}$ on
$\tilde{\eps}_{2|\mu}$ and $\eps_{k|\mu}^{(j)}$
 is not free, but it satisfies relations of Pollack type~\cite{Pollack,Hopper:2021ell,lv2:progress} associated with $\Gamma(2)$ cusp forms. In the representation $\rho$, the only potentially non-trivial remnant is the condition 
\beq
\big[\rho(\eps_{4|2}),\rho(\eps_{4|3})\big]+\big[\rho(\eps_{4|3}),\rho(\eps_{4|4})\big]+\big[\rho(\eps_{4|4}),\rho(\eps_{4|2})\big]=0\,,
\eeq
which  can be verified in the representation~\eqref{eq:epsrep2F1}.

\paragraph{\bf Initial conditions and zeta generators.}

For the period matrix~\eqref{eq:pertau} 
with an $\dimeps$-factorised differential equation, we can also determine a decomposition into the general normalised solution $\bUcan(\tau,\dimeps)$ and initial conditions similar to~\eqref{eq:initMQ}. We now take the boundary point $\tau_0\to i \infty$ and wish to isolate the matrices $\bM(\dimeps)$ and $\bQ(\dimeps)$ in (cf.~\eqref{eq:N_fac})
\begin{equation}
\bPtilde(\tau,\dimeps) = \bUcan(\tau,\dimeps) \bReqv(i\infty,\dimeps)^{-1}  \,\bM(\dimeps)\,\bQ(\dimeps)\,,
\quad \text{where}\quad  \bReqv(i\infty,\dimeps)^{-1}= \mathds{1}
\label{qqmm}
\end{equation}
is the tangentially regularised limit of~\eqref{eq:Reqv2F1} for $\tau \to i \infty$ since $\frac{\pi \dimeps \tau}{2i} =  \log (q^{-\frac14\dimeps }) $, which regulates to zero. 
One can show that $\bM(\dimeps)$ is a power series in $\dimeps$ that contains only rational numbers, $\log 2$ and odd zeta values, while $\bQ(\dimeps)$ is a Laurent series in $i \pi \dimeps$ with a simple pole and that only contains rational numbers. 
At the lowest orders, we have
\begin{align}
\label{eq:MQ2F1}
\bM(\dimeps) &=  \mathds{1}  + \dimeps \log 2\,  \bM_1 + \frac12 \dimeps^2 \log^2 2 \,\bM_1^2  + \dimeps^3 \left(\zeta_3 \bM_3 +\frac16 \log^3 2 \,\bM_1^3\right) + \mathcal{O}(\dimeps^4)\,, \nonumber\\
\bQ(\dimeps) &= \frac1{i \pi \dimeps} \bQ_{-1} + \bQ_0 + i \pi \dimeps \bQ_1 +  \mathcal{O}(\dimeps^2)\,,
\end{align}
with the coefficients of $\bM(\dimeps)$ given by 
\begin{align}
\bM_1 &=\begin{pmatrix}2(\alpha+\gamma)&0\\ 0 & 2(\alpha+\gamma)\end{pmatrix}\,,\quad \nonumber\\
\bM_w & =
\frac{2^w-1}{2w}
\Bigl(
\alpha^w-(2\alpha+\gamma)^w
+\beta^w+(\alpha+\gamma-\beta)^w
\Bigr)\mathds{1}
\\[2mm]
&\quad +
\frac{
(2^w-1)
\Bigl(
\alpha^w-(2\alpha+\gamma)^w
-\beta^w-(\alpha+\gamma-\beta)^w
\Bigr)
+2(\alpha+\gamma)^w
}{
2w(\alpha+\gamma)
}
\begin{pmatrix}
0 & 1\\
(\alpha+\gamma)^2 & 0
\end{pmatrix} \,,\nonumber
\end{align}
for odd $w\geq 3$. Since the matrices $\bM_k$, $\bM_\ell$ commute for any odd $k,\ell$ one can also write $\bM(\dimeps)$ as
\begin{equation}
\label{eq:Mclosed}
\bM(\dimeps)=4^{(\alpha+\gamma)\dimeps}
\exp\!\bigg(\sum_{\substack{w\geq3\\w\ {\rm odd}}}
 \zeta_w\dimeps^w\bM_w
\bigg)
\,.
\end{equation}
The first few orders of $\bQ(\dimeps)$ in (\ref{qqmm}) are
\begin{equation}
\bQ_{-1} = \begin{pmatrix}0&0\\0&-2\end{pmatrix}\,,\quad
\bQ_0=\begin{pmatrix}1&1\\-(\alpha+\gamma)&\alpha-2\beta+\gamma\end{pmatrix}\,,\quad
\bQ_1 = \begin{pmatrix}0&-\frac12(2\alpha-\beta+\gamma)\\[1mm]
0&-\frac43\alpha^2+\frac32\alpha\beta-\beta^2-\frac{13}{6}\alpha\gamma+\frac32\beta\gamma-\frac56\gamma^2\
\end{pmatrix}\,.
\end{equation}
The closed formula~\eqref{eq:Mclosed} makes it manifest that $\bM(\dimeps)$ is group-like with commuting and non-vanishing coefficients $\bM_1$ and $\bM_{2k+1}$ of $\log 2$ and $\zeta_{2k+1}$, respectively, and given by an exponential of the primitive elements. This is again consistent with the coaction formula~\eqref{knownco}.
We also note that the matrices $\bM_w \dimeps^w=\rho(\sigma_w)$ are matrix representations of (genus-one) zeta generators $\sigma_w$ (for odd $w$) that appear in the abstract construction of iterated Eisenstein integrals for $\mathrm{SL}_2(\mathbb{Z})$~\cite{EnriquezEllAss,Brown:depth3,hainmatsumoto2020}, where the extension to $\Gamma(2)$~\cite{Hopper:2022ell, Hirose:2023,lv2:progress} introduces the additional generator $\sigma_1$ associated with $\log 2$.

The final step is to verify that the commutation relations between the zeta generators $\sigma_w$ and the derivations $\tilde{\eps}_{2|\mu}$ and $\eps_{k|\mu}$ are respected. We shall study them in much more detail in~\cite{lv2:progress}, but record here that all relations that one can access with this unfaithful representation $\rho$ of $\sigma_w$, $\tilde{\eps}_{2|\mu}$ and $\eps_{k|\mu}$ are indeed satisfied, many of them trivially, since $\bM_1$ and $\tilde\eps_{2|\mu}$ are proportional to the identity matrix. A non-trivial check is obtained from commutators of the form $[\sigma_3, \eps_{4|\mu}]$ that do not vanish in the representation. In~\cite{lv2:progress} we shall exhibit the abstract algebra expressions for these commutators (in a different ordering convention for $\operatorname{Pexp}$, coming from considering the differential equation of the inverse period matrix) and we have verified explicitly that the above matrices are consistent with this change of convention. In the present convention, this commutation relation reads,
\beq
\big[\rho(\sigma_3), \rho(\eps_{4|\mu})\big] = -\frac38 \big[\rho(\eps_{4|2}^{(2)}),\rho(\eps_{4|\mu})\big]\,.
\eeq
Zeta generators have a decomposition into `arithmetic' and `geometric' parts~\cite{Brown:2014pnb}, $\sigma_w= z_w+ \sigma_w^{\mathrm{g}}$ with $\sigma_w^{\mathrm{g}}\in\mathfrak{u}$ made out of $\tilde\eps_{2|\mu}$ and $\eps_{k|\mu}^{(j)}$. Computing the arithmetic parts 
$z_1$ and $z_w$ with $w \geq 3$ odd in the present representation we find
\begin{equation}\label{eq:rho(z)}
    \rho(z_1)  = 0
    \,,\quad\quad
   \rho(z_w) =  \frac{1}{2} \dimeps^w \operatorname{tr}(\bM_w)\mathds{1} = 
 \frac{2^w-1}{2w}    \dimeps^w 
\Bigl(
\alpha^w-(2\alpha+\gamma)^w
+\beta^w+(\alpha+\gamma-\beta)^w
\Bigr)  \mathds{1}
\, ,
\end{equation}
where the proportionality to the unit matrix reflects the $\mathfrak{sl}_2$-singlet property of $z_w$.

\paragraph{\bf Single-valued and equivariant analogues.} 
The period matrix in equivariant gauge is $\bP(\tau,\dimeps) = \bU(\tau,\dimeps)\bM(\dimeps)\bQ(\dimeps)$, where $\bU(\tau,\dimeps)$ is the path-ordered exponential obtained by exponentiating the connection matrix $\bA_{\textrm{eqv}}$. The Laurent expansion of $\bP(\tau,\dimeps)$ can be expressed order by order in $\dimeps$ in terms of iterated Eisenstein integrals for the congruence subgroup $\Gamma(2)$.

Using~\eqref{eq:Usv} and~\eqref{eq:Psv}, we can construct the single-valued analogue $\bP^{\sv}$ of $\bP$,
\beq
\bP^{\sv}(\tau,\dimeps) = \bU(\tau,\dimeps)\bM^{\sv}(\dimeps)\overline{\bU(\tau,-\dimeps)^{-1}}\,,
\label{psveq}
\eeq
where we recall  $\bM^\sv(\dimeps)=\bM(\dimeps)\overline{\bM(-\dimeps)}$.
The single-valued analogue $\bP^{\sv}$ inherits the modular transformation properties of the connection matrix $\bA_{\textrm{eqv}}$ from~\eqref{eq:A_eqv_trafo}, and thus transforms equivariantly under modular transformations
\beq\bP^{\sv}(\Gamma\cdot\tau,\dimeps) =  \bZ_\Gamma(\dimeps) \bP^{\sv}(\tau,\dimeps) \bZ_\Gamma(\dimeps)^{-1}\,,\qquad \Gamma\in\Gamma(2)\,.
\label{eq:Psv_equivariance}
\eeq
The equivariance property implies that the Laurent expansion of $\bP^{\sv}$ involves \emph{equivariant}  iterated Eisenstein integrals for $\Gamma(2)$ accompanied by rational polynomials in $\varepsilon\log 2$, $\varepsilon^{2k+1}\zeta_{2k+1}$ and $\alpha$, $\beta$, $\gamma$. We have explicitly checked up to and including order $\dimeps^{6}$ 
that we can express the Laurent coefficients in terms of the equivariant analogues of iterated integrals of Eisenstein series for $\Gamma(2)$ from~\cite{Duhr:2025lvr,lv2:progress}. At this point a comment is in order: 
Following \cite{Brown:2020mhp, Dorigoni:2024oft}, the companion paper \cite{lv2:progress} to this work introduces the \emph{equivariant} analogues of iterated integral of Eisenstein series of level 2 which are obtained from the matrix
\beq\bP^{\textrm{eqv}}(\tau,\dimeps) = \bU(\tau,\dimeps)\bM^{\sv}(\dimeps)\overline{\bU(\tau,-\dimeps)^{-1}}\bM^{\rm sv}_z(\dimeps)^{-1} = \bP^{\sv}(\tau,\dimeps)\bM^{\rm sv}_z(\dimeps)^{-1}\,,
\label{peqveq}
\eeq
where $\bM^{\rm sv}_z(\dimeps)$ is obtained from $\bM^{\sv}(\dimeps)$ by replacing the matrices representing the zeta generators, $\dimeps^w\,\bM_w = \rho(\sigma_w)$ by the representations $\rho(z_w)$ of the arithmetic parts only,
\beq
\bM^{\rm sv}_z(\eps) = \bM^{\sv}(\eps)|_{\rho(\sigma_w)\to \rho(z_w)}\,.
\eeq
Then $\bP^{\textrm{eqv}}$ has the same equivariance properties as in~\eqref{eq:Psv_equivariance}, but with $^{\sv}$ replaced by $^{\textrm{eqv}}$. For this to be true, we must have that $\bZ_{\Gamma}(\dimeps)$ and $\bM^{\rm sv}_z(\dimeps)$ commute. From~\eqref{eq:Z2F1} one can easily see that $\bZ_{\Gamma}(\dimeps)$ is an irreducible representation of $\Gamma(2)$ (in fact, of $\operatorname{SL}_2(\mathbb{Z}))$, and so by Schur's lemma, $\bM^{\rm sv}_z(\dimeps)$ must be proportional to the identity matrix, in agreement with our explicit results in~\eqref{eq:rho(z)}.

In the abstract setting, the motivation for the multiplication by $\bM^{\rm sv}_z(\dimeps)^{-1}$ in (\ref{peqveq}) is to attain a series $\bP^{\textrm{eqv}}(\tau,\dimeps)$ which preserves equivariance and solely involves the generators $\eps_{k|\mu}^{(j)}$ of $\mathfrak{u}$ in its expansion, without any separate coefficients of
(arithmetic) zeta generators $z_w$. One may instead
pass to the single-valued version (\ref{eq:Usv}) of the path-ordered exponential by replacing the rightmost factor in (\ref{peqveq}) by $\bM^{\rm sv}_z(\dimeps)^{-1} \mapsto \bM^{\rm sv}(\dimeps)^{-1}$ in terms of the full zeta generators. This still yields a series $\bUeqv^{\textrm{sv}}(\tau,\dimeps) =  \bP^{\sv}(\tau,\dimeps)\bM^{\rm sv}(\dimeps)^{-1}$ in a matrix representation of the Lie group of $\mathfrak{u}$, but spoils the equivariance property (\ref{eq:Psv_equivariance}) and in fact illustrates a general feature: Trivialising the boundary condition $\tau \rightarrow i \infty$ of $\bP^{\sv}(\tau,\dimeps)$ in (\ref{psveq})
via right-multiplication by $\bM^{\rm sv}(\dimeps)^{-1}$ 
is incompatible with the equivariant transformation (\ref{eq:Psv_equivariance}), a trade-off which in an abstract setting was described in \cite{Brown:2020mhp, Dorigoni:2024oft} for iterated Eisenstein integrals for ${\rm SL}_2(\mathbb Z)$. Our example suggests that, more generally, modular properties of Feynman integrals in the kinematic variables $\bx$ rely on the interplay between the iterated integrals of $\bUeqv(\bx,\dimeps)$ in (\ref{toeqvg}) and the matrix representations of zeta generators in the initial condition $\bM(\dimeps)$.

\paragraph{\bf Coactions on hypergeometric functions.}
Using the same reasoning as in the derivation of~\eqref{eq:Feynman_coaction}, we can immediately write down the coaction on the entries of (the motivic uplifts of) the period matrix $\bP$ in equivariant gauge, 
\beq\label{eq:hypgeo_coaction_1}
\Delta\big(\bP^{\mathfrak{m}}(\tau,\dimeps)\big) = \bP^{\nil}(\tau,\dimeps)\bP^{\mathfrak{m}}(\tau,\dimeps)\,,
\eeq
where $\bP^{\nil}$ is the de Rham analogue of the period matrix. Explicitly, we have for $1\le i,j\le2$, 
\beq
\Delta\big(\bP^{\mathfrak{m}}(\tau,\dimeps)_{ij}\big) = \bP^{\mathfrak{m}}(\tau,\dimeps)_{1j}\otimes\bP^{\nil}(\tau,\dimeps)_{i1}+\bP^{\mathfrak{m}}(\tau,\dimeps)_{2j}\otimes\bP^{\nil}(\tau,\dimeps)_{i2}\,.
\label{eq:hypgeo_coaction_2}
\eeq
In passing from~\eqref{eq:hypgeo_coaction_1} to~\eqref{eq:hypgeo_coaction_2} we reinstated the $\otimes$ symbol according to~\eqref{eq:XuYm}. 
The motivic uplifts can be represented in terms of the entries of $\bP$ via the period map,
\beq
\bP^{\mathfrak{m}}(\tau,\dimeps)_{ij} \longrightarrow \bP(\tau,\dimeps)_{ij}\,.
\eeq
The period map is not available for de Rham periods. Instead, they can be represented via the single-valued analogues 
\beq
\bP^{\nil}(\tau,\dimeps)_{ij} \longrightarrow \bP^{\sv}(\tau,\dimeps)_{ij}\,.
\eeq
In the same way as the equivariance property (\ref{eq:Psv_equivariance}) of $\bP^{\sv}(\tau,\dimeps)$ hinges on the initial conditions, the simple form (\ref{eq:hypgeo_coaction_1}) of the coaction of $\bP(\tau,\dimeps) = \bU(\tau,\dimeps)\bM(\dimeps)\bQ(\dimeps)$ is sensitive to the $\bM(\dimeps)$ factor of the initial conditions which carries matrix representations of zeta generators. When isolating the series $\bU(\tau,\dimeps)$ of iterated Eisenstein integrals, one arrives at the coaction formula
$\Delta\big(\bU^\mot(\tau,\dimeps)\big) = \bUeqv^{\nil}(\tau,\dimeps)\bM^{\nil}(\dimeps) \bUeqv^{\mot}(\tau,\dimeps)\bM^{\nil}(\dimeps)^{-1}$ of (\ref{eq:coaction}) where $\bUeqv^{\mot}(\tau,\dimeps)$ is additionally twisted by the zeta generators of the initial condition.

The coaction in~\eqref{eq:hypgeo_coaction_1} and~\eqref{eq:hypgeo_coaction_2} is structurally very similar to the coaction on hypergeometric functions whose Laurent expansions involve MPLs 
~\cite{Abreu:2019wzk,Brown:2019jng,Kamlesh:2024cvo}. In particular, we expect to find that our coaction formula is consistent with expansions in $\dimeps$ and the proposals for the coaction on the iterated integrals of Eisenstein series from~\cite{Kleinschmidt:2025dtk,mat:progress}. However, we note that, unlike for the previous example where the Laurent coefficients only involve MPLs, in this case we
do not attempt to represent the de Rham analogues as images of motivic periods under a de Rham projection. The representation as single-valued analogues, however, should always be available, cf.~\cite{Brown:2015ylf, Brown:2018omk, Brown:2019jng}. We expect this to be one of the key changes required to extend the diagrammatic coaction in dimensional regularisation as in~\cite{Abreu:2017enx,Abreu:2017mtm,Abreu:2021vhb} to Feynman integrals beyond polylogarithms, like the two-loop sunrise integral with three massive propagators. It would be interesting to understand if our coaction is related to the one appearing in
\cite{Tapuskovic:2023xiu} and more generally in
the context of the theory of Mellin motives currently being developed~\cite{Mellin:progress}.


\begin{thebibliography}{100}

\bibitem{tHooft:1972tcz}
G.~'t~Hooft and M.~J.~G. Veltman, {\it {Regularization and Renormalization of
  Gauge Fields}},  {\em Nucl. Phys. B} {\bf 44} (1972) 189--213.

\bibitem{Cicuta:1972jf}
G.~M. Cicuta and E.~Montaldi, {\it {Analytic renormalization via continuous
  space dimension}},  {\em Lett. Nuovo Cim.} {\bf 4} (1972) 329--332.

\bibitem{Bollini:1972ui}
C.~G. Bollini and J.~J. Giambiagi, {\it {Dimensional Renormalization: The
  Number of Dimensions as a Regularizing Parameter}},  {\em Nuovo Cim. B} {\bf
  12} (1972) 20--26.

\bibitem{AomotoKita}
K.~Aomoto and M.~Kita, {\em Theory of Hypergeometric Functions}.
\newblock Springer Monographs in Mathematics. Springer Japan, 2011.

\bibitem{Mastrolia:2018uzb}
P.~Mastrolia and S.~Mizera, {\it {Feynman Integrals and Intersection Theory}},
  {\em JHEP} {\bf 02} (2019) 139, [\href{http://arxiv.org/abs/1810.03818}{{\tt
  arXiv:1810.03818}}].

\bibitem{MR1852188}
M.~Kontsevich and D.~Zagier, {\it Periods},  in {\em Mathematics
  unlimited---2001 and beyond}, pp.~771--808.
\newblock Springer, Berlin, 2001.

\bibitem{Bogner:2007mn}
C.~Bogner and S.~Weinzierl, {\it {Periods and Feynman integrals}},  {\em J.
  Math. Phys.} {\bf 50} (2009) 042302,
  [\href{http://arxiv.org/abs/0711.4863}{{\tt arXiv:0711.4863}}].

\bibitem{Lappo:1953m}
J.~A. Lappo-Danilevsky, {\em M\'emoires sur la th\'eorie des syst\`emes des
  \'equations diff\'erentielles lin\'eaires}.
\newblock Chelsea Publishing Co., New York, 1953.

\bibitem{Remiddi:1999ew}
E.~Remiddi and J.~A.~M. Vermaseren, {\it {Harmonic polylogarithms}},  {\em Int.
  J. Mod. Phys. A} {\bf 15} (2000) 725--754,
  [\href{http://arxiv.org/abs/hep-ph/9905237}{{\tt hep-ph/9905237}}].

\bibitem{Goncharov:1998kja}
A.~B. Goncharov, {\it {Multiple polylogarithms, cyclotomy and modular
  complexes}},  {\em Math. Res. Lett.} {\bf 5} (1998) 497--516,
  [\href{http://arxiv.org/abs/1105.2076}{{\tt arXiv:1105.2076}}].

\bibitem{Goncharov:2001iea}
A.~B. Goncharov, {\it {Multiple polylogarithms and mixed Tate motives}},
  \href{http://arxiv.org/abs/math/0103059}{{\tt math/0103059}}.

\bibitem{Waldschmidt:2002}
M.~Waldschmidt, {\it Multiple polylogarithms: an introduction},  in {\em Number
  theory and discrete mathematics ({C}handigarh, 2000)}, Trends Math.,
  pp.~1--12.
\newblock Birkh\"auser, Basel, 2002.

\bibitem{Deligne1989TheGR}
P.~P. Deligne, {\it Le groupe fondamental de la droite projective moins trois
  points},  in {\em Galois Groups over $\mathbb{Q}$}, (New York, NY),
  pp.~79--297, Springer US, 1989.

\bibitem{Brown:2014pnb}
F.~Brown, {\it {Multiple Modular Values and the relative completion of the
  fundamental group of $M_{1,1}$}},  \href{http://arxiv.org/abs/1407.5167}{{\tt
  arXiv:1407.5167}}.

\bibitem{Panzer:2015ida}
E.~Panzer, {\em {Feynman integrals and hyperlogarithms}}.
\newblock PhD thesis, Humboldt U., 2015.
\newblock \href{http://arxiv.org/abs/1506.07243}{{\tt arXiv:1506.07243}}.

\bibitem{Abreu:2022mfk}
S.~Abreu, R.~Britto, and C.~Duhr, {\it {The SAGEX review on scattering
  amplitudes Chapter 3: Mathematical structures in Feynman integrals}},  {\em
  J. Phys. A} {\bf 55} (2022), no.~44 443004,
  [\href{http://arxiv.org/abs/2203.13014}{{\tt arXiv:2203.13014}}].

\bibitem{Gehrmann:2001pz}
T.~Gehrmann and E.~Remiddi, {\it {Numerical evaluation of harmonic
  polylogarithms}},  {\em Comput. Phys. Commun.} {\bf 141} (2001) 296--312,
  [\href{http://arxiv.org/abs/hep-ph/0107173}{{\tt hep-ph/0107173}}].

\bibitem{Gehrmann:2001jv}
T.~Gehrmann and E.~Remiddi, {\it {Numerical evaluation of two-dimensional
  harmonic polylogarithms}},  {\em Comput. Phys. Commun.} {\bf 144} (2002)
  200--223, [\href{http://arxiv.org/abs/hep-ph/0111255}{{\tt hep-ph/0111255}}].

\bibitem{Vollinga:2004sn}
J.~Vollinga and S.~Weinzierl, {\it {Numerical evaluation of multiple
  polylogarithms}},  {\em Comput. Phys. Commun.} {\bf 167} (2005) 177,
  [\href{http://arxiv.org/abs/hep-ph/0410259}{{\tt hep-ph/0410259}}].

\bibitem{Buehler:2011ev}
S.~Buehler and C.~Duhr, {\it {CHAPLIN - Complex Harmonic Polylogarithms in
  Fortran}},  {\em Comput. Phys. Commun.} {\bf 185} (2014) 2703--2713,
  [\href{http://arxiv.org/abs/1106.5739}{{\tt arXiv:1106.5739}}].

\bibitem{Frellesvig:2016ske}
H.~Frellesvig, D.~Tommasini, and C.~Wever, {\it {On the reduction of
  generalized polylogarithms to $\text{Li}_n$ and $\text{Li}_{2,2}$ and on the
  evaluation thereof}},  {\em JHEP} {\bf 03} (2016) 189,
  [\href{http://arxiv.org/abs/1601.02649}{{\tt arXiv:1601.02649}}].

\bibitem{Ablinger:2018sat}
J.~Ablinger, J.~Bl{\"u}mlein, M.~Round, and C.~Schneider, {\it {Numerical
  Implementation of Harmonic Polylogarithms to Weight w = 8}},  {\em Comput.
  Phys. Commun.} {\bf 240} (2019) 189--201,
  [\href{http://arxiv.org/abs/1809.07084}{{\tt arXiv:1809.07084}}].

\bibitem{Naterop:2019xaf}
L.~Naterop, A.~Signer, and Y.~Ulrich, {\it {handyG {\textemdash}Rapid numerical
  evaluation of generalised polylogarithms in Fortran}},  {\em Comput. Phys.
  Commun.} {\bf 253} (2020) 107165,
  [\href{http://arxiv.org/abs/1909.01656}{{\tt arXiv:1909.01656}}].

\bibitem{Ablinger:2013cf}
J.~Ablinger, J.~Bl{\"u}mlein, and C.~Schneider, {\it {Analytic and Algorithmic
  Aspects of Generalized Harmonic Sums and Polylogarithms}},  {\em J. Math.
  Phys.} {\bf 54} (2013) 082301, [\href{http://arxiv.org/abs/1302.0378}{{\tt
  arXiv:1302.0378}}].

\bibitem{Duhr:2014woa}
C.~Duhr, {\it {Mathematical aspects of scattering amplitudes}},  in {\em
  {Theoretical Advanced Study Institute in Elementary Particle Physics}:
  {Journeys Through the Precision Frontier: Amplitudes for Colliders}},
  pp.~419--476, 2015.
\newblock \href{http://arxiv.org/abs/1411.7538}{{\tt arXiv:1411.7538}}.

\bibitem{Panzer:2014caa}
E.~Panzer, {\it {Algorithms for the symbolic integration of hyperlogarithms
  with applications to Feynman integrals}},  {\em Comput. Phys. Commun.} {\bf
  188} (2015) 148--166, [\href{http://arxiv.org/abs/1403.3385}{{\tt
  arXiv:1403.3385}}].

\bibitem{Duhr:2019tlz}
C.~Duhr and F.~Dulat, {\it {PolyLogTools {\textemdash} polylogs for the
  masses}},  {\em JHEP} {\bf 08} (2019) 135,
  [\href{http://arxiv.org/abs/1904.07279}{{\tt arXiv:1904.07279}}].

\bibitem{Kardos:2025klp}
A.~Kardos, S.-O. Moch, and O.~Schnetz, {\it {HyperFORM: a FORM package for
  parametric integration with hyperlogarithms}},  {\em Eur. Phys. J. C} {\bf
  86} (2026), no.~6 659, [\href{http://arxiv.org/abs/2511.19992}{{\tt
  arXiv:2511.19992}}].

\bibitem{hyperlogprocedures}
O.~Schnetz, ``{H}yper{L}og{P}rocedures -- {M}aple procedures available on the
  homepage of the author.''
\newblock https://www.math.fau.de/person/oliver-schnetz/.

\bibitem{Goncharov:2005sla}
A.~Goncharov, {\it {Galois symmetries of fundamental groupoids and
  noncommutative geometry}},  {\em Duke Math.~J.} {\bf 128} (2005) 209,
  [\href{http://arxiv.org/abs/math/0208144}{{\tt math/0208144}}].

\bibitem{Goncharov:2010jf}
A.~B. Goncharov, M.~Spradlin, C.~Vergu, and A.~Volovich, {\it {Classical
  Polylogarithms for Amplitudes and Wilson Loops}},  {\em Phys. Rev. Lett.}
  {\bf 105} (2010) 151605, [\href{http://arxiv.org/abs/1006.5703}{{\tt
  arXiv:1006.5703}}].

\bibitem{Brown:2011ik}
F.~Brown, {\it {On the decomposition of motivic multiple zeta values}},  in
  {\em Galois-Teichm\"uller theory and arithmetic geometry}, vol.~68 of {\em
  Adv. Studies in Pure Math.}, pp.~31--58, Math. Soc. Japan, 2012.
\newblock \href{http://arxiv.org/abs/1102.1310}{{\tt arXiv:1102.1310}}.

\bibitem{brownmixedZ}
F.~Brown, {\it {Mixed Tate motives over $\mathbb{Z}$}},  {\em Annals of
  Mathematics} {\bf 175} (2012) 949--976,
  [\href{http://arxiv.org/abs/1102.1312}{{\tt arXiv:1102.1312}}].

\bibitem{Duhr:2011zq}
C.~Duhr, H.~Gangl, and J.~R. Rhodes, {\it {From polygons and symbols to
  polylogarithmic functions}},  {\em JHEP} {\bf 10} (2012) 075,
  [\href{http://arxiv.org/abs/1110.0458}{{\tt arXiv:1110.0458}}].

\bibitem{Duhr:2012fh}
C.~Duhr, {\it {Hopf algebras, coproducts and symbols: an application to Higgs
  boson amplitudes}},  {\em JHEP} {\bf 08} (2012) 043,
  [\href{http://arxiv.org/abs/1203.0454}{{\tt arXiv:1203.0454}}].

\bibitem{Brown:2008um}
F.~Brown, {\it {The Massless higher-loop two-point function}},  {\em Commun.
  Math. Phys.} {\bf 287} (2009) 925--958,
  [\href{http://arxiv.org/abs/0804.1660}{{\tt arXiv:0804.1660}}].

\bibitem{Bogner:2014mha}
C.~Bogner and F.~Brown, {\it {Feynman integrals and iterated integrals on
  moduli spaces of curves of genus zero}},  {\em Commun. Num. Theor. Phys.}
  {\bf 09} (2015) 189--238, [\href{http://arxiv.org/abs/1408.1862}{{\tt
  arXiv:1408.1862}}].

\bibitem{Anastasiou:2013srw}
C.~Anastasiou, C.~Duhr, F.~Dulat, and B.~Mistlberger, {\it {Soft triple-real
  radiation for Higgs production at N3LO}},  {\em JHEP} {\bf 07} (2013) 003,
  [\href{http://arxiv.org/abs/1302.4379}{{\tt arXiv:1302.4379}}].

\bibitem{Ablinger:2014yaa}
J.~Ablinger, J.~Bl{\"u}mlein, C.~Raab, C.~Schneider, and F.~Wi{\ss}brock, {\it
  {Calculating Massive 3-loop Graphs for Operator Matrix Elements by the Method
  of Hyperlogarithms}},  {\em Nucl. Phys. B} {\bf 885} (2014) 409--447,
  [\href{http://arxiv.org/abs/1403.1137}{{\tt arXiv:1403.1137}}].

\bibitem{Caron-Huot:2020bkp}
S.~Caron-Huot, L.~J. Dixon, J.~M. Drummond, F.~Dulat, J.~Foster,
  O.~G\"urdo\u{g}an, M.~von Hippel, A.~J. McLeod, and G.~Papathanasiou, {\it
  {The Steinmann Cluster Bootstrap for $\mathcal{N}$ = 4 Super Yang-Mills
  Amplitudes}},  {\em PoS} {\bf CORFU2019} (2020) 003,
  [\href{http://arxiv.org/abs/2005.06735}{{\tt arXiv:2005.06735}}].

\bibitem{Kotikov:1990kg}
A.~V. Kotikov, {\it {Differential equations method: New technique for massive
  Feynman diagrams calculation}},  {\em Phys. Lett. B} {\bf 254} (1991)
  158--164.

\bibitem{Kotikov:1991pm}
A.~V. Kotikov, {\it {Differential equation method: The Calculation of N point
  Feynman diagrams}},  {\em Phys. Lett. B} {\bf 267} (1991) 123--127. [Erratum:
  Phys.Lett.B 295, 409--409 (1992)].

\bibitem{Kotikov:1991hm}
A.~V. Kotikov, {\it {Differential equations method: The Calculation of vertex
  type Feynman diagrams}},  {\em Phys. Lett. B} {\bf 259} (1991) 314--322.

\bibitem{Remiddi:1997ny}
E.~Remiddi, {\it {Differential equations for Feynman graph amplitudes}},  {\em
  Nuovo Cim. A} {\bf 110} (1997) 1435--1452,
  [\href{http://arxiv.org/abs/hep-th/9711188}{{\tt hep-th/9711188}}].

\bibitem{Gehrmann:1999as}
T.~Gehrmann and E.~Remiddi, {\it {Differential equations for two loop four
  point functions}},  {\em Nucl. Phys. B} {\bf 580} (2000) 485--518,
  [\href{http://arxiv.org/abs/hep-ph/9912329}{{\tt hep-ph/9912329}}].

\bibitem{Henn:2013pwa}
J.~M. Henn, {\it {Multiloop integrals in dimensional regularization made
  simple}},  {\em Phys. Rev. Lett.} {\bf 110} (2013) 251601,
  [\href{http://arxiv.org/abs/1304.1806}{{\tt arXiv:1304.1806}}].

\bibitem{Pogel:2022yat}
S.~P\"ogel, X.~Wang, and S.~Weinzierl, {\it {The three-loop equal-mass banana
  integral in \ensuremath{\varepsilon}-factorised form with meromorphic modular
  forms}},  {\em JHEP} {\bf 09} (2022) 062,
  [\href{http://arxiv.org/abs/2207.12893}{{\tt arXiv:2207.12893}}].

\bibitem{Pogel:2022ken}
S.~P{\"o}gel, X.~Wang, and S.~Weinzierl, {\it {Taming Calabi-Yau Feynman
  Integrals: The Four-Loop Equal-Mass Banana Integral}},  {\em Phys. Rev.
  Lett.} {\bf 130} (2023), no.~10 101601,
  [\href{http://arxiv.org/abs/2211.04292}{{\tt arXiv:2211.04292}}].

\bibitem{Pogel:2022vat}
S.~P\"ogel, X.~Wang, and S.~Weinzierl, {\it {Bananas of equal mass: any loop,
  any order in the dimensional regularisation parameter}},  {\em JHEP} {\bf 04}
  (2023) 117, [\href{http://arxiv.org/abs/2212.08908}{{\tt arXiv:2212.08908}}].

\bibitem{Gorges:2023zgv}
L.~G\"orges, C.~Nega, L.~Tancredi, and F.~J. Wagner, {\it {On a procedure to
  derive \ensuremath{\epsilon}-factorised differential equations beyond
  polylogarithms}},  {\em JHEP} {\bf 07} (2023) 206,
  [\href{http://arxiv.org/abs/2305.14090}{{\tt arXiv:2305.14090}}].

\bibitem{Duhr:2025lbz}
C.~Duhr, S.~Maggio, C.~Nega, B.~Sauer, L.~Tancredi, and F.~J. Wagner, {\it
  {Aspects of canonical differential equations for Calabi-Yau geometries and
  beyond}},  {\em JHEP} {\bf 06} (2025) 128,
  [\href{http://arxiv.org/abs/2503.20655}{{\tt arXiv:2503.20655}}].

\bibitem{Maggio:2025jel}
S.~Maggio and Y.~Sohnle, {\it {On canonical differential equations for
  Calabi-Yau multi-scale Feynman integrals}},  {\em JHEP} {\bf 10} (2025) 202,
  [\href{http://arxiv.org/abs/2504.17757}{{\tt arXiv:2504.17757}}].

\bibitem{e-collaboration:2025frv}
{\bf {\ensuremath{\varepsilon}}-\!} Collaboration, I.~Bree et~al., {\it
  {Geometric Bookkeeping Guide to Feynman Integral Reduction and
  {\ensuremath{\epsilon}}-Factorized Differential Equations}},  {\em Phys. Rev.
  Lett.} {\bf 136} (2026), no.~24 241602,
  [\href{http://arxiv.org/abs/2506.09124}{{\tt arXiv:2506.09124}}].

\bibitem{Bree:2025tug}
{\bf {\ensuremath{\varepsilon}}-\!} Collaboration, I.~Bree et~al., {\it {New
  algorithms for Feynman integral reduction and epsilon-factorized differential
  equations}},  {\em Phys. Rev. D} {\bf 113} (2026), no.~11 116019,
  [\href{http://arxiv.org/abs/2511.15381}{{\tt arXiv:2511.15381}}].

\bibitem{e-collaboration:2026aup}
{\bf {\ensuremath{\varepsilon}}-\!} Collaboration, I.~Bree et~al., {\it
  {Improving integration-by-parts and differential equations}},
  \href{http://arxiv.org/abs/2602.10651}{{\tt arXiv:2602.10651}}.

\bibitem{Forner:2026vby}
F.~Forner, C.~C. Mella, C.~Nega, L.~Tancredi, and F.~J. Wagner, {\it {Integrand
  analysis, leading singularities and canonical bases beyond polylogarithms}},
  {\em JHEP} {\bf 10} (2026) 213, [\href{http://arxiv.org/abs/2604.25270}{{\tt
  arXiv:2604.25270}}].

\bibitem{Usyukina:1992jd}
N.~I. Usyukina and A.~I. Davydychev, {\it {An Approach to the evaluation of
  three and four point ladder diagrams}},  {\em Phys. Lett. B} {\bf 298} (1993)
  363--370.

\bibitem{Usyukina:1993ch}
N.~I. Usyukina and A.~I. Davydychev, {\it {Exact results for three and four
  point ladder diagrams with an arbitrary number of rungs}},  {\em Phys. Lett.
  B} {\bf 305} (1993) 136--143.

\bibitem{Dixon:2012yy}
L.~J. Dixon, C.~Duhr, and J.~Pennington, {\it {Single-valued harmonic
  polylogarithms and the multi-Regge limit}},  {\em JHEP} {\bf 10} (2012) 074,
  [\href{http://arxiv.org/abs/1207.0186}{{\tt arXiv:1207.0186}}].

\bibitem{Drummond:2012bg}
J.~M. Drummond, {\it {Generalised ladders and single-valued polylogarithms}},
  {\em JHEP} {\bf 02} (2013) 092, [\href{http://arxiv.org/abs/1207.3824}{{\tt
  arXiv:1207.3824}}].

\bibitem{Chavez:2012kn}
F.~Chavez and C.~Duhr, {\it {Three-mass triangle integrals and single-valued
  polylogarithms}},  {\em JHEP} {\bf 11} (2012) 114,
  [\href{http://arxiv.org/abs/1209.2722}{{\tt arXiv:1209.2722}}].

\bibitem{Drummond:2013nda}
J.~Drummond, C.~Duhr, B.~Eden, P.~Heslop, J.~Pennington, and V.~A. Smirnov,
  {\it {Leading singularities and off-shell conformal integrals}},  {\em JHEP}
  {\bf 08} (2013) 133, [\href{http://arxiv.org/abs/1303.6909}{{\tt
  arXiv:1303.6909}}].

\bibitem{Brown:2015ztw}
F.~Brown and O.~Schnetz, {\it {Single-valued multiple polylogarithms and a
  proof of the zig{\textendash}zag conjecture}},  {\em J. Number Theor.} {\bf
  148} (2015) 478--506.

\bibitem{Schnetz:2016fhy}
O.~Schnetz, {\it {Numbers and Functions in Quantum Field Theory}},  {\em Phys.
  Rev. D} {\bf 97} (2018), no.~8 085018,
  [\href{http://arxiv.org/abs/1606.08598}{{\tt arXiv:1606.08598}}].

\bibitem{DelDuca:2016lad}
V.~Del~Duca, S.~Druc, J.~Drummond, C.~Duhr, F.~Dulat, R.~Marzucca,
  G.~Papathanasiou, and B.~Verbeek, {\it {Multi-Regge kinematics and the moduli
  space of Riemann spheres with marked points}},  {\em JHEP} {\bf 08} (2016)
  152, [\href{http://arxiv.org/abs/1606.08807}{{\tt arXiv:1606.08807}}].

\bibitem{Schnetz:2021ebf}
O.~Schnetz, {\it {Generalized single-valued hyperlogarithms}},
  \href{http://arxiv.org/abs/2111.11246}{{\tt arXiv:2111.11246}}.

\bibitem{Haug:2025ava}
J.~Haug and F.~Wunder, {\it {Single-valued representation of unpolarized and
  polarized semi-inclusive deep inelastic scattering at next-to-next-to-leading
  order}},  {\em Phys. Rev. D} {\bf 112} (2025), no.~11 114036,
  [\href{http://arxiv.org/abs/2505.18109}{{\tt arXiv:2505.18109}}].

\bibitem{brownSVHPLs}
F.~Brown, {\it {Single-valued multiple polylogarithms in one variable}},  {\em
  C. R. Acad. Sci. Paris, Ser. I} {\bf 338} (2004) 527.

\bibitem{brownSV}
F.~Brown, {\it {Single-valued hyperlogarithms and unipotent differential
  equations}},  2004.

\bibitem{Schnetz:2013hqa}
O.~Schnetz, {\it {Graphical functions and single-valued multiple
  polylogarithms}},  {\em Commun. Num. Theor. Phys.} {\bf 08} (2014) 589--675,
  [\href{http://arxiv.org/abs/1302.6445}{{\tt arXiv:1302.6445}}].

\bibitem{Brown:2013gia}
F.~Brown, {\it Single-valued motivic periods and multiple zeta values},  {\em
  Forum Math. Sigma} {\bf 2} (2014) Paper No. e25, 37,
  [\href{http://arxiv.org/abs/1309.5309}{{\tt arXiv:1309.5309}}].

\bibitem{Broedel:2016kls}
J.~Broedel, M.~Sprenger, and A.~Torres~Orjuela, {\it {Towards single-valued
  polylogarithms in two variables for the seven-point remainder function in
  multi-Regge kinematics}},  {\em Nucl. Phys. B} {\bf 915} (2017) 394--413,
  [\href{http://arxiv.org/abs/1606.08411}{{\tt arXiv:1606.08411}}].

\bibitem{Bourjaily:2022bwx}
J.~L. Bourjaily et~al., {\it {Functions Beyond Multiple Polylogarithms for
  Precision Collider Physics}},  in {\em {Snowmass 2021}}, 3, 2022.
\newblock \href{http://arxiv.org/abs/2203.07088}{{\tt arXiv:2203.07088}}.

\bibitem{Sabry}
A.~Sabry, {\it {Fourth order spectral functions for the electron propagator}},
  {\em Nucl. Phys.} {\bf 33} (1962), no.~17 401--430.

\bibitem{Bauberger:1994by}
S.~Bauberger, F.~A. Berends, M.~Bohm, and M.~Buza, {\it {Analytical and
  numerical methods for massive two loop selfenergy diagrams}},  {\em Nucl.
  Phys. B} {\bf 434} (1995) 383--407,
  [\href{http://arxiv.org/abs/hep-ph/9409388}{{\tt hep-ph/9409388}}].

\bibitem{Laporta:2004rb}
S.~Laporta and E.~Remiddi, {\it {Analytic treatment of the two loop equal mass
  sunrise graph}},  {\em Nucl. Phys. B} {\bf 704} (2005) 349--386,
  [\href{http://arxiv.org/abs/hep-ph/0406160}{{\tt hep-ph/0406160}}].

\bibitem{Bloch:2013tra}
S.~Bloch and P.~Vanhove, {\it {The elliptic dilogarithm for the sunset graph}},
   {\em J. Number Theor.} {\bf 148} (2015) 328--364,
  [\href{http://arxiv.org/abs/1309.5865}{{\tt arXiv:1309.5865}}].

\bibitem{Brown:2011wfj}
F.~Brown and A.~Levin, {\it {Multiple Elliptic Polylogarithms}},
  \href{http://arxiv.org/abs/1110.6917}{{\tt arXiv:1110.6917}}.

\bibitem{Broedel:2014vla}
J.~Broedel, C.~R. Mafra, N.~Matthes, and O.~Schlotterer, {\it {Elliptic
  multiple zeta values and one-loop superstring amplitudes}},  {\em JHEP} {\bf
  07} (2015) 112, [\href{http://arxiv.org/abs/1412.5535}{{\tt
  arXiv:1412.5535}}].

\bibitem{Broedel:2017kkb}
J.~Broedel, C.~Duhr, F.~Dulat, and L.~Tancredi, {\it {Elliptic polylogarithms
  and iterated integrals on elliptic curves. Part I: general formalism}},  {\em
  JHEP} {\bf 05} (2018) 093, [\href{http://arxiv.org/abs/1712.07089}{{\tt
  arXiv:1712.07089}}].

\bibitem{Broedel:2017siw}
J.~Broedel, C.~Duhr, F.~Dulat, and L.~Tancredi, {\it {Elliptic polylogarithms
  and iterated integrals on elliptic curves II: an application to the sunrise
  integral}},  {\em Phys. Rev. D} {\bf 97} (2018), no.~11 116009,
  [\href{http://arxiv.org/abs/1712.07095}{{\tt arXiv:1712.07095}}].

\bibitem{Enriquez2023EllipticH}
B.~Enriquez and F.~Zerbini, {\it Elliptic hyperlogarithms},  {\em Canad. J.
  Math.} {\bf 78} (2026), no.~3 733--768,
  [\href{http://arxiv.org/abs/2307.01833}{{\tt arXiv:2307.01833}}].

\bibitem{ManinModular}
Y.~I. Manin, {\it {Iterated integrals of modular forms and noncommutative
  modular symbols}},  in {\em Algebraic geometry and number theory}, vol.~253
  of {\em Progr. Math.}, (Boston), pp.~565--597, Birkh\"auser Boston, 2006.
\newblock \href{http://arxiv.org/abs/math/0502576}{{\tt math/0502576}}.

\bibitem{Broedel:2015hia}
J.~Broedel, N.~Matthes, and O.~Schlotterer, {\it {Relations between elliptic
  multiple zeta values and a special derivation algebra}},  {\em J. Phys. A}
  {\bf 49} (2016), no.~15 155203, [\href{http://arxiv.org/abs/1507.02254}{{\tt
  arXiv:1507.02254}}].

\bibitem{Ablinger:2017bjx}
J.~Ablinger, J.~Bl{\"u}mlein, A.~De~Freitas, M.~van Hoeij, E.~Imamoglu, C.~G.
  Raab, C.~S. Radu, and C.~Schneider, {\it {Iterated Elliptic and
  Hypergeometric Integrals for Feynman Diagrams}},  {\em J. Math. Phys.} {\bf
  59} (2018), no.~6 062305, [\href{http://arxiv.org/abs/1706.01299}{{\tt
  arXiv:1706.01299}}].

\bibitem{Adams:2017ejb}
L.~Adams and S.~Weinzierl, {\it {Feynman integrals and iterated integrals of
  modular forms}},  {\em Commun. Num. Theor. Phys.} {\bf 12} (2018) 193--251,
  [\href{http://arxiv.org/abs/1704.08895}{{\tt arXiv:1704.08895}}].

\bibitem{Duhr:2019rrs}
C.~Duhr and L.~Tancredi, {\it {Algorithms and tools for iterated Eisenstein
  integrals}},  {\em JHEP} {\bf 02} (2020) 105,
  [\href{http://arxiv.org/abs/1912.00077}{{\tt arXiv:1912.00077}}].

\bibitem{Matthes2022}
N.~Matthes, {\it Iterated primitives of meromorphic quasimodular forms for
  sl$_2(\mathbb{Z})$},  {\em Transactions of the American Mathematical Society}
  {\bf 375} (2022), no.~2 1443--1460.

\bibitem{Broedel:2021zij}
J.~Broedel, C.~Duhr, and N.~Matthes, {\it {Meromorphic modular forms and the
  three-loop equal-mass banana integral}},  {\em JHEP} {\bf 02} (2022) 184,
  [\href{http://arxiv.org/abs/2109.15251}{{\tt arXiv:2109.15251}}].

\bibitem{Brown:2010bw}
F.~Brown and O.~Schnetz, {\it {A K3 in $\phi^4$}},  {\em Duke Math. J.} {\bf
  161} (2012), no.~10 1817--1862, [\href{http://arxiv.org/abs/1006.4064}{{\tt
  arXiv:1006.4064}}].

\bibitem{Huang:2013kh}
R.~Huang and Y.~Zhang, {\it {On Genera of Curves from High-loop Generalized
  Unitarity Cuts}},  {\em JHEP} {\bf 04} (2013) 080,
  [\href{http://arxiv.org/abs/1302.1023}{{\tt arXiv:1302.1023}}].

\bibitem{Bloch:2014qca}
S.~Bloch, M.~Kerr, and P.~Vanhove, {\it {A Feynman integral via higher normal
  functions}},  {\em Compos. Math.} {\bf 151} (2015), no.~12 2329--2375,
  [\href{http://arxiv.org/abs/1406.2664}{{\tt arXiv:1406.2664}}].

\bibitem{Hauenstein:2014mda}
J.~D. Hauenstein, R.~Huang, D.~Mehta, and Y.~Zhang, {\it {Global Structure of
  Curves from Generalized Unitarity Cut of Three-loop Diagrams}},  {\em JHEP}
  {\bf 02} (2015) 136, [\href{http://arxiv.org/abs/1408.3355}{{\tt
  arXiv:1408.3355}}].

\bibitem{Bloch:2016izu}
S.~Bloch, M.~Kerr, and P.~Vanhove, {\it {Local mirror symmetry and the sunset
  Feynman integral}},  {\em Adv. Theor. Math. Phys.} {\bf 21} (2017)
  1373--1453, [\href{http://arxiv.org/abs/1601.08181}{{\tt arXiv:1601.08181}}].

\bibitem{Primo:2017ipr}
A.~Primo and L.~Tancredi, {\it {Maximal cuts and differential equations for
  Feynman integrals. An application to the three-loop massive banana graph}},
  {\em Nucl. Phys. B} {\bf 921} (2017) 316--356,
  [\href{http://arxiv.org/abs/1704.05465}{{\tt arXiv:1704.05465}}].

\bibitem{Bourjaily:2018ycu}
J.~L. Bourjaily, Y.-H. He, A.~J. Mcleod, M.~Von~Hippel, and M.~Wilhelm, {\it
  {Traintracks through Calabi-Yau Manifolds: Scattering Amplitudes beyond
  Elliptic Polylogarithms}},  {\em Phys. Rev. Lett.} {\bf 121} (2018), no.~7
  071603, [\href{http://arxiv.org/abs/1805.09326}{{\tt arXiv:1805.09326}}].

\bibitem{Bourjaily:2018yfy}
J.~L. Bourjaily, A.~J. McLeod, M.~von Hippel, and M.~Wilhelm, {\it {Bounded
  Collection of Feynman Integral Calabi-Yau Geometries}},  {\em Phys. Rev.
  Lett.} {\bf 122} (2019), no.~3 031601,
  [\href{http://arxiv.org/abs/1810.07689}{{\tt arXiv:1810.07689}}].

\bibitem{Bourjaily:2019hmc}
J.~L. Bourjaily, A.~J. McLeod, C.~Vergu, M.~Volk, M.~Von~Hippel, and
  M.~Wilhelm, {\it {Embedding Feynman Integral (Calabi-Yau) Geometries in
  Weighted Projective Space}},  {\em JHEP} {\bf 01} (2020) 078,
  [\href{http://arxiv.org/abs/1910.01534}{{\tt arXiv:1910.01534}}].

\bibitem{Klemm:2019dbm}
A.~Klemm, C.~Nega, and R.~Safari, {\it {The $l$-loop Banana Amplitude from GKZ
  Systems and relative Calabi-Yau Periods}},  {\em JHEP} {\bf 04} (2020) 088,
  [\href{http://arxiv.org/abs/1912.06201}{{\tt arXiv:1912.06201}}].

\bibitem{Bonisch:2020qmm}
K.~B{\"o}nisch, F.~Fischbach, A.~Klemm, C.~Nega, and R.~Safari, {\it {Analytic
  structure of all loop banana integrals}},  {\em JHEP} {\bf 05} (2021) 066,
  [\href{http://arxiv.org/abs/2008.10574}{{\tt arXiv:2008.10574}}].

\bibitem{Bonisch:2021yfw}
K.~B{\"o}nisch, C.~Duhr, F.~Fischbach, A.~Klemm, and C.~Nega, {\it {Feynman
  integrals in dimensional regularisation and extensions of Calabi-Yau
  motives}},  {\em JHEP} {\bf 09} (2022) 156,
  [\href{http://arxiv.org/abs/2108.05310}{{\tt arXiv:2108.05310}}].

\bibitem{Duhr:2022pch}
C.~Duhr, A.~Klemm, F.~Loebbert, C.~Nega, and F.~Porkert, {\it
  {Yangian-Invariant Fishnet Integrals in Two Dimensions as Volumes of
  Calabi-Yau Varieties}},  {\em Phys. Rev. Lett.} {\bf 130} (2023), no.~4
  041602, [\href{http://arxiv.org/abs/2209.05291}{{\tt arXiv:2209.05291}}].

\bibitem{Marzucca:2023gto}
R.~Marzucca, A.~J. McLeod, B.~Page, S.~P{\"o}gel, and S.~Weinzierl, {\it {Genus
  drop in hyperelliptic Feynman integrals}},  {\em Phys. Rev. D} {\bf 109}
  (2024), no.~3 L031901, [\href{http://arxiv.org/abs/2307.11497}{{\tt
  arXiv:2307.11497}}].

\bibitem{Duhr:2023eld}
C.~Duhr, A.~Klemm, F.~Loebbert, C.~Nega, and F.~Porkert, {\it {The Basso-Dixon
  formula and Calabi-Yau geometry}},  {\em JHEP} {\bf 03} (2024) 177,
  [\href{http://arxiv.org/abs/2310.08625}{{\tt arXiv:2310.08625}}].

\bibitem{Frellesvig:2023bbf}
H.~Frellesvig, R.~Morales, and M.~Wilhelm, {\it {Calabi-Yau Meets Gravity: A
  Calabi-Yau Threefold at Fifth Post-Minkowskian Order}},  {\em Phys. Rev.
  Lett.} {\bf 132} (2024), no.~20 201602,
  [\href{http://arxiv.org/abs/2312.11371}{{\tt arXiv:2312.11371}}].

\bibitem{Klemm:2024wtd}
A.~Klemm, C.~Nega, B.~Sauer, and J.~Plefka, {\it {Calabi-Yau periods for black
  hole scattering in classical general relativity}},  {\em Phys. Rev. D} {\bf
  109} (2024), no.~12 124046, [\href{http://arxiv.org/abs/2401.07899}{{\tt
  arXiv:2401.07899}}].

\bibitem{Duhr:2024hjf}
C.~Duhr, A.~Klemm, F.~Loebbert, C.~Nega, and F.~Porkert, {\it {Geometry from
  integrability: multi-leg fishnet integrals in two dimensions}},  {\em JHEP}
  {\bf 07} (2024) 008, [\href{http://arxiv.org/abs/2402.19034}{{\tt
  arXiv:2402.19034}}].

\bibitem{Jockers:2024uan}
H.~Jockers, S.~Kotlewski, P.~Kuusela, A.~J. McLeod, S.~P{\"o}gel, M.~Sarve,
  X.~Wang, and S.~Weinzierl, {\it {A Calabi-Yau-to-curve correspondence for
  Feynman integrals}},  {\em JHEP} {\bf 01} (2025) 030,
  [\href{http://arxiv.org/abs/2404.05785}{{\tt arXiv:2404.05785}}].

\bibitem{Frellesvig:2024zph}
H.~Frellesvig, R.~Morales, and M.~Wilhelm, {\it {Classifying post-Minkowskian
  geometries for gravitational waves via loop-by-loop Baikov}},  {\em JHEP}
  {\bf 08} (2024) 243, [\href{http://arxiv.org/abs/2405.17255}{{\tt
  arXiv:2405.17255}}].

\bibitem{Driesse:2024feo}
M.~Driesse, G.~U. Jakobsen, A.~Klemm, G.~Mogull, C.~Nega, J.~Plefka, B.~Sauer,
  and J.~Usovitsch, {\it {Emergence of Calabi{\textendash}Yau manifolds in
  high-precision black-hole scattering}},  {\em Nature} {\bf 641} (2025),
  no.~8063 603--607, [\href{http://arxiv.org/abs/2411.11846}{{\tt
  arXiv:2411.11846}}].

\bibitem{Duhr:2024uid}
C.~Duhr, F.~Porkert, and S.~F. Stawinski, {\it {Canonical differential
  equations beyond genus one}},  {\em JHEP} {\bf 02} (2025) 014,
  [\href{http://arxiv.org/abs/2412.02300}{{\tt arXiv:2412.02300}}].

\bibitem{Frellesvig:2024rea}
H.~Frellesvig, R.~Morales, S.~P{\"o}gel, S.~Weinzierl, and M.~Wilhelm, {\it
  {Calabi-Yau Feynman integrals in gravity:
  {\ensuremath{\varepsilon}}-factorized form for apparent singularities}},
  {\em JHEP} {\bf 02} (2025) 209, [\href{http://arxiv.org/abs/2412.12057}{{\tt
  arXiv:2412.12057}}].

\bibitem{Schimmrigk:2024xid}
R.~Schimmrigk, {\it {Special Fano geometry from Feynman integrals}},  {\em
  Phys. Lett. B} {\bf 864} (2025) 139420,
  [\href{http://arxiv.org/abs/2412.20236}{{\tt arXiv:2412.20236}}].

\bibitem{Duhr:2025tdf}
C.~Duhr, {\it {Modular forms for three-loop banana integrals}},  {\em JHEP}
  {\bf 08} (2025) 218, [\href{http://arxiv.org/abs/2502.15325}{{\tt
  arXiv:2502.15325}}].

\bibitem{Duhr:2025ppd}
C.~Duhr and S.~Maggio, {\it {Feynman integrals, elliptic integrals and
  two-parameter K3 surfaces}},  {\em JHEP} {\bf 06} (2025) 250,
  [\href{http://arxiv.org/abs/2502.15326}{{\tt arXiv:2502.15326}}].

\bibitem{Brammer:2025rqo}
D.~Brammer, H.~Frellesvig, R.~Morales, and M.~Wilhelm, {\it {Classification of
  Feynman integral geometries for black-hole scattering at 5PM order}},  {\em
  JHEP} {\bf 10} (2025) 212, [\href{http://arxiv.org/abs/2505.10274}{{\tt
  arXiv:2505.10274}}].

\bibitem{Duhr:2025kkq}
C.~Duhr, S.~Maggio, F.~Porkert, C.~Semper, and S.~F. Stawinski, {\it
  {Three-loop banana integrals with four unequal masses}},  {\em JHEP} {\bf 12}
  (2025) 034, [\href{http://arxiv.org/abs/2507.23061}{{\tt arXiv:2507.23061}}].

\bibitem{Pogel:2025bca}
S.~P{\"o}gel, T.~Teschke, X.~Wang, and S.~Weinzierl, {\it {The unequal-mass
  three-loop banana integral}},  {\em JHEP} {\bf 01} (2026) 021,
  [\href{http://arxiv.org/abs/2507.23594}{{\tt arXiv:2507.23594}}].

\bibitem{Duhr:2025ouy}
C.~Duhr and S.~Maggio, {\it {Three-loop banana integrals with three equal
  masses}},  {\em JHEP} {\bf 04} (2026) 187,
  [\href{http://arxiv.org/abs/2511.19245}{{\tt arXiv:2511.19245}}].

\bibitem{delaCruz:2025szs}
L.~de~la Cruz, P.~P. Novichkov, and P.~Vanhove, {\it {Fano and reflexive
  polytopes from Feynman integrals}},  {\em JHEP} {\bf 06} (2026) 062,
  [\href{http://arxiv.org/abs/2512.10518}{{\tt arXiv:2512.10518}}].

\bibitem{Bargiela:2025vwl}
P.~Bargiela, H.~Frellesvig, R.~Marzucca, R.~Morales, F.~Seefeld, M.~Wilhelm,
  and T.-Z. Yang, {\it {The spectrum of Feynman-integral geometries at two
  loops}},  {\em JHEP} {\bf 05} (2026) 057,
  [\href{http://arxiv.org/abs/2512.13794}{{\tt arXiv:2512.13794}}].

\bibitem{Yang:2026rgb}
F.~Yang, J.~Gong, and Y.~Zhang, {\it {Genus drop involving nonhyperelliptic
  curves in Feynman integrals}},  {\em Phys. Rev. D} {\bf 114} (2026), no.~5
  056013, [\href{http://arxiv.org/abs/2605.07729}{{\tt arXiv:2605.07729}}].

\bibitem{Brown:2015fyf}
F.~Brown, {\it {Feynman amplitudes, coaction principle, and cosmic Galois
  group}},  {\em Commun. Num. Theor. Phys.} {\bf 11} (2017) 453--556,
  [\href{http://arxiv.org/abs/1512.06409}{{\tt arXiv:1512.06409}}].

\bibitem{Brown:2015ylf}
F.~Brown, {\it Notes on motivic periods},  {\em Commun. Number Theory Phys.}
  {\bf 11} (2017), no.~3 557--655, [\href{http://arxiv.org/abs/1512.06410}{{\tt
  arXiv:1512.06410}}].

\bibitem{Berkovits:2022ivl}
N.~Berkovits, E.~D'Hoker, M.~B. Green, H.~Johansson, and O.~Schlotterer, {\it
  {Snowmass White Paper: String Perturbation Theory}},  in {\em {Snowmass
  2021}}, 3, 2022.
\newblock \href{http://arxiv.org/abs/2203.09099}{{\tt arXiv:2203.09099}}.

\bibitem{Schlotterer:2012ny}
O.~Schlotterer and S.~Stieberger, {\it {Motivic Multiple Zeta Values and
  Superstring Amplitudes}},  {\em J. Phys. A} {\bf 46} (2013) 475401,
  [\href{http://arxiv.org/abs/1205.1516}{{\tt arXiv:1205.1516}}].

\bibitem{Stieberger:2013wea}
S.~Stieberger, {\it {Closed superstring amplitudes, single-valued multiple zeta
  values and the Deligne associator}},  {\em J. Phys. A} {\bf 47} (2014)
  155401, [\href{http://arxiv.org/abs/1310.3259}{{\tt arXiv:1310.3259}}].

\bibitem{Stieberger:2014hba}
S.~Stieberger and T.~R. Taylor, {\it {Closed String Amplitudes as Single-Valued
  Open String Amplitudes}},  {\em Nucl. Phys. B} {\bf 881} (2014) 269--287,
  [\href{http://arxiv.org/abs/1401.1218}{{\tt arXiv:1401.1218}}].

\bibitem{Schlotterer:2018zce}
O.~Schlotterer and O.~Schnetz, {\it {Closed strings as single-valued open
  strings: A genus-zero derivation}},  {\em J. Phys. A} {\bf 52} (2019), no.~4
  045401, [\href{http://arxiv.org/abs/1808.00713}{{\tt arXiv:1808.00713}}].

\bibitem{Brown:2018omk}
F.~Brown and C.~Dupont, {\it {Single-valued integration and double copy}},
  {\em J. Reine Angew. Math.} {\bf 2021} (2021), no.~775 145--196,
  [\href{http://arxiv.org/abs/1810.07682}{{\tt arXiv:1810.07682}}].

\bibitem{Vanhove:2018elu}
P.~Vanhove and F.~Zerbini, {\it {Single-valued hyperlogarithms, correlation
  functions and closed string amplitudes}},  {\em Adv. Theor. Math. Phys.} {\bf
  26} (2022) 455--530, [\href{http://arxiv.org/abs/1812.03018}{{\tt
  arXiv:1812.03018}}].

\bibitem{Brown:2019wna}
F.~Brown and C.~Dupont, {\it {Single-valued integration and superstring
  amplitudes in genus zero}},  {\em Commun. Math. Phys.} {\bf 382} (2021),
  no.~2 815--874, [\href{http://arxiv.org/abs/1910.01107}{{\tt
  arXiv:1910.01107}}].

\bibitem{Mafra:2022wml}
C.~R. Mafra and O.~Schlotterer, {\it {Tree-level amplitudes from the pure
  spinor superstring}},  {\em Phys. Rept.} {\bf 1020} (2023) 1--162,
  [\href{http://arxiv.org/abs/2210.14241}{{\tt arXiv:2210.14241}}].

\bibitem{Baune:2024uwj}
K.~Baune, J.~Broedel, and F.~Zerbini, {\it {Closed-string amplitude recursions
  from the Deligne associator}},  \href{http://arxiv.org/abs/2412.17579}{{\tt
  arXiv:2412.17579}}.

\bibitem{Broedel:2017jdo}
J.~Broedel, N.~Matthes, G.~Richter, and O.~Schlotterer, {\it {Twisted elliptic
  multiple zeta values and non-planar one-loop open-string amplitudes}},  {\em
  J. Phys. A} {\bf 51} (2018), no.~28 285401,
  [\href{http://arxiv.org/abs/1704.03449}{{\tt arXiv:1704.03449}}].

\bibitem{DHoker:2023khh}
E.~D'Hoker, M.~Hidding, and O.~Schlotterer, {\it {Cyclic Products of
  Higher-Genus Szeg{\"o} Kernels, Modular Tensors, and Polylogarithms}},  {\em
  Phys. Rev. Lett.} {\bf 133} (2024), no.~2 021602,
  [\href{http://arxiv.org/abs/2308.05044}{{\tt arXiv:2308.05044}}].

\bibitem{DHoker:2025jgb}
E.~D'Hoker and O.~Schlotterer, {\it {Worldsheet fermion correlators, modular
  tensors and higher genus integration kernels}},  {\em SciPost Phys.} {\bf 19}
  (2025), no.~4 091, [\href{http://arxiv.org/abs/2505.07947}{{\tt
  arXiv:2505.07947}}].

\bibitem{DHoker:2015gmr}
E.~D'Hoker, M.~B. Green, and P.~Vanhove, {\it {On the modular structure of the
  genus-one Type II superstring low energy expansion}},  {\em JHEP} {\bf 08}
  (2015) 041, [\href{http://arxiv.org/abs/1502.06698}{{\tt arXiv:1502.06698}}].

\bibitem{DHoker:2015wxz}
E.~D'Hoker, M.~B. Green, {\"O}.~G{\"u}rdogan, and P.~Vanhove, {\it {Modular
  Graph Functions}},  {\em Commun. Num. Theor. Phys.} {\bf 11} (2017) 165--218,
  [\href{http://arxiv.org/abs/1512.06779}{{\tt arXiv:1512.06779}}].

\bibitem{DHoker:2016mwo}
E.~D'Hoker and M.~B. Green, {\it {Identities between Modular Graph Forms}},
  {\em J. Number Theor.} {\bf 189} (2018) 25--80,
  [\href{http://arxiv.org/abs/1603.00839}{{\tt arXiv:1603.00839}}].

\bibitem{DHoker:2017pvk}
E.~D'Hoker, M.~B. Green, and B.~Pioline, {\it {Higher genus modular graph
  functions, string invariants, and their exact asymptotics}},  {\em Commun.
  Math. Phys.} {\bf 366} (2019), no.~3 927--979,
  [\href{http://arxiv.org/abs/1712.06135}{{\tt arXiv:1712.06135}}].

\bibitem{DHoker:2020uid}
E.~D'Hoker and O.~Schlotterer, {\it {Identities among higher genus modular
  graph tensors}},  {\em Commun. Num. Theor. Phys.} {\bf 16} (2022), no.~1
  35--74, [\href{http://arxiv.org/abs/2010.00924}{{\tt arXiv:2010.00924}}].

\bibitem{Zerbini:2015rss}
F.~Zerbini, {\it {Single-valued multiple zeta values in genus 1 superstring
  amplitudes}},  {\em Commun. Num. Theor. Phys.} {\bf 10} (2016) 703--737,
  [\href{http://arxiv.org/abs/1512.05689}{{\tt arXiv:1512.05689}}].

\bibitem{Broedel:2018izr}
J.~Broedel, O.~Schlotterer, and F.~Zerbini, {\it {From elliptic multiple zeta
  values to modular graph functions: open and closed strings at one loop}},
  {\em JHEP} {\bf 01} (2019) 155, [\href{http://arxiv.org/abs/1803.00527}{{\tt
  arXiv:1803.00527}}].

\bibitem{Panzertalk}
E.~Panzer, ``{Talk ``Modular graph functions as iterated Eisenstein integrals''
  given at the workshop ``Elliptic Integrals in Mathematics and Physics''
  (Ascona, Switzerland)}.''
  \url{https://indico.cern.ch/event/700233/contributions/3112451/attachments/1712442/2761239/elliptic.pdf},
  2018.

\bibitem{Gerken:2020xfv}
J.~E. Gerken, A.~Kleinschmidt, C.~R. Mafra, O.~Schlotterer, and B.~Verbeek,
  {\it {Towards closed strings as single-valued open strings at genus one}},
  {\em J. Phys. A} {\bf 55} (2022), no.~2 025401,
  [\href{http://arxiv.org/abs/2010.10558}{{\tt arXiv:2010.10558}}].

\bibitem{Gerken:2020yii}
J.~E. Gerken, A.~Kleinschmidt, and O.~Schlotterer, {\it {Generating series of
  all modular graph forms from iterated Eisenstein integrals}},  {\em JHEP}
  {\bf 07} (2020), no.~07 190, [\href{http://arxiv.org/abs/2004.05156}{{\tt
  arXiv:2004.05156}}].

\bibitem{Dorigoni:2022npe}
D.~Dorigoni, M.~Doroudiani, J.~Drewitt, M.~Hidding, A.~Kleinschmidt,
  N.~Matthes, O.~Schlotterer, and B.~Verbeek, {\it {Modular graph forms from
  equivariant iterated Eisenstein integrals}},  {\em JHEP} {\bf 12} (2022) 162,
  [\href{http://arxiv.org/abs/2209.06772}{{\tt arXiv:2209.06772}}].

\bibitem{Dorigoni:2024oft}
D.~Dorigoni, M.~Doroudiani, J.~Drewitt, M.~Hidding, A.~Kleinschmidt,
  O.~Schlotterer, L.~Schneps, and B.~Verbeek, {\it {Non-holomorphic modular
  forms from zeta generators}},  {\em JHEP} {\bf 10} (2024) 053,
  [\href{http://arxiv.org/abs/2403.14816}{{\tt arXiv:2403.14816}}].

\bibitem{Brown:2017qwo}
F.~Brown, {\it A class of non-holomorphic modular forms {I}},  {\em Research in
  the Mathematical Sciences} {\bf 5} (2018) 7,
  [\href{http://arxiv.org/abs/1707.01230}{{\tt arXiv:1707.01230}}].

\bibitem{Brown:2020mhp}
F.~Brown, {\it {A class of non-holomorphic modular forms II: equivariant
  iterated Eisenstein integrals}},  {\em Forum~of~Mathematics,~Sigma} {\bf 8}
  (2020) 1, [\href{http://arxiv.org/abs/1708.03354}{{\tt arXiv:1708.03354}}].

\bibitem{Tkachov:1981wb}
F.~V. Tkachov, {\it {A Theorem on Analytical Calculability of Four Loop
  Renormalization Group Functions}},  {\em Phys. Lett. B} {\bf 100} (1981)
  65--68.

\bibitem{Chetyrkin:1981qh}
K.~G. Chetyrkin and F.~V. Tkachov, {\it {Integration by Parts: The Algorithm to
  Calculate beta Functions in 4 Loops}},  {\em Nucl. Phys. B} {\bf 192} (1981)
  159--204.

\bibitem{Frellesvig:2019kgj}
H.~Frellesvig, F.~Gasparotto, S.~Laporta, M.~K. Mandal, P.~Mastrolia,
  L.~Mattiazzi, and S.~Mizera, {\it {Decomposition of Feynman Integrals on the
  Maximal Cut by Intersection Numbers}},  {\em JHEP} {\bf 05} (2019) 153,
  [\href{http://arxiv.org/abs/1901.11510}{{\tt arXiv:1901.11510}}].

\bibitem{Smirnov:2010hn}
A.~V. Smirnov and A.~V. Petukhov, {\it {The Number of Master Integrals is
  Finite}},  {\em Lett. Math. Phys.} {\bf 97} (2011) 37--44,
  [\href{http://arxiv.org/abs/1004.4199}{{\tt arXiv:1004.4199}}].

\bibitem{Bitoun:2017nre}
T.~Bitoun, C.~Bogner, R.~P. Klausen, and E.~Panzer, {\it {Feynman integral
  relations from parametric annihilators}},  {\em Lett. Math. Phys.} {\bf 109}
  (2019), no.~3 497--564, [\href{http://arxiv.org/abs/1712.09215}{{\tt
  arXiv:1712.09215}}].

\bibitem{Henn:2014qga}
J.~M. Henn, {\it {Lectures on differential equations for Feynman integrals}},
  {\em J. Phys. A} {\bf 48} (2015) 153001,
  [\href{http://arxiv.org/abs/1412.2296}{{\tt arXiv:1412.2296}}].

\bibitem{Lee:2014ioa}
R.~N. Lee, {\it {Reducing differential equations for multiloop master
  integrals}},  {\em JHEP} {\bf 04} (2015) 108,
  [\href{http://arxiv.org/abs/1411.0911}{{\tt arXiv:1411.0911}}].

\bibitem{Meyer:2016slj}
C.~Meyer, {\it {Transforming differential equations of multi-loop Feynman
  integrals into canonical form}},  {\em JHEP} {\bf 04} (2017) 006,
  [\href{http://arxiv.org/abs/1611.01087}{{\tt arXiv:1611.01087}}].

\bibitem{Prausa:2017ltv}
M.~Prausa, {\it {epsilon: A tool to find a canonical basis of master
  integrals}},  {\em Comput. Phys. Commun.} {\bf 219} (2017) 361--376,
  [\href{http://arxiv.org/abs/1701.00725}{{\tt arXiv:1701.00725}}].

\bibitem{Gituliar:2017vzm}
O.~Gituliar and V.~Magerya, {\it {Fuchsia: a tool for reducing differential
  equations for Feynman master integrals to epsilon form}},  {\em Comput. Phys.
  Commun.} {\bf 219} (2017) 329--338,
  [\href{http://arxiv.org/abs/1701.04269}{{\tt arXiv:1701.04269}}].

\bibitem{Meyer:2017joq}
C.~Meyer, {\it {Algorithmic transformation of multi-loop master integrals to a
  canonical basis with CANONICA}},  {\em Comput. Phys. Commun.} {\bf 222}
  (2018) 295--312, [\href{http://arxiv.org/abs/1705.06252}{{\tt
  arXiv:1705.06252}}].

\bibitem{Dlapa:2020cwj}
C.~Dlapa, J.~Henn, and K.~Yan, {\it {Deriving canonical differential equations
  for Feynman integrals from a single uniform weight integral}},  {\em JHEP}
  {\bf 05} (2020) 025, [\href{http://arxiv.org/abs/2002.02340}{{\tt
  arXiv:2002.02340}}].

\bibitem{Lee:2020zfb}
R.~N. Lee, {\it {Libra: A package for transformation of differential systems
  for multiloop integrals}},  {\em Comput. Phys. Commun.} {\bf 267} (2021)
  108058, [\href{http://arxiv.org/abs/2012.00279}{{\tt arXiv:2012.00279}}].

\bibitem{Henn:2020lye}
J.~Henn, B.~Mistlberger, V.~A. Smirnov, and P.~Wasser, {\it {Constructing d-log
  integrands and computing master integrals for three-loop four-particle
  scattering}},  {\em JHEP} {\bf 04} (2020) 167,
  [\href{http://arxiv.org/abs/2002.09492}{{\tt arXiv:2002.09492}}].

\bibitem{Chen:2020uyk}
J.~Chen, X.~Jiang, X.~Xu, and L.~L. Yang, {\it {Constructing canonical Feynman
  integrals with intersection theory}},  {\em Phys. Lett. B} {\bf 814} (2021)
  136085, [\href{http://arxiv.org/abs/2008.03045}{{\tt arXiv:2008.03045}}].

\bibitem{Dlapa:2022wdu}
C.~Dlapa, J.~M. Henn, and F.~J. Wagner, {\it {An algorithmic approach to
  finding canonical differential equations for elliptic Feynman integrals}},
  {\em JHEP} {\bf 08} (2023) 120, [\href{http://arxiv.org/abs/2211.16357}{{\tt
  arXiv:2211.16357}}].

\bibitem{Yang:2025ofz}
L.~L. Yang and Y.~Zhang, {\it {From $\mathrm{d} \! \log$ to $\mathrm{d}
  \mathcal{E}$: Canonical Elliptic Integrands and Modular Symbol Letters with
  Pure eMPLs}},  \href{http://arxiv.org/abs/2512.19370}{{\tt
  arXiv:2512.19370}}.

\bibitem{Chaubey:2025adn}
E.~Chaubey and V.~Sotnikov, {\it {Elliptic Leading Singularities and Canonical
  Integrands}},  {\em Phys. Rev. Lett.} {\bf 135} (2025), no.~10 101903,
  [\href{http://arxiv.org/abs/2504.20897}{{\tt arXiv:2504.20897}}].

\bibitem{Chaubey:2026egr}
E.~Chaubey and V.~Sotnikov, {\it {How to choose a good rational basis for
  elliptic Feynman integrals?}},  \href{http://arxiv.org/abs/2609.08875}{{\tt
  arXiv:2609.08875}}.

\bibitem{Henn:2016jdu}
J.~M. Henn and B.~Mistlberger, {\it {Four-Gluon Scattering at Three Loops,
  Infrared Structure, and the Regge Limit}},  {\em Phys. Rev. Lett.} {\bf 117}
  (2016), no.~17 171601, [\href{http://arxiv.org/abs/1608.00850}{{\tt
  arXiv:1608.00850}}].

\bibitem{chen1977iterated}
K.-T. Chen, {\it Iterated path integrals},  {\em Bulletin of the American
  Mathematical Society} {\bf 83} (1977), no.~5 831--879.

\bibitem{Aomoto:1994}
M.~Kita and M.~Yoshida, {\it Intersection theory for twisted cycles},  {\em
  Mathematische Nachrichten} {\bf 166} (1994), no.~1 287--304.

\bibitem{Knizhnik:1984nr}
V.~G. Knizhnik and A.~B. Zamolodchikov, {\it {Current Algebra and Wess-Zumino
  Model in Two-Dimensions}},  {\em Nucl. Phys. B} {\bf 247} (1984) 83--103.

\bibitem{Etingof:1998}
P.~I. Etingof, I.~B. Frenkel, and A.~A. Kirillov, Jr., {\em Lectures on
  representation theory and {K}nizhnik-{Z}amolodchikov equations}, vol.~58 of
  {\em Mathematical Surveys and Monographs}.
\newblock American Mathematical Society, Providence, RI, 1998.

\bibitem{Drinfeld:1991}
V.~G. Drinfeld, {\it On quasitriangular quasi-hopf algebras and on a group that
  is closely connected with gal$(\bar{\mathbb q}/\mathbb q)$},  {\em Leningrad
  Math. J.} {\bf 2} (1991), no.~4 829--860.

\bibitem{Levrac}
A.~Levin and G.~Racinet, {\it Towards multiple elliptic polylogarithms},
  \href{http://arxiv.org/abs/math/0703237}{{\tt math/0703237}}.

\bibitem{Calaque:2009}
D.~Calaque, B.~Enriquez, and P.~Etingof, {\it Universal {KZB} equations: the
  elliptic case},  in {\em Algebra, arithmetic, and geometry: in honor of {Y}u.
  {I}. {M}anin. {V}ol. {I}}, vol.~269 of {\em Progr. Math.}, pp.~165--266.
\newblock Birkh\"auser Boston, Boston, MA, 2009.
\newblock \href{http://arxiv.org/abs/math/0702670}{{\tt math/0702670}}.

\bibitem{EnriquezEllAss}
B.~Enriquez, {\it Elliptic associators},  {\em Selecta Math. (N.S.)} {\bf 20}
  (2014), no.~2 491--584, [\href{http://arxiv.org/abs/1003.1012}{{\tt
  arXiv:1003.1012}}].

\bibitem{Bernard:1988}
D.~Bernard, {\it On the Wess-Zumino-Witten models on Riemann surfaces},  {\em
  Nuclear Phys. B} {\bf 309} (1988), no.~1 145-174.

\bibitem{Enriquez:2011}
B.~Enriquez, {\it Flat connections on configuration spaces and braid groups of
  surfaces},  {\em Advances in Mathematics} {\bf 252} (2014) 204--226,
  [\href{http://arxiv.org/abs/1112.0864}{{\tt arXiv:1112.0864}}].

\bibitem{DHoker:2023vax}
E.~D'Hoker, M.~Hidding, and O.~Schlotterer, {\it {Constructing polylogarithms
  on higher-genus Riemann surfaces}},  {\em Commun. Num. Theor. Phys.} {\bf 19}
  (2025), no.~2 355--413, [\href{http://arxiv.org/abs/2306.08644}{{\tt
  arXiv:2306.08644}}].

\bibitem{Baune:2024biq}
K.~Baune, J.~Broedel, E.~Im, A.~Lisitsyn, and F.~Zerbini, {\it
  {Schottky{\textendash}Kronecker forms and hyperelliptic polylogarithms}},
  {\em J. Phys. A} {\bf 57} (2024), no.~44 445202,
  [\href{http://arxiv.org/abs/2406.10051}{{\tt arXiv:2406.10051}}].

\bibitem{DHoker:2026lgg}
E.~D'Hoker and O.~Schlotterer, {\it {Single-valued flat connections in several
  variables on arbitrary Riemann surfaces}},
  \href{http://arxiv.org/abs/2602.01461}{{\tt arXiv:2602.01461}}.

\bibitem{DHoker:2026mxk}
E.~D'Hoker, B.~Enriquez, O.~Schlotterer, and F.~Zerbini, {\it {Flat connections
  on moduli spaces I: Local (1,0)-extension of the DHS connection}},
  \href{http://arxiv.org/abs/2608.29169}{{\tt arXiv:2608.29169}}.

\bibitem{jimbo1989introduction}
M.~Jimbo, {\it Introduction to the {Y}ang-{B}axter equation},  {\em
  International Journal of Modern Physics A} {\bf 4} (1989), no.~15 3759--3777.

\bibitem{kassel2012quantum}
C.~Kassel, {\em Quantum groups}, vol.~155 of {\em Graduate Texts in
  Mathematics}.
\newblock Springer-Verlag, New York, 1995.

\bibitem{Bezru:1994}
R.~Bezrukavnikov, {\it {Koszul DG-algebras arising from configuration spaces}},
   {\em Geom. Funct. Anal.} {\bf 4} (1994), no.~2 119-135.

\bibitem{Henn:2013tua}
J.~M. Henn, A.~V. Smirnov, and V.~A. Smirnov, {\it {Analytic results for planar
  three-loop four-point integrals from a Knizhnik-Zamolodchikov equation}},
  {\em JHEP} {\bf 07} (2013) 128, [\href{http://arxiv.org/abs/1306.2799}{{\tt
  arXiv:1306.2799}}].

\bibitem{Calaque:2019ell}
D.~Calaque and M.~Gonzalez, {\it On the universal ellipsitomic {KZB}
  connection},  {\em Selecta Math. (N.S.)} {\bf 26} (2020), no.~5 Paper No. 73,
  59, [\href{http://arxiv.org/abs/1908.03887}{{\tt arXiv:1908.03887}}].

\bibitem{Calaque:2020ell}
D.~Calaque and M.~Gonzalez, {\it Ellipsitomic associators},  {\em M\'em. Soc.
  Math. Fr. (N.S.)} (2023), no.~179 x+79,
  [\href{http://arxiv.org/abs/2004.07271}{{\tt arXiv:2004.07271}}].

\bibitem{Hopper:2021ell}
E.~Hopper, {\it The universal elliptic {KZB} connection in higher level},  {\em
  Michigan Math. J.} {\bf 74} (2024), no.~4 863--896,
  [\href{http://arxiv.org/abs/2107.14320}{{\tt arXiv:2107.14320}}].

\bibitem{Hopper:2022ell}
E.~Hopper, {\it {Polylogarithm Variations and Motivic Extensions of $\mathbb Q$
  by $\mathbb Q(m)$}},  \href{http://arxiv.org/abs/2208.01153}{{\tt
  arXiv:2208.01153}}.

\bibitem{Hain:1998}
R.~M. Hain, {\it The {H}odge de {R}ham theory of relative {M}alcev completion},
   {\em Ann. Sci. \'Ecole Norm. Sup. (4)} {\bf 31} (1998), no.~1 47--92.

\bibitem{Hain:2014}
R.~Hain, {\it The {H}odge--de {R}ham theory of modular groups},  in {\em Recent
  advances in {H}odge theory}, vol.~427 of {\em London Math. Soc. Lecture Note
  Ser.}, pp.~422--514.
\newblock Cambridge Univ. Press, Cambridge, 2016.

\bibitem{Arkani-Hamed:2010pyv}
N.~Arkani-Hamed, J.~L. Bourjaily, F.~Cachazo, and J.~Trnka, {\it {Local
  Integrals for Planar Scattering Amplitudes}},  {\em JHEP} {\bf 06} (2012)
  125, [\href{http://arxiv.org/abs/1012.6032}{{\tt arXiv:1012.6032}}].

\bibitem{Broedel:2018qkq}
J.~Broedel, C.~Duhr, F.~Dulat, B.~Penante, and L.~Tancredi, {\it {Elliptic
  Feynman integrals and pure functions}},  {\em JHEP} {\bf 01} (2019) 023,
  [\href{http://arxiv.org/abs/1809.10698}{{\tt arXiv:1809.10698}}].

\bibitem{Furusho2000TheMZ}
H.~Furusho, {\it The multiple zeta value algebra and the stable derivation
  algebra},  {\em Publ. Res. Inst. Math. Sci.} {\bf 39} (2003), no.~4 695--720,
  [\href{http://arxiv.org/abs/math/0011261}{{\tt math/0011261}}].

\bibitem{DG:2005}
P.~Deligne and A.~B. Goncharov, {\it Groupes fondamentaux motiviques de {T}ate
  mixte},  {\em Ann. Sci. \'{E}cole Norm. Sup. (4)} {\bf 38} (2005), no.~1
  1--56, [\href{http://arxiv.org/abs/math/0302267}{{\tt math/0302267}}].

\bibitem{Brown:depth3}
F.~Brown, {\it Zeta elements in depth 3 and the fundamental {L}ie algebra of
  the infinitesimal {T}ate curve},  {\em Forum Math. Sigma} {\bf 5} (2017)
  Paper No. e1, 56, [\href{http://arxiv.org/abs/1504.04737}{{\tt
  arXiv:1504.04737}}].

\bibitem{hainmatsumoto2020}
R.~Hain and M.~Matsumoto, {\it Universal mixed elliptic motives},  {\em Journal
  of the Institute of Mathematics of Jussieu} {\bf 19} (2020), no.~3 663--766,
  [\href{http://arxiv.org/abs/1512.03975}{{\tt arXiv:1512.03975}}].

\bibitem{Furusho:2020}
H.~Furusho, {\it Galois action on knots {II}: {P}roalgebraic string links and
  knots},  in {\em Periods in quantum field theory and arithmetic}, vol.~314 of
  {\em Springer Proc. Math. Stat.}, pp.~541--591.
\newblock Springer, Cham, 2020.

\bibitem{Dorigoni:2024iyt}
D.~Dorigoni, M.~Doroudiani, J.~Drewitt, M.~Hidding, A.~Kleinschmidt,
  O.~Schlotterer, L.~Schneps, and B.~Verbeek, {\it {Canonicalizing Zeta
  Generators: Genus Zero and Genus One}},  {\em Commun. Math. Phys.} {\bf 407}
  (2026), no.~1 12, [\href{http://arxiv.org/abs/2406.05099}{{\tt
  arXiv:2406.05099}}].

\bibitem{Gonch:cyclo}
A.~B. Goncharov, {\it {The dihedral Lie algebras and Galois symmetries of
  $\pi_1^{(l)}(\mathbb{P}-(\{0,\infty\}\cup \mu_N))$}},  {\em Duke Mathematical
  Journal} {\bf 110} (2001), no.~3 397 -- 487,
  [\href{http://arxiv.org/abs/math/0009121}{{\tt math/0009121}}].

\bibitem{Glanois:2016}
C.~Glanois, {\em {Periods of the motivic fundamental groupoid of $\mathbb{P}^1
  \setminus \{0,\mu_N,\infty \}$}}.
\newblock PhD thesis, Universit\'e Pierre et Marie Curie, 2016.
\newblock \href{http://arxiv.org/abs/1603.05155}{{\tt arXiv:1603.05155}}.

\bibitem{Hirose:2023}
M.~Hirose, {\it {The cyclotomic Grothendieck-Teichm\"uller group and the
  motivic Galois group}},  \href{http://arxiv.org/abs/2301.04064}{{\tt
  arXiv:2301.04064}}.

\bibitem{Panzer:2016snt}
E.~Panzer and O.~Schnetz, {\it {The Galois coaction on $\phi^4$ periods}},
  {\em Commun. Num. Theor. Phys.} {\bf 11} (2017) 657--705,
  [\href{http://arxiv.org/abs/1603.04289}{{\tt arXiv:1603.04289}}].

\bibitem{Dupont:2024}
C.~Dupont, {\it {An introduction to mixed Tate motives}},
  \href{http://arxiv.org/abs/2404.03770}{{\tt arXiv:2404.03770}}.

\bibitem{mat:progress}
E.~Claasen, C.~Duhr, M.~Doroudiani, A.~Kleinschmidt, F.~Lippert,
  O.~Schlotterer, and Y.~Sohnle, {\em {work in progress}}.

\bibitem{Abreu:2017mtm}
S.~Abreu, R.~Britto, C.~Duhr, and E.~Gardi, {\it {Diagrammatic Hopf algebra of
  cut Feynman integrals: the one-loop case}},  {\em JHEP} {\bf 12} (2017) 090,
  [\href{http://arxiv.org/abs/1704.07931}{{\tt arXiv:1704.07931}}].

\bibitem{Abreu:2021vhb}
S.~Abreu, R.~Britto, C.~Duhr, E.~Gardi, and J.~Matthew, {\it {The diagrammatic
  coaction beyond one loop}},  {\em JHEP} {\bf 10} (2021) 131,
  [\href{http://arxiv.org/abs/2106.01280}{{\tt arXiv:2106.01280}}].

\bibitem{Frost:2023stm}
H.~Frost, M.~Hidding, D.~Kamlesh, C.~Rodriguez, O.~Schlotterer, and B.~Verbeek,
  {\it {Motivic coaction and single-valued map of polylogarithms from zeta
  generators}},  {\em J. Phys. A} {\bf 57} (2024), no.~31 31LT01,
  [\href{http://arxiv.org/abs/2312.00697}{{\tt arXiv:2312.00697}}].

\bibitem{Frost:2025lre}
H.~Frost, M.~Hidding, D.~Kamlesh, C.~Rodriguez, O.~Schlotterer, and B.~Verbeek,
  {\it {Deriving motivic coactions and single-valued maps at genus zero from
  zeta generators}},  {\em Commun. Num. Theor. Phys.} {\bf 20} (2026), no.~2
  391--460, [\href{http://arxiv.org/abs/2503.02096}{{\tt arXiv:2503.02096}}].

\bibitem{Schlotterer:2025qjv}
O.~Schlotterer, Y.~Sohnle, and Y.-X. Tao, {\it {Elliptic modular graph forms,
  equivariant iterated integrals and single-valued elliptic polylogarithms}},
  \href{http://arxiv.org/abs/2511.15883}{{\tt arXiv:2511.15883}}.

\bibitem{Broedel:2018rwm}
J.~Broedel, C.~Duhr, F.~Dulat, B.~Penante, and L.~Tancredi, {\it {From modular
  forms to differential equations for Feynman integrals}},  in {\em {KMPB
  Conference}: {Elliptic Integrals, Elliptic Functions and Modular Forms in
  Quantum Field Theory}}, pp.~107--131, 2019.
\newblock \href{http://arxiv.org/abs/1807.00842}{{\tt arXiv:1807.00842}}.

\bibitem{Adams:2018yfj}
L.~Adams and S.~Weinzierl, {\it {The $\varepsilon$-form of the differential
  equations for Feynman integrals in the elliptic case}},  {\em Phys. Lett. B}
  {\bf 781} (2018) 270--278, [\href{http://arxiv.org/abs/1802.05020}{{\tt
  arXiv:1802.05020}}].

\bibitem{Drewitt:2023con}
J.~Drewitt and J.~Pimm, {\it {Real-analytic modular forms for $\Gamma_0(N)$ and
  their $L$-series}},  {\em Journal of Number Theory} {\bf 271} (2025) 1--32,
  [\href{http://arxiv.org/abs/2310.01127}{{\tt arXiv:2310.01127}}].

\bibitem{Duhr:2025lvr}
C.~Duhr and F.~Lippert, {\it {Equivariant primitives of Eisenstein series for
  congruence subgroups}},  {\em Commun. Num. Theor. Phys.} {\bf 20} (2026),
  no.~1 193--228, [\href{http://arxiv.org/abs/2502.04752}{{\tt
  arXiv:2502.04752}}].

\bibitem{lv2:progress}
E.~Claasen, C.~Duhr, M.~Doroudiani, A.~Kleinschmidt, F.~Lippert,
  O.~Schlotterer, and Y.~Sohnle, {\em {Equivariant iterated Eisenstein
  integrals for congruence subgroups of level 2}}.

\bibitem{Kleinschmidt:2025dtk}
A.~Kleinschmidt, F.~Porkert, and O.~Schlotterer, {\it {Towards motivic
  coactions at genus one from zeta generators}},  {\em JHEP} {\bf 05} (2026)
  105, [\href{http://arxiv.org/abs/2508.02800}{{\tt arXiv:2508.02800}}].

\bibitem{Drummond:2013vz}
J.~M. Drummond and E.~Ragoucy, {\it {Superstring amplitudes and the
  associator}},  {\em JHEP} {\bf 08} (2013) 135,
  [\href{http://arxiv.org/abs/1301.0794}{{\tt arXiv:1301.0794}}].

\bibitem{Abreu:2017enx}
S.~Abreu, R.~Britto, C.~Duhr, and E.~Gardi, {\it {Algebraic Structure of Cut
  Feynman Integrals and the Diagrammatic Coaction}},  {\em Phys. Rev. Lett.}
  {\bf 119} (2017), no.~5 051601, [\href{http://arxiv.org/abs/1703.05064}{{\tt
  arXiv:1703.05064}}].

\bibitem{Abreu:2019wzk}
S.~Abreu, R.~Britto, C.~Duhr, E.~Gardi, and J.~Matthew, {\it {From positive
  geometries to a coaction on hypergeometric functions}},  {\em JHEP} {\bf 02}
  (2020) 122, [\href{http://arxiv.org/abs/1910.08358}{{\tt arXiv:1910.08358}}].

\bibitem{Britto:2021prf}
R.~Britto, S.~Mizera, C.~Rodriguez, and O.~Schlotterer, {\it {Coaction and
  double-copy properties of configuration-space integrals at genus zero}},
  {\em JHEP} {\bf 05} (2021) 053, [\href{http://arxiv.org/abs/2102.06206}{{\tt
  arXiv:2102.06206}}].

\bibitem{Kamlesh:2024cvo}
D.~Kamlesh, {\em {Motivic coaction on generalized hypergeometric functions}}.
\newblock PhD thesis, Oxford University, 2024.

\bibitem{Mellin:progress}
F.~Brown, C.~Dupont, J.~Fr\'esan, and M.~Tapu{\v{s}}kovi{\'c}, {\em {work in
  progress}}.

\bibitem{Brown:2019jng}
F.~Brown and C.~Dupont, {\it {Lauricella Hypergeometric Functions, Unipotent
  Fundamental Groups of the Punctured Riemann Sphere, and Their Motivic
  Coactions}},  {\em Nagoya Math. J.} {\bf 249} (2023) 148--220,
  [\href{http://arxiv.org/abs/1907.06603}{{\tt arXiv:1907.06603}}].

\bibitem{Broedel:2018iwv}
J.~Broedel, C.~Duhr, F.~Dulat, B.~Penante, and L.~Tancredi, {\it {Elliptic
  symbol calculus: from elliptic polylogarithms to iterated integrals of
  Eisenstein series}},  {\em JHEP} {\bf 08} (2018) 014,
  [\href{http://arxiv.org/abs/1803.10256}{{\tt arXiv:1803.10256}}].

\bibitem{Wilhelm:2022wow}
M.~Wilhelm and C.~Zhang, {\it {Symbology for elliptic multiple polylogarithms
  and the symbol prime}},  {\em JHEP} {\bf 01} (2023) 089,
  [\href{http://arxiv.org/abs/2206.08378}{{\tt arXiv:2206.08378}}].

\bibitem{Forum:2022lpz}
A.~Forum and M.~von Hippel, {\it {A symbol and coaction for higher-loop sunrise
  integrals}},  {\em SciPost Phys. Core} {\bf 6} (2023) 050,
  [\href{http://arxiv.org/abs/2209.03922}{{\tt arXiv:2209.03922}}].

\bibitem{Schnetz:2017bko}
O.~Schnetz, {\it {The Galois coaction on the electron anomalous magnetic
  moment}},  {\em Commun. Num. Theor. Phys.} {\bf 12} (2018) 335--354,
  [\href{http://arxiv.org/abs/1711.05118}{{\tt arXiv:1711.05118}}].

\bibitem{Abreu:2014cla}
S.~Abreu, R.~Britto, C.~Duhr, and E.~Gardi, {\it {From multiple unitarity cuts
  to the coproduct of Feynman integrals}},  {\em JHEP} {\bf 10} (2014) 125,
  [\href{http://arxiv.org/abs/1401.3546}{{\tt arXiv:1401.3546}}].

\bibitem{Abreu:2015zaa}
S.~Abreu, R.~Britto, and H.~Gr{\"o}nqvist, {\it {Cuts and coproducts of massive
  triangle diagrams}},  {\em JHEP} {\bf 07} (2015) 111,
  [\href{http://arxiv.org/abs/1504.00206}{{\tt arXiv:1504.00206}}].

\bibitem{Tapuskovic:2019cpr}
M.~Tapu{\v{s}}kovi{\'c}, {\it {Motivic Galois coaction and one-loop Feynman
  graphs}},  {\em Commun. Num. Theor. Phys.} {\bf 15} (2021), no.~2 221--278,
  [\href{http://arxiv.org/abs/1911.01540}{{\tt arXiv:1911.01540}}].

\bibitem{Mounoud:2026kgu}
U.~Mounoud, {\it {Motivic Galois theory for one-loop Feynman integrals in
  momentum space}},  \href{http://arxiv.org/abs/2605.20106}{{\tt
  arXiv:2605.20106}}.

\bibitem{McLeod:2026jpz}
A.~McLeod, A.~Pokraka, and L.~Ren, {\it {A Graphical Coaction for FRW
  Wavefunction Coefficients}},  \href{http://arxiv.org/abs/2603.25703}{{\tt
  arXiv:2603.25703}}.

\bibitem{McLeod:2026kpo}
A.~J. McLeod, A.~Pokraka, and L.~Ren, {\it {A Graphical Coaction for FRW
  Integrals from Partial/Relative Twisted (Co)homology}},
  \href{http://arxiv.org/abs/2606.13627}{{\tt arXiv:2606.13627}}.

\bibitem{Kreimer:2021jni}
D.~Kreimer and K.~Yeats, {\it {Algebraic interplay between renormalization and
  monodromy}},  {\em Adv. Theor. Math. Phys.} {\bf 27} (2023), no.~1 87--191,
  [\href{http://arxiv.org/abs/2105.05948}{{\tt arXiv:2105.05948}}].

\bibitem{Anastasiou:2002yz}
C.~Anastasiou and K.~Melnikov, {\it {Higgs boson production at hadron colliders
  in NNLO QCD}},  {\em Nucl. Phys. B} {\bf 646} (2002) 220--256,
  [\href{http://arxiv.org/abs/hep-ph/0207004}{{\tt hep-ph/0207004}}].

\bibitem{Anastasiou:2003yy}
C.~Anastasiou, L.~J. Dixon, K.~Melnikov, and F.~Petriello, {\it {Dilepton
  rapidity distribution in the Drell-Yan process at NNLO in QCD}},  {\em Phys.
  Rev. Lett.} {\bf 91} (2003) 182002,
  [\href{http://arxiv.org/abs/hep-ph/0306192}{{\tt hep-ph/0306192}}].

\bibitem{Hidding:2021xot}
M.~Hidding, {\em {Computational and mathematical aspects of Feynman
  integrals}}.
\newblock PhD thesis, Trinity Coll., Dublin, 2021.

\bibitem{Tapuskovic:2023xiu}
M.~Tapu{\v{s}}kovi{\'c}, {\it {The cosmic Galois group, the sunrise Feynman
  integral, and the relative completion of $\Gamma^1(6)$}},  {\em Commun. Num.
  Theor. Phys.} {\bf 18} (2024), no.~2 261--326,
  [\href{http://arxiv.org/abs/2303.17534}{{\tt arXiv:2303.17534}}].

\bibitem{Moch:2001zr}
S.~Moch, P.~Uwer, and S.~Weinzierl, {\it {Nested sums, expansion of
  transcendental functions and multiscale multiloop integrals}},  {\em J. Math.
  Phys.} {\bf 43} (2002) 3363--3386,
  [\href{http://arxiv.org/abs/hep-ph/0110083}{{\tt hep-ph/0110083}}].

\bibitem{Mimachi:2002gi}
K.~Mimachi and M.~Yoshida, {\it {Intersection numbers of twisted cycles and the
  correlation functions of the conformal field theory. 2.}},  {\em Commun.
  Math. Phys.} {\bf 234} (2003) 339--358,
  [\href{http://arxiv.org/abs/math/0208097}{{\tt math/0208097}}].

\bibitem{Huber:2007dx}
T.~Huber and D.~Ma{\^\i}tre, {\it {HypExp 2, Expanding hypergeometric functions
  about half-integer parameters}},  {\em Comput. Phys. Commun.} {\bf 178}
  (2008) 755--776, [\href{http://arxiv.org/abs/0708.2443}{{\tt
  arXiv:0708.2443}}].

\bibitem{Kalmykov:2008ofy}
M.~Y. Kalmykov, V.~V. Bytev, B.~A. Kniehl, B.~F.~L. Ward, and S.~A. Yost, {\it
  {Feynman Diagrams, Differential Reduction, and Hypergeometric Functions}},
  {\em PoS} {\bf ACAT08} (2008) 125,
  [\href{http://arxiv.org/abs/0901.4716}{{\tt arXiv:0901.4716}}].

\bibitem{Derkachov:2018rot}
S.~Derkachov, V.~Kazakov, and E.~Olivucci, {\it {Basso-Dixon Correlators in
  Two-Dimensional Fishnet CFT}},  {\em JHEP} {\bf 04} (2019) 032,
  [\href{http://arxiv.org/abs/1811.10623}{{\tt arXiv:1811.10623}}].

\bibitem{Derkachev:2020xib}
S.~E. Derkachev, A.~V. Ivanov, and L.~A. Shumilov, {\it
  {Mellin{\textendash}Barnes Transformation for Two-Loop Master-Diagram}},
  {\em Zap. Nauchn. Semin.} {\bf 494} (2020), no.~0 144--167,
  [\href{http://arxiv.org/abs/2303.09203}{{\tt arXiv:2303.09203}}].

\bibitem{Blumlein:2021hbq}
J.~Bl{\"u}mlein, M.~Saragnese, and C.~Schneider, {\it {Hypergeometric
  structures in Feynman integrals}},  {\em Ann. Math. Artif. Intell.} {\bf 91}
  (2023), no.~5 591--649, [\href{http://arxiv.org/abs/2111.15501}{{\tt
  arXiv:2111.15501}}].

\bibitem{Duhr:2023bku}
C.~Duhr and F.~Porkert, {\it {Feynman integrals in two dimensions and
  single-valued hypergeometric functions}},  {\em JHEP} {\bf 02} (2024) 179,
  [\href{http://arxiv.org/abs/2309.12772}{{\tt arXiv:2309.12772}}].

\bibitem{Porkert:2025gjd}
F.~M. Porkert, {\em {Novel Mathematical Techniques for Scattering Amplitudes}}.
\newblock PhD thesis, U. Bonn (main), 2025.

\bibitem{Sohnle:2026ivr}
Y.~Sohnle, {\em {Feynman integrals and modular forms}}.
\newblock PhD thesis, Uppsala University, 2026.

\bibitem{Virasoro:1969me}
M.~A. Virasoro, {\it {Alternative constructions of crossing-symmetric
  amplitudes with regge behavior}},  {\em Phys. Rev.} {\bf 177} (1969)
  2309--2311.

\bibitem{tHooft:1978jhc}
G.~'t~Hooft and M.~J.~G. Veltman, {\it {Scalar One Loop Integrals}},  {\em
  Nucl. Phys. B} {\bf 153} (1979) 365--401.

\bibitem{Levine}
M.~Levine, {\it Tate motives and the vanishing conjectures for algebraic
  {$K$}-theory},  in {\em Algebraic {$K$}-theory and algebraic topology ({L}ake
  {L}ouise, {AB}, 1991)}, vol.~407 of {\em NATO Adv. Sci. Inst. Ser. C: Math.
  Phys. Sci.}, pp.~167--188.
\newblock Kluwer Acad. Publ., Dordrecht, 1993.

\bibitem{intersection}
K.~Cho and K.~Matsumoto, {\it {Intersection theory for twisted cohomologies and
  twisted Riemann's period relations I}},  {\em Nagoya Math. J.} {\bf 139}
  (1995) 67--86.

\bibitem{Mizera:2019gea}
S.~Mizera, {\em {Aspects of Scattering Amplitudes and Moduli Space
  Localization}}.
\newblock PhD thesis, Princeton, Inst. Advanced Study, 2020.
\newblock \href{http://arxiv.org/abs/1906.02099}{{\tt arXiv:1906.02099}}.

\bibitem{Zagier:123}
J.~H. Bruinier, G.~van~der Geer, G.~Harder, and D.~Zagier, {\em The 1-2-3 of
  modular forms}.
\newblock Universitext. Springer-Verlag, Berlin, 2008.
\newblock Lectures from the Summer School on Modular Forms and their
  Applications held in Nordfjordeid, June 2004.

\bibitem{Pollack}
A.~Pollack, ``{Relations between derivations arising from modular forms}.''
  \url{https://dukespace.lib.duke.edu/dspace/handle/10161/1281}, 2009.
\newblock Undergraduate thesis, Duke University.

\end{thebibliography}
\end{document}